\documentclass[twocolumn]{aastex62}
\usepackage{xspace}
\usepackage{amsmath}

\newcommand{\kms}{km~s$^{-1}$\xspace}
\newcommand{\ms}{m~s$^{-1}$\xspace}

\newcommand{\msun}{$M_\odot$\xspace}

\newcommand{\mjup}{$M_\mathrm{Jup}$\xspace}
\newcommand{\rjup}{$R_\mathrm{Jup}$\xspace}
\newcommand{\mearth}{$M_\oplus$\xspace}
\newcommand{\rearth}{$R_\oplus$\xspace}

\received{February 25, 2019}
\revised{May 16, 2019 \& June 6, 2019}
\submitjournal{AAS Journals}

\shorttitle{V1298 Tau b}
\shortauthors{David et al.}

\begin{document}

\title{A WARM JUPITER-SIZED PLANET TRANSITING THE PRE-MAIN SEQUENCE STAR V1298 TAU}

\correspondingauthor{Trevor J.\ David}
\email{trevor.j.david@jpl.nasa.gov}

\author[0000-0001-6534-6246]{Trevor J.\ David}
\affil{Jet Propulsion Laboratory, California Institute of Technology, 4800 Oak Grove Drive, Pasadena, CA 91109, USA}

\author[0000-0002-3656-6706]{Ann Marie Cody}
\affil{NASA Ames Research Center, Moffett Field, CA 94035, USA}

\author[0000-0002-3385-8391]{Christina L. Hedges}
\affil{Bay Area Environmental Research Institute, P.O. Box 25, Moffett Field, CA 94035, USA}

\author[0000-0003-2008-1488]{Eric E.\ Mamajek}
\affil{Jet Propulsion Laboratory, California Institute of Technology, 4800 Oak Grove Drive, Pasadena, CA 91109, USA}
\affil{Department of Physics \& Astronomy, University of Rochester, Rochester, NY 14627, USA}

\author{Lynne A.\ Hillenbrand}
\affil{Department of Astronomy, California Institute of Technology, Pasadena, CA 91125, USA}

\author[0000-0002-5741-3047]{David R.\ Ciardi}
\affil{Caltech/IPAC-NASA Exoplanet Science Institute, Pasadena, CA 91125, USA}

\author[0000-0002-5627-5471]{Charles A. Beichman}
\affil{Jet Propulsion Laboratory, California Institute of Technology, 4800 Oak Grove Drive, Pasadena, CA 91109, USA}
\affil{Caltech/IPAC-NASA Exoplanet Science Institute, Pasadena, CA 91125, USA}

\author[0000-0003-0967-2893]{Erik A.\ Petigura}
\altaffiliation{NASA Hubble Fellow}
\affil{Department of Astronomy, California Institute of Technology, Pasadena, CA 91125, USA}

\author[0000-0003-3504-5316]{Benjamin J.\ Fulton}
\affil{Caltech/IPAC-NASA Exoplanet Science Institute, Pasadena, CA 91125, USA}

\author[0000-0002-0531-1073]{Howard T.\ Isaacson}
\affil{Astronomy Department, University of California, Berkeley, CA 94720, USA}

\author[0000-0001-8638-0320]{Andrew W.\ Howard}
\affil{Department of Astronomy, California Institute of Technology, Pasadena, CA 91125, USA}

\author[0000-0002-2592-9612]{Jonathan Gagn{\'e}}
\altaffiliation{Banting Fellow}
\affil{Institut de Recherche sur les Exoplan{\`e}tes, D{\'e}partement de Physique, Universit{\'e} de Montr{\'e}al, Montr{\'e}al QC, H3C 3J7, Canada}

\author{Nicholas K. Saunders}
\affil{NASA Ames Research Center, Moffett Field, CA 94035, USA}

\author[0000-0001-6381-515X]{Luisa M.\ Rebull}
\affil{Caltech/IPAC-IRSA, Pasadena, CA 91125, USA}

\author[0000-0003-3595-7382]{John R.\ Stauffer}
\affil{Caltech/IPAC-SSC, Pasadena, CA 91125, USA}

\author[0000-0002-1871-6264]{Gautam Vasisht}
\affil{Jet Propulsion Laboratory, California Institute of Technology, 4800 Oak Grove Drive, Pasadena, CA 91109, USA}

\author[0000-0001-8074-2562]{Sasha Hinkley}
\affil{University of Exeter, Physics Department, Stocker Road, Exeter EX4 4QL, UK}

\begin{abstract}
We report the detection of V1298 Tau b, a warm Jupiter-sized planet ($R_P$ = 0.91 $\pm$ 0.05~ $R_\mathrm{Jup}$, $P = 24.1$ days) transiting a young solar analog with an estimated age of 23 million years. The star and its planet belong to Group 29, a young association in the foreground of the Taurus-Auriga star-forming region. While hot Jupiters have been previously reported around young stars, those planets are non-transiting and near-term atmospheric characterization is not feasible. The V1298 Tau system is a compelling target for follow-up study through transmission spectroscopy and Doppler tomography owing to the transit depth (0.5\%), host star brightness ($K_s$ = 8.1 mag), and rapid stellar rotation ($v\sin{i}$ = 23 \kms). Although the planet is Jupiter-sized, its mass is presently unknown due to high-amplitude radial velocity jitter. Nevertheless, V1298 Tau b may help constrain formation scenarios for at least one class of close-in exoplanets, providing a window into the nascent evolution of planetary interiors and atmospheres.
\end{abstract}


\section{Introduction} \label{sec:intro}
The properties and demographics of young exoplanets may potentially reveal valuable insights into the physics of planet formation. For example, theoretical models predict that the radii of super-Earths and mini-Neptunes were significantly larger at ages $\lesssim$100~Myr \citep{OwenWu2013}. The valley observed in the radius distribution of small planets \citep{Fulton2017, Fulton:Petigura:2018} is then expected to form gradually through some combination of photo-evaporation \citep{OwenWu2013, LopezFortney2013}, planetary impacts \citep{Schlichting2015}, and core-envelope interactions \citep{Ginzburg2018, GuptaSchlichting2018}. All of the aforementioned processes result in net atmospheric loss, but it is not clear which mechanism dominates. 

For giant planets, the post-formation radii, temperatures, and luminosities may help to constrain how planets accrete gaseous envelopes and perhaps whether or not they possess rocky cores \citep[e.g.][]{Burrows1997, Baraffe2003, Marley2007, Fortney2007, Fortney2010, SpiegelBurrows2012, Mordasini2013, Mordasini2017}. 

Discerning which physical mechanisms are important to the evolution of exoplanets, and which are not, requires studying planetary systems of a variety of ages. For this reason, the task of identifying exoplanets in stellar associations with well-defined ages has attracted significant attention lately \citep{Ciardi2018, Curtis2018, Libralato2016, Livingston2018, Livingston2019, Mann2016a, Mann2017, Mann2018, Obermeier2016, Pepper2017, Quinn2012, Quinn2014, Rizzuto2018, Vanderburg2018}. 

At the youngest ages ($<$100 Myr), when observed exoplanet properties might place the strongest constraints on formation theories, there are only a small number of known exoplanets; several young giant planets have been discovered via direct imaging \citep[see][for a review]{Bowler2016} while there are only five validated or candidate close-in exoplanets with securely measured ages. Three of the close-in planets are hot Jupiters detected through radial velocity monitoring of T Tauri stars \citep{Donati2016, JohnsKrull2016, Yu2017}, one is a candidate transiting hot Jupiter of controversial nature \citep{vanEyken2012, Yu2015}, and the other is a Neptune-sized planet transiting a pre-main sequence star \citep{David2016, Mann2016}.

\begin{figure*}
    \centering
    \includegraphics[width=\textwidth]{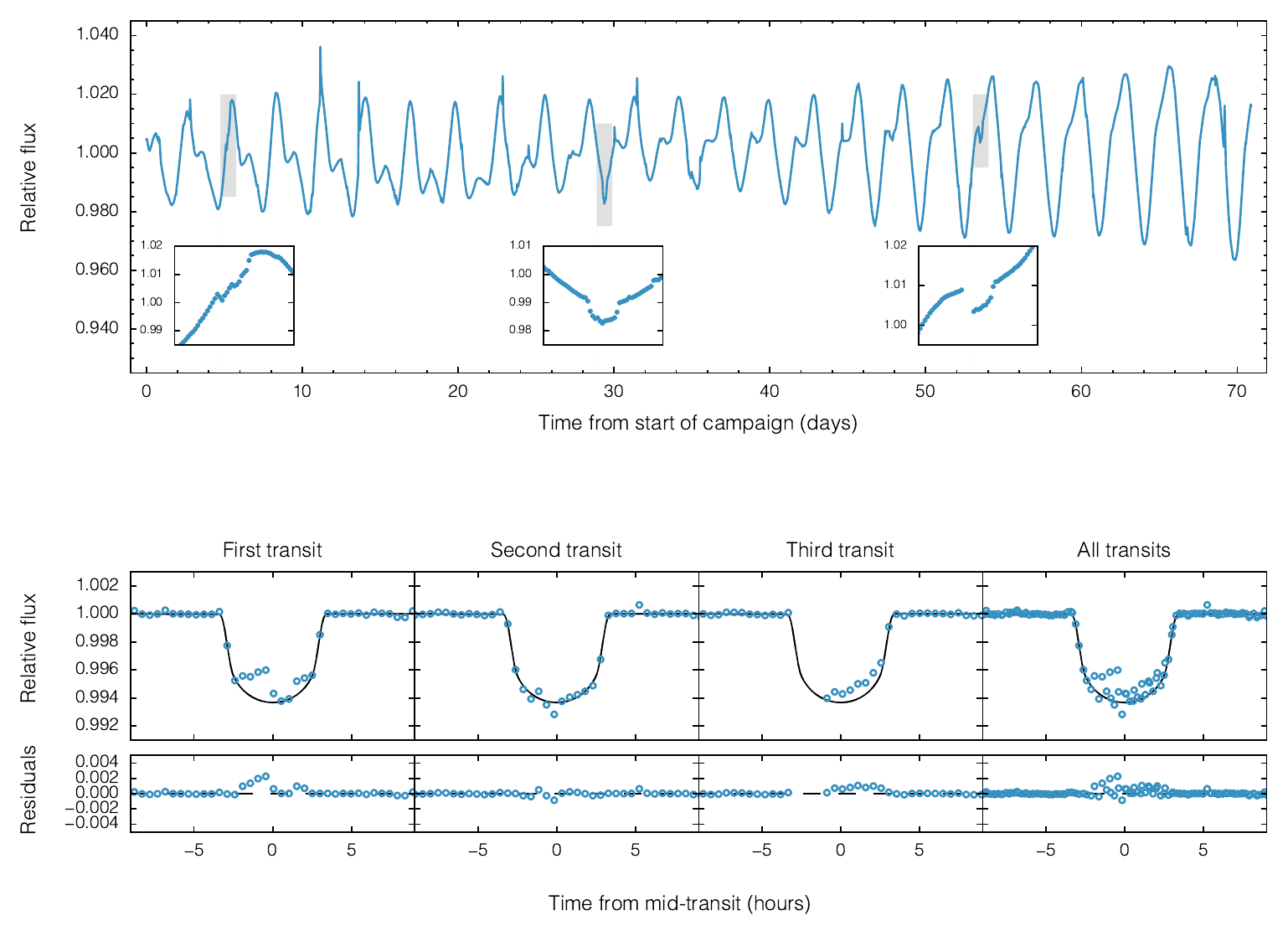}
    \caption{K2 time series photometry of V1298 Tau. \textit{Top:} Systematics-corrected \textit{K2} photometry of V1298 Tau from the \textsc{everest 2.0} pipeline. Inset panels show day-long windows around individual transits, also indicated by the grey shaded regions. \textit{Bottom:} Folded and individual transits of V1298 Tau b (points) with a transit model fit (solid line) found from excluding anomalous observations. Residuals of this fit (Fit 1, as described in \S~\ref{subsec:transitfit}) are shown below each panel. Observations missing from the ingress of the third transit are due to an interruption in data acquisition which affected all targets in Campaign 4.}
    \label{fig:lightcurve}
\end{figure*}

NASA's \textit{Kepler Space Telescope} observed thousands of young stars during its extended \textit{K2} mission \citep{Howell2014}. For most of these stars, \textit{K2} provided time series photometry with unprecedented precision, cadence, and duration. Using astrometry from the second data release of the \textit{Gaia} mission \citep{Gaia2018} and a Bayesian membership algorithm \citep{Gagne2018}, we searched the entire \textit{K2} source catalog for high-probability members of young moving groups and stellar associations. Our search yielded 432 candidate members of the Taurus-Auriga star-forming region, including V1298 Tau. In an automated search for periodic transits or eclipses within the the \textit{K2} photometry for these candidate members we detected 0.5\% dimming events which last 6.4 hours and repeat every 24.1 days within the light curve of V1298 Tau (Figure~\ref{fig:lightcurve}). Follow-on observations confirm the periodic transits are most probably due to a Jovian-sized planet orbiting at a separation of 0.17~AU from V1298 Tau. 

In \S\ref{sec:observations} we describe the \textit{K2} time series photometry from which the planet V1298 Tau b was detected and the follow-on observations used to validate the planet. Our analysis of these observations and assessment of false-positive scenarios is described in \S\ref{sec:analysis}. We place the planet V1298 Tau b in context in \S\ref{sec:discussion}, and present our conclusions in \S\ref{sec:conclusions}. 

\section{Observations \\and Characterization of the Data} \label{sec:observations}
\subsection{\textit{K2} time series photometry} \label{photometry}
The \textit{Kepler} space telescope observed V1298 Tau during Campaign 4 of the \textit{K2} mission \citep{Howell2014}, between 2015 February 7 and 2015 April 23 UTC. The star was included in the following Guest Observer programs: GO4020 (Stello), GO4057 (Pr\v{s}a), GO4060 (Coughlin), GO4090 (Caldwell), GO4092 (Brown), and GO4104 (Dragomir). Telescope pointing drift combined with intra-pixel detector sensitivity variations imprint systematic artifacts upon \textit{K2} time series photometry. We corrected for these systematic effects using the \textsc{everest 2.0} pipeline \citep{Luger2018}, which employs a variant of the pixel-level decorrelation (PLD) method \citep{Deming2013}. We also confirmed the transits are present in the raw uncorrected photometry using the interactive \textsc{lightkurve} tool \citep{lightkurve}, as well as in independent reductions of the data using the \textsc{k2sc} \citep{Aigrain2016} and \textsc{k2phot} \citep{Petigura2018} pipelines. 

The \textit{K2} time series photometry is characterized by quasi-periodic variations with a peak-to-trough amplitude of $\sim$6\% and a period of 2.851~$\pm$~0.050 days (\S\ref{subsec:activity}), values typical of similarly young stars \citep{Rebull2018}. We attribute this variability to stellar rotation and a non-axisymmetric distribution of spots on the stellar surface. The photometry also reveal several flares over the 70.8 day observing baseline, confirming that the star is magnetically active.  Prior to fitting transit models to the photometry, we removed the intrinsic stellar variability through Gaussian process (GP) regression. For the GP we used a Mat{\'e}rn-3/2 kernel with an additional white noise term, using a white noise amplitude of 243 ppm, red noise amplitude of 1.215 $\times$ 10$^5$ counts, and red noise timescale of 29.17 days. The transits were masked prior to both the PLD detrending and the final GP regression to prevent under- or over-fitting the transit signal.

\begin{figure*}
    \centering
    \includegraphics[width=\textwidth]{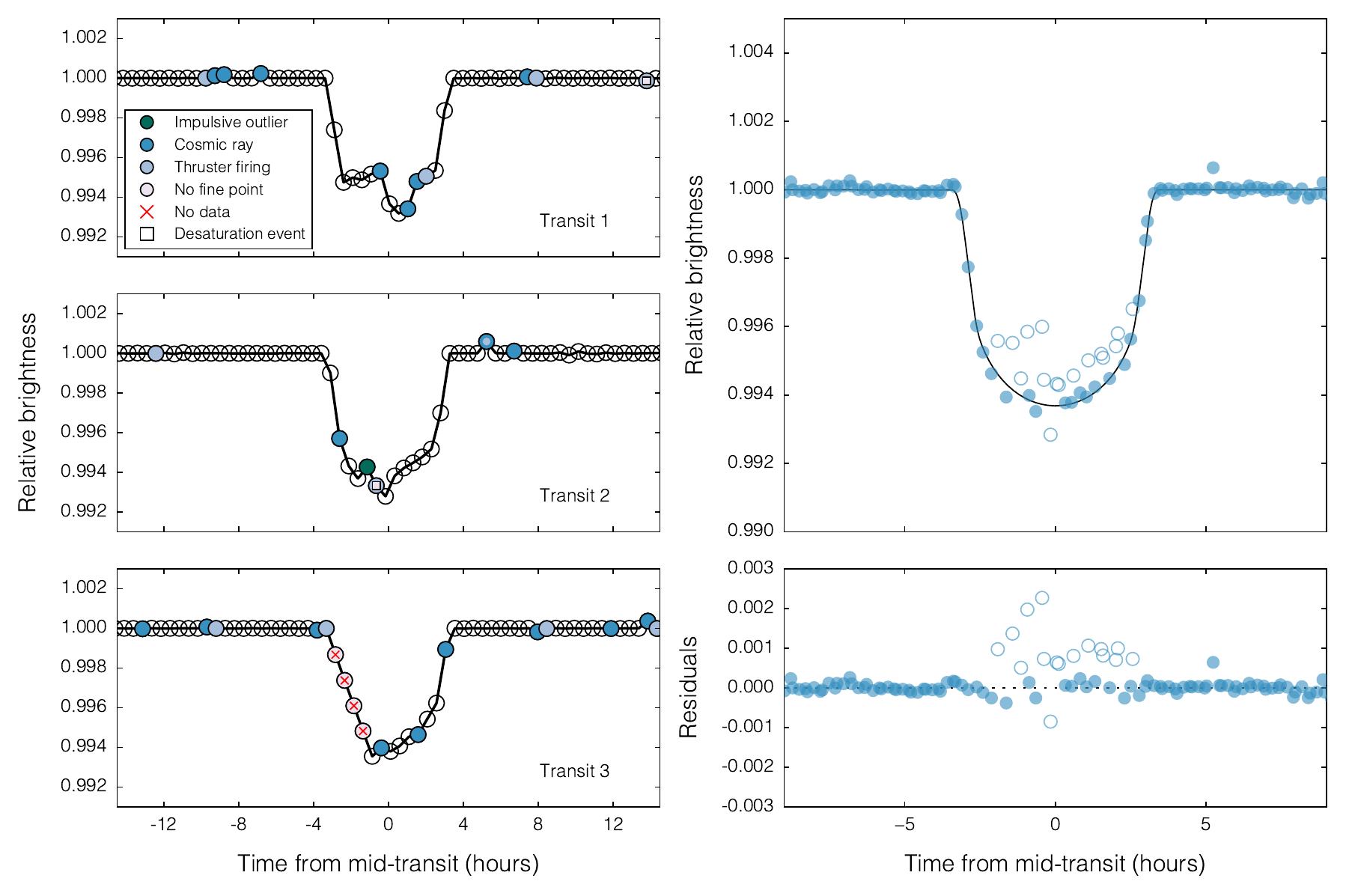}
    \caption{\textit{Left:} Individual transits of V1298 Tau observed with \textit{K2}. The light curve shown is one possible realization from the \textsc{everest 2.0} pipeline. Measurements affected by various telescope systematics are as indicated by the figure legend. Hybrid symbols indicate when multiple quality flags were raised. For example, the pale blue circle with an inset lavender square indicates a thruster firing, no fine point, and a desaturation event. The lavender circles with red x's represent cadences with no fine point and no data. \textit{Right:} Transit fit (black line) with observations included in the final fit shown as filled circles, and excluded observations as open circles (Fit 1 is shown, as described in \S~\ref{subsec:transitfit}).}
    \label{fig:outliers}
\end{figure*}

From inspection of systematics-corrected light curves using both the \textsc{everest 2.0} and \textsc{k2sc} pipelines we find that there are several outlying or missing in-transit measurements (Figure \ref{fig:outliers}). To investigate the potential roles of known systematic effects in causing these outliers, we examined the quality flags\footnote{Quality flag meanings are summarized in Table 2.3 of the \textit{Kepler} archive manual (\url{https://archive.stsci.edu/kepler/manuals/archive_manual.pdf}), discussed in detail in the \textit{Kepler} Data Characteristics Manual (\url{https://archive.stsci.edu/kepler/manuals/Data_Characteristics.pdf}), and may be interpreted with software available at \url{https://gist.github.com/barentsen/}.} for each in-transit observation. There are six unique data quality flags which are raised during the in-transit observations of V1298 Tau. We summarize the flag meanings here.

\begin{itemize}
    \item Impulsive outlier: The \textit{Kepler} PDC pipeline \citep{Stumpe:etal:2012, Smith:etal:2012} identifies outliers before cotrending. These ``impulsive outliers'' may be astrophysical (such as deep transits, or cosmic ray hits) or systematic.
    \item Cosmic ray: This quality flag indicates that a cosmic ray impinged on at least one of the pixels contained in the photometric aperture. Cosmic rays can affect pixels in the short-term (depositing excess charge, which is cleared on the subsequent destructive readout), medium-term (temporarily changing a pixel's sensitivity), or long-term (sometimes changing a pixel's sensitivity permanently).
    \item Thruster firing: During the \textit{K2} mission, spacecraft thrusters were fired every $\sim$6 hours to mitigate pointing drift. This strategy constrains the excursions of a star on the detector, but introduces sawtooth-like systematics in the photometry due to intra-pixel sensitivity variations. Observations acquired during a thruster firing may thus appear as outliers, depending on how the data are de-trended.
    \item No fine point: Occasionally the \textit{Kepler} spacecraft loses fine pointing control. A degradation of pointing precision may be accompanied by increased image motion and thus reduced photometric precision.
    \item No data: As implied, this quality flag indicates the spacecraft was not collecting data.
    \item Desaturation event: When angular momentum builds up in the reaction wheels (due to solar radiation torques), a thruster firing is required to ``desaturate'' the reaction wheels and keep their speeds within operating limits. During the \textit{K2} mission, this quality flag was also used to indicate a ``resaturation event,'' in which thrusters were fired to keep reaction wheels safely away from zero angular velocity. Fine pointing control is not maintained during desaturation/resaturation events, and so the subsequent motion of the star on the detector can lead to a loss in photometric precision. Desaturation/resaturation events can affect a single long cadence observation.
\end{itemize}

For V1298 Tau, most notably, observations of the third transit ingress (cadences 106358 through 106361), are missing due to the fact that the \textit{Kepler} spacecraft was not in fine-pointing mode and did not acquire data. The outlying observations during the first transit, which are present in both the \textsc{everest 2.0} and \textsc{k2sc} light curves, are not as easily explained. Transit profile distortions may be systematic or astrophysical in nature, and possibly the result of both effects. Quality flags from the \textit{K2} photometry indicate that some observations taken during the first transit were affected by cosmic rays or thruster firings (dark blue and pale blue circles, respectively, in Fig.~\ref{fig:outliers}). However, observations closely spaced in time which also appear to be outliers were not affected by such systematics and thruster firings in particular are naturally detrended with the pixel-level decorrelation approach of the \textsc{everest 2.0} pipeline. It is possible that the outliers are instead astrophysical in nature, arising from the planet occulting active regions on the stellar surface. The transit depth is reduced when the planet crosses a dark starspot and increased if the planet crosses a bright plage or flaring region. With only three transits, we can not conclusively attribute any outliers to spot-crossing events, but the light curve of V1298 Tau is indeed suggestive of a non-axisymmetric spotted stellar surface. 

\subsection{High-resolution imaging} \label{subsec:imaging}
We observed V1298~Tau with infrared high-resolution adaptive optics (AO) imaging at Keck Observatory.  The Keck Observatory observations were made with the NIRC2 instrument on Keck-II behind the natural guide star AO system.  The observations were made on 2018~Nov~22 UT in the standard 3-point dither pattern that is used with NIRC2 to avoid the left lower quadrant of the detector which is typically noisier than the other three quadrants. The dither pattern step size was 3$\arcsec$ and was repeated twice, with each dither offset from the previous dither by 0.5$\arcsec$.  

The observations were made in the narrow-band Br-$\gamma$ filter $(\lambda_o = 2.1686; \Delta\lambda = 0.0326\mu$m) with an integration time of 1.25 seconds with one coadd per frame for a total of 11.25 seconds on target and in $J$-cont $(\lambda_o = 1.2132; \Delta\lambda = 0.0198\mu$m) with an integration time of 5 seconds with one coadd per frame for a total of 45 seconds on target.  The camera was in the narrow-angle mode with a full field of view of $\sim$10$\arcsec$ and a pixel scale of approximately 9.9442~mas per pixel. The Keck AO observations show no additional stellar companions were detected to within a resolution $\sim$0.05$\arcsec$ full width at half maximum (FWHM).

The sensitivities of the final combined AO image were determined by injecting simulated sources azimuthally around the primary target every 45$^\circ $ at separations of integer multiples of the central source's FWHM \citep{Furlan2017}. The brightness of each injected source was scaled until standard aperture photometry detected it with $5\sigma$ significance. The resulting brightness of the injected sources relative to the target set the contrast limits at that injection location. The final $5\sigma $ limit at each separation was determined from the average of all of the determined limits at that separation; the uncertainty on the 5$\sigma$ limit was set by the root mean square dispersion of the azimuthal slices at a given radial distance. 

\subsection{Spectroscopic observations and radial velocities.}
We observed V1298 Tau on 9 nights between 3 November 2018 and 26 January 2019 UTC using the High Resolution Spectrograph (HIRES) on the Keck-I telescope \citep{Vogt1994}. We first acquired a spectrum for characterization purposes on 3 November 2018, before commencing precision radial velocity (PRV) monitoring with the iodine cell. Our characterization spectrum has a resolution of $R~\approx~36,000$ between $\sim$4800--9200~\AA. For the PRV observations, a template spectrum with resolution of $R~\approx~80,000$ was obtained for use with the California Planet Survey RV pipeline \citep{Howard2010}. The PRV observations themselves have a resolution of $R~\approx$~50,000 between $\sim$3600--8000~\AA. The PRVs for V1298 Tau are presented in Table~\ref{table:prvs}. From those data we measured the optical radial velocity jitter to be $\sigma_{RV} = $~216, 71, and 5~\ms over 5.2 days, 10 hours, and 30 minutes, respectively. While a measurement of the planet mass is precluded by the large stellar variability, we derive an upper limit in \S\ref{subsec:masslimit}.

For the first epoch of spectroscopic observations, the radial velocity was derived by cross-correlating the spectrum with a radial velocity standard (Table~\ref{table:params}). For this measurement we used the G2 type standard HD 3765 \citep{nidever2002}, which we observed with HIRES on the same night and in the same spectrograph configuration. Uncertainty was quantified from the dispersion among measurements relative to different standards, and over 24 different spectral orders. The sky-projected rotational velocity was also estimated from this spectrum by artificially broadening the spectral standard Gl 651 (SpT = G8) using the standard Gray broadening profile \citep{Gray2005} with $\epsilon= 0.6$. Velocities between 9 and 50~\kms were sampled and the best-fit value for $v\sin{i}$ was determined by minimizing residuals, suggesting $23 \pm 2$~\kms.

\begin{deluxetable*}{lllrl}
\tablecaption{Keck/HIRES precision radial velocities of V1298 Tau \label{table:prvs}}
\tablecolumns{5}
\tablenum{1}
\tablewidth{0pt}
\tablehead{
\colhead{Observation} &
\colhead{UTC Date} &
\colhead{JD} & \colhead{RV} & \colhead{$\sigma_\mathrm{RV}$} \\
\colhead{} & \colhead{} & \colhead{} &
\colhead{(m~s$^{-1}$)} & \colhead{(m~s$^{-1}$)}
}
\startdata
rj310.68  & 2018-11-16 & 2458438.94547 & +459.27 &  6.23 \\
rj311.73  & 2018-11-21 & 2458443.81966 & -233.24 &  4.73 \\
rj311.113 & 2018-11-21 & 2458443.95650 & -67.81 &  3.91 \\
rj311.114 & 2018-11-21 & 2458443.96733 & -56.12 &  4.26 \\
rj311.115 & 2018-11-21 & 2458443.97835 & -64.60 &  4.00 \\
rj311.172 & 2018-11-21 & 2458444.15574 & -37.38 &  5.58 \\
rj314.102 & 2018-12-24 & 2458476.80809 & +147.66 & 8.48 \\
rj315.63  & 2018-12-26 & 2458479.00517 & +14.88  & 7.80 \\
rj316.90  & 2019-01-07 & 2458490.79036 & +191.34 & 8.39 \\
rj316.376 & 2019-01-08 & 2458491.75402 & -101.69 & 7.34 \\
rj317.81  & 2019-01-25 & 2458508.91881 & -136.91 & 8.72 \\
rj317.398 & 2019-01-26 & 2458509.78912 & +127.77 & 8.76 \\
\enddata
\end{deluxetable*}

\section{Analysis} \label{sec:analysis}

\subsection{Association membership and kinematics} \label{subsec:membership}

V1298 Tau was first proposed as a young star and candidate Taurus-Auriga member based on a detection of its X-ray emission from the \textit{ROSAT} All-Sky Survey \citep{WKS1996}. The star's relative youth (age $\lesssim$100 Myr) was then verified on the basis of strong lithium 6708~\AA\ absorption \citep{Wichmann2000}. Taurus-Auriga, at $\sim$3 Myr, contains both classical T Tauri stars in the process of forming planets as well as weak-line T Tauri stars lacking accretion signatures. V1298 Tau, an early K-type star which has no significant infrared excess and exhibits H$\alpha$ in absorption, could fit into the latter category. 

Alternatively, while V1298 Tau is young, it may be significantly older than 3 Myr. The existence of an older and more spatially distributed pre-main sequence population in the vicinity of Taurus has been recognized \citep[e.g.][]{Hartmann1991, Slesnick2006, Kraus2017, Zhang2018}, but a recent kinematic analysis enabled by \textit{Gaia} suggests the older population is physically unrelated to the Taurus-Auriga association \citep{Luhman2018}. Many of the more spatially distributed candidate young stars likely belong to a newly identified, older association preliminarily named Group 29 \citep{Luhman2018}, to which V1298 Tau was proposed to belong \citep{Oh2017}.

The \textit{Gaia} astrometry and barycentric radial velocity of V1298 Tau were used to estimate the star's probability of membership to the Taurus-Auriga (Tau-Aur) star-forming region. Using the BANYAN $\Sigma$ classification tool \citep{Gagne2018}, we found that V1298 Tau has a 99.8\% probability of belonging to Taurus and a 0.2\% probability of being a field star. However, BANYAN $\Sigma$ utilizes an inclusive model for Taurus-Auriga and does not yet take into account the aforementioned analysis of kinematic substructure in the association. 

\begin{figure}
    \centering
    \includegraphics[width=\linewidth]{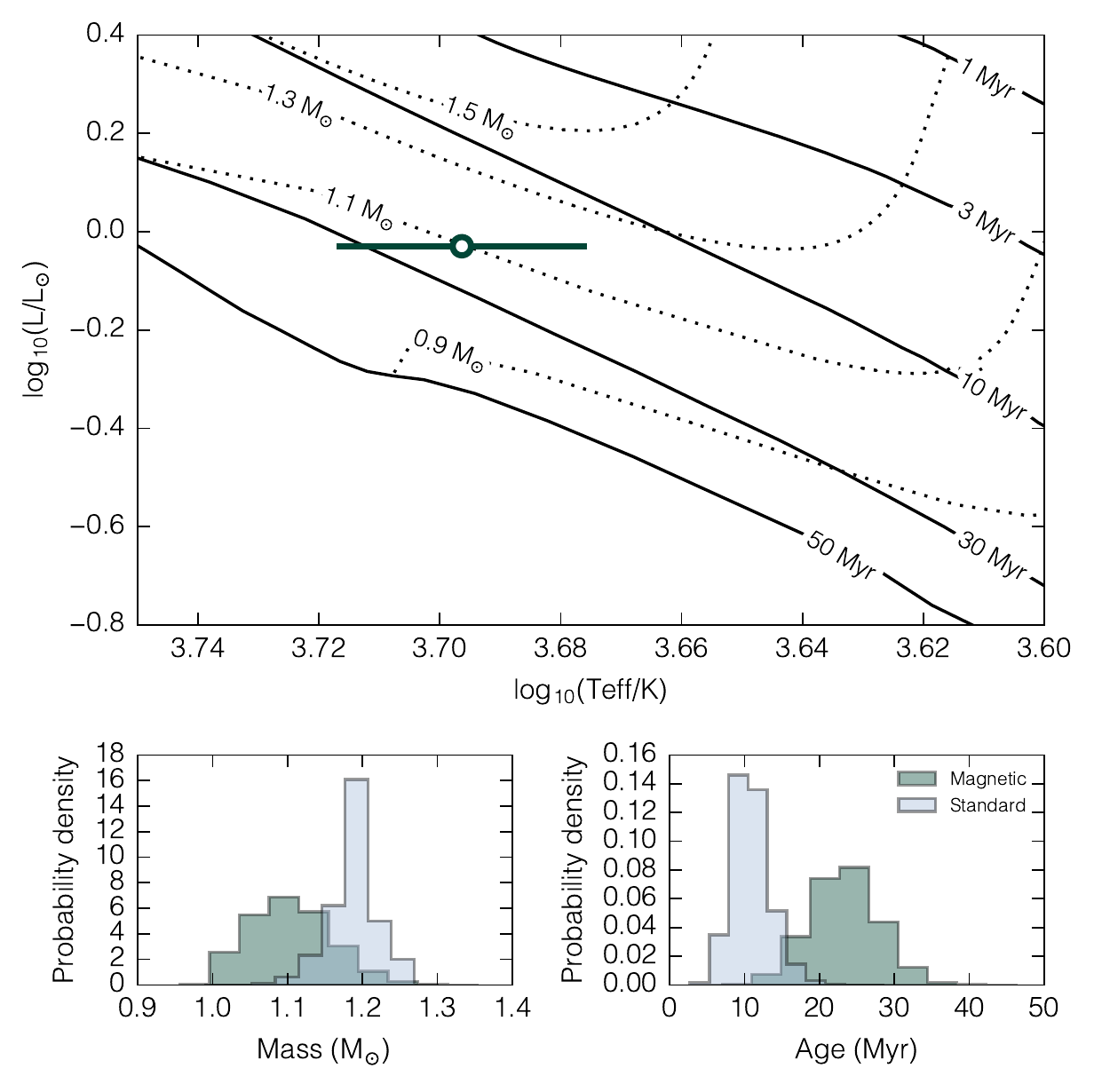}
    \caption{\textit{Top:} V1298 Tau (blue point) in a theoretical H-R diagram. Isochrones (solid lines) and mass tracks (dotted lines) from Dartmouth models including the effects of magnetic fields \citep{Feiden2016} are shown. \textit{Bottom:} Mass and age distributions from Monte Carlo simulations for V1298 Tau according to magnetic (green) and standard \citep[blue,][]{Dotter2008} Dartmouth evolutionary models. Given the precise parallax and near-infrared photometry, the uncertainties in mass and age are dominated by the error in $T_\mathrm{eff}$ and model systematics.}
    \label{fig:hrd}
\end{figure}

\begin{figure*}
    \centering
    \includegraphics[height=6cm]{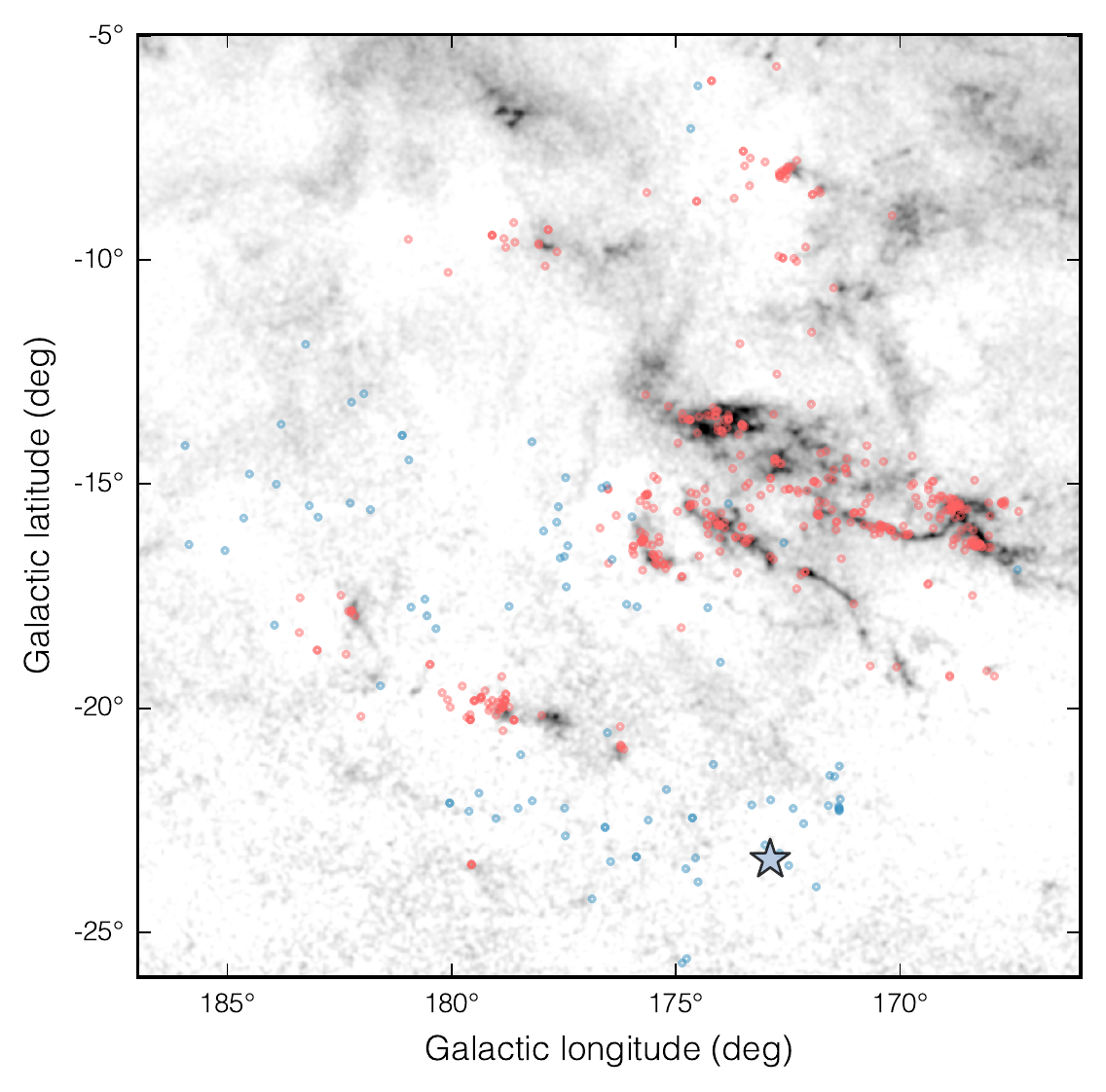}
    \includegraphics[height=6cm]{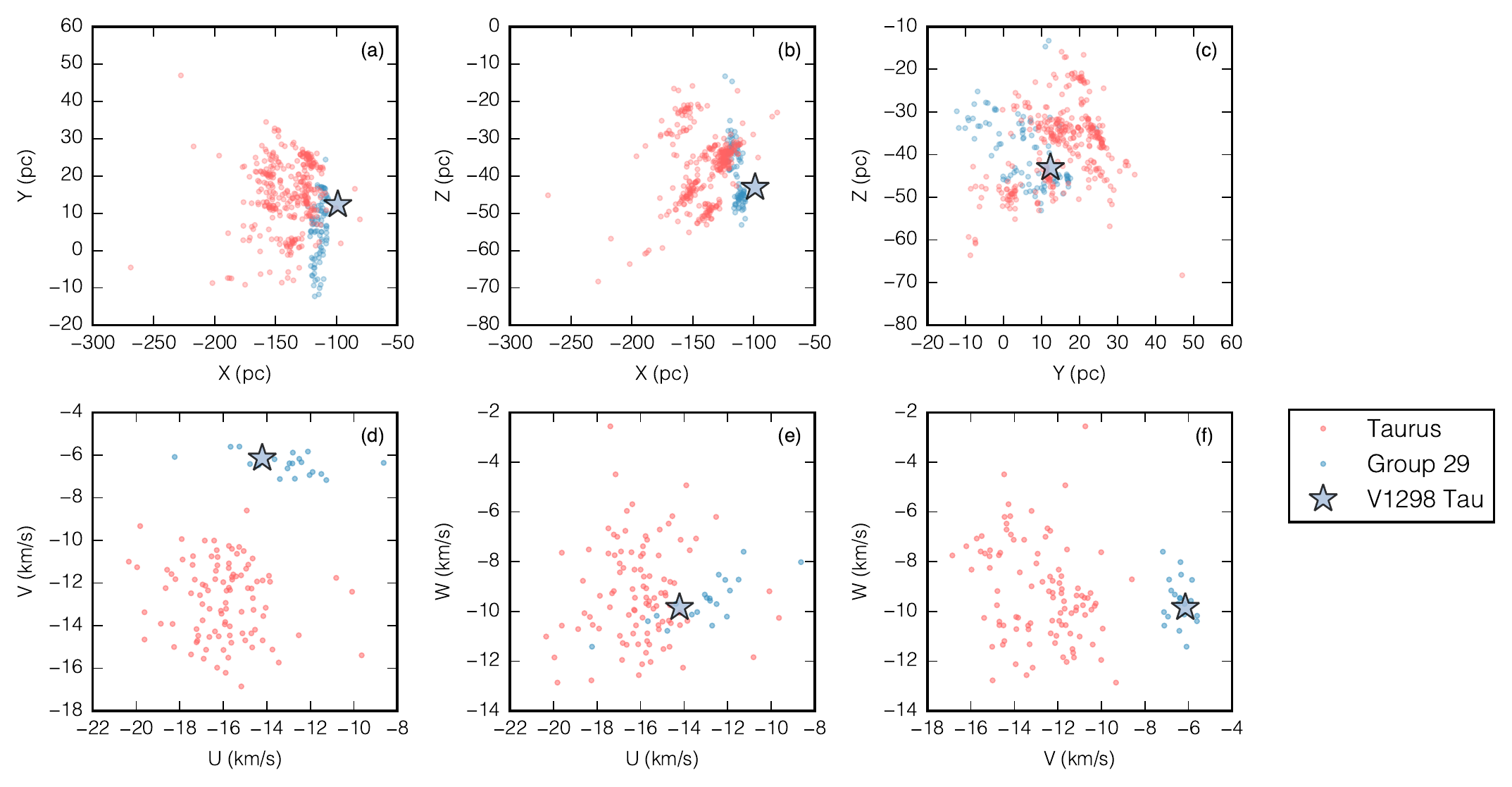}
    \caption{\textit{Left:} Galactic coordinates of V1298 Tau (blue star) relative to Taurus-Auriga members (red circles) and proposed members of Group 29 (blue circles). Stellar positions are overlaid on an extinction map \citep{Dobashi2005}. \textit{Right:} Galactic positions and kinematics of V1298 Tau, Taurus, and Group 29. \textit{(a-c)} The galactic positions of Taurus members and Group 29 candidate members relative to V1298 Tau. The narrow range in $X$ spanned by Group 29 candidates is a result of a stringent parallax cut, and there are likely more unidentified members in the foreground. \textit{(d-f)} Galactic kinematics of Taurus members and Group 29 candidate members relative to V1298 Tau.}
    \label{fig:membership}
\end{figure*}

\begin{figure}
    \centering
    \includegraphics[width=\linewidth]{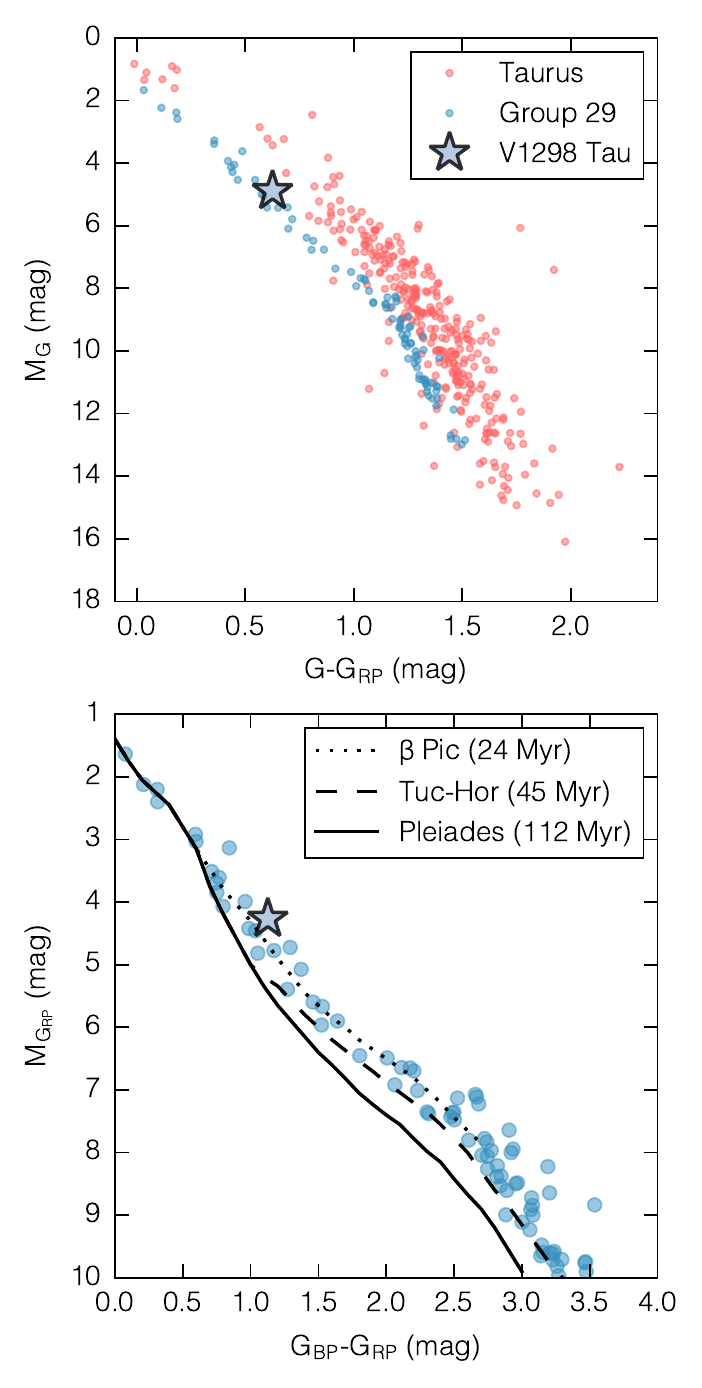}
    \caption{\textit{Top:} V1298 Tau (blue star) compared to Taurus-Auriga members (red circles) and proposed members of Group 29 (blue circles) in a color-absolute magnitude diagram. \textit{Bottom:} V1298 Tau in comparison to empirical fits to the single star sequences of two moving groups and one open cluster \citep{Luhman2018}.}
    \label{fig:camd}
\end{figure}

\begin{figure*}
    \centering
    \includegraphics[width=\textwidth]{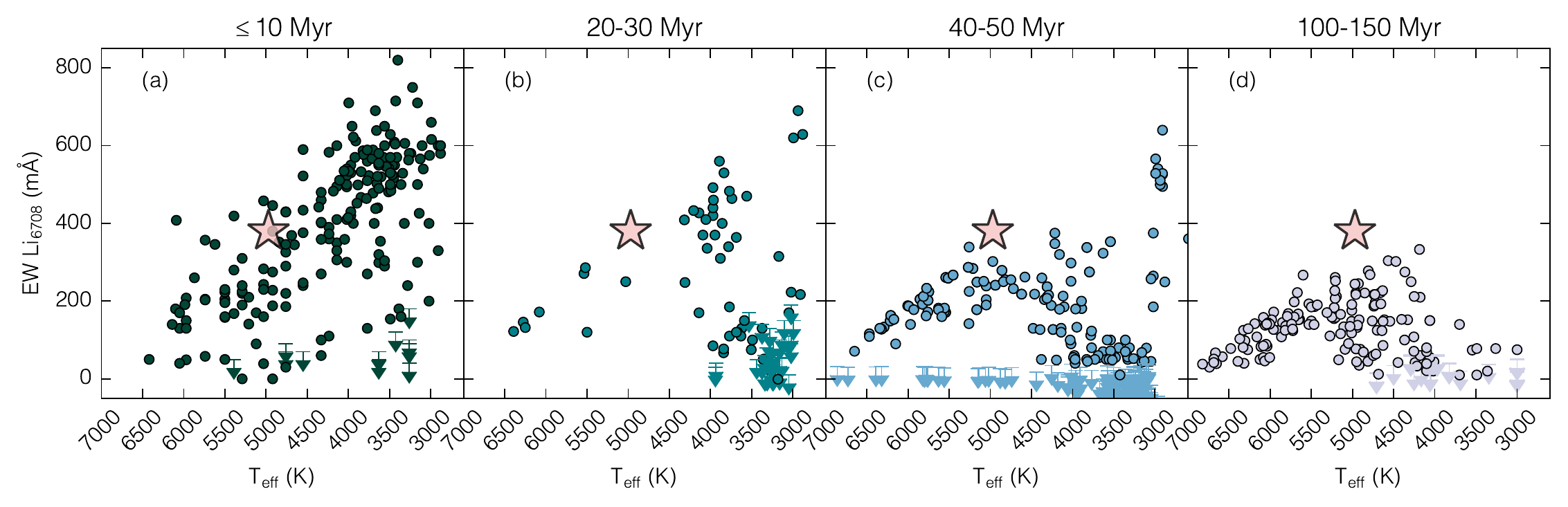}
    \caption{Lithium depletion in young associations. Relationship between effective temperature and Li~6708~\AA\ equivalent width for members of various young associations, moving groups, and clusters. V1298 Tau (pink star) shown for comparison. \textit{(a)} Taurus-Auriga, $\eta$ Chamaeleontis, and TW Hydrae. \textit{(b)} $\beta$ Pictoris moving group and NGC 1960. \textit{(c)} Tucanae-Horologium, Carina, Columba, IC2602, and IC2391. \textit{(d)} Pleiades and AB Doradus moving group.}
    \label{fig:lithium}
\end{figure*}

From the Hertzsprung-Russell diagram position (Figure~\ref{fig:hrd}), stellar kinematics (Figure~\ref{fig:membership}), and spatial location relative to the Taurus molecular clouds (Figure~\ref{fig:membership}), we conclude that V1298 Tau is a member of Group 29, an older association in the foreground of Taurus-Auriga. The age of Group 29 is estimated to be less than 45~Myr by comparison with known moving groups in a color-absolute magnitude diagram \citep[][see also Figure~\ref{fig:camd}]{Luhman2018}. We note, however, that the precise age and substructure of this group has not been fully explored. A comparison of Li I 6708~\AA\ equivalent widths among young stars compiled from the literature\footnote{Li 6708~\AA\ equivalent widths and temperatures/spectral types were compiled from \citet{Soderblom1993, Jones1996, Barrado2001, Randich2001, Mentuch2008, YeeJensen2010, Malo2013, Jeffries2013, BinksJeffries2014, Kraus2014, Kraus2017, Bouvier2018}.} supports an upper limit of 45~Myr for the age of V1298 Tau (Figure~\ref{fig:lithium}).

\subsection{Stellar parameters} \label{subsec:stellarparams}
We determined the stellar effective temperature, luminosity, and radius from 2MASS near-infrared photometry, the trigonometric parallax, and empirical relations for pre-main sequence stars \citep{Pecaut2013}. From a three-dimensional map of the local interstellar medium, we determined the color excess along the line-of-sight to V1298 Tau to be $E(B-V)$ = 0.024 $\pm$ 0.015~mag \citep{Lallement2014, Capitanio2017}. Assuming an extinction law \citep{Yuan2013}, we determined the $J$-band extinction to be $A_J$ = 0.019 $\pm$ 0.010 mag and the star's intrinsic $J-K_s$ color $(J-K_s)_0$ = 0.582 $\pm$ 0.032 mag. From the intrinsic $(J-K_s)_0$ color and linear interpolation between empirical pre-main sequence relations, we determined the effective temperature and the appropriate $J$-band bolometric correction. Using the extinction-corrected $J$-band magnitude we then calculated the bolometric luminosity. From the luminosity and effective temperature, we determined the stellar radius from the Stefan-Boltzmann law. Our adopted temperature is consistent with previous determinations, which range from 4920--5080 K \citep{Davies2014,Wichmann2000,Wahhaj2010,PallaStahler2002}. The spectral type has been reported as K1 \citep{WKS1996} and K1.5 \citep{Kraus2017}, and we find the HIRES spectrum to be most consistent with K0. The mass, estimated below, suggests V1298 Tau will evolve to become a late F-type or early G-type main sequence star.

We used stellar evolution models to estimate the mass and age of V1298 Tau in a theoretical Hertzsprung-Russell diagram. The pre-main sequence phase of evolution is particularly uncertain in theoretical models due to a dearth of calibrators. One major uncertainty regards the importance of magnetic fields. To estimate the magnitude of the model uncertainty for V1298 Tau we present two determinations of mass and age using models that neglect and account for magnetic fields \citep{Dotter2008, Feiden2016}. Standard models produce a mass and age of $M_*$~=~1.19~$\pm$~0.03~\msun and $\tau$~=~11~$\pm$~3~Myr, respectively. By comparison, magnetic models produce a mass of $M_*$~=~1.10~$\pm$~0.05~\msun and an age of $\tau$~=~23~$\pm$~4~Myr, which we ultimately adopt (Figure~\ref{fig:hrd}). Uncertainties in all derived stellar parameters were calculated from Monte Carlo simulations modeling input parameters as normal distributions with widths corresponding to the errors in the photometry, parallax, and extinction. The derived stellar parameters are summarized in Table~\ref{table:params}.

As mentioned earlier, V1298 Tau lacks either spectroscopic accretion signatures or an infrared excess out to 24~\micron, with an upper limit to the fractional disk luminosity of $L_\mathrm{disk}/L_\star < 1.5 \times 10^{-4}$ \citep{Wahhaj2010}. The star's spectral energy distribution is well-described by a photosphere with excess emission in the FUV and NUV (see \S~\ref{subsec:activity}). The lack of any significant circumstellar disk is consistent with the age we infer for V1298 Tau.

\begin{deluxetable*}{lc}[b!]
\tablecaption{Properties of V1298 Tau \label{table:params}}
\tablecolumns{2}
\tablenum{2}
\tablewidth{0pt}
\tablehead{
\colhead{Parameter} &
\colhead{Value} \\
}
\startdata
Designations & EPIC 210818897\\
& [WKS96] 4\\
& RX J0405.3+2009\\
Right ascension (J2000.0) &
04$^\mathrm{h}$05$^\mathrm{m}$19.6$^\mathrm{s}$ \\
Declination (J2000.0) & +20$^\circ$09'25.6'' \\
Parallax, $\varpi$ (mas) & 9.214 $\pm$ 0.059 \\ 
Proper motion R.A., $\mathrm{\mu_\alpha}$ (mas yr$^\mathrm{-1}$) & 5.23 $\pm$ 0.13 \\
Proper motion Dec., $\mathrm{\mu_\delta}$ (mas yr$^\mathrm{-1}$) & -16.077 $\pm$ 0.048 \\
Distance, $d$ (pc) & 108.5 $\pm$ 0.7 \\
Spectral type & K0--K1.5 \\
Stellar age, $\mathrm{\tau_\star}$ (Myr) &  23 $\pm$ 4 \\
Stellar mass, $M_\star$ ($M_\odot$ ) & 1.10 $\pm$ 0.05 \\
Stellar radius, $R_\star$ ($R_\odot$) &  1.305 $\pm$ 0.070 \\
Effective temperature, $T_\mathrm{eff}$ (K) & 4970 $\pm$ 120 \\
Luminosity, $L_\star$ ($L_\odot$) & 0.934 $\pm$ 0.044 \\
Mean stellar density, $\rho_\star$ (g cm$^\mathrm{-3}$) & 0.697 $\pm$ 0.075 \\
Surface gravity, $\log{g_\star}$ (dex) & 4.246 $\pm$ 0.034 \\
Stellar rotation period, $P_\mathrm{rot}$ (d) & 2.865 $\pm$ 0.012 \\
Projected rotational velocity, $v\sin{i}$ (km~s$^\mathrm{-1}$) & 23 $\pm$ 2  \\
Barycentric radial velocity, $\mathrm{\gamma}$ (km~s$^\mathrm{-1}$) & 16.15 $\pm$ 0.38 \\
Li~I~6708\AA\ equivalent width (m\AA) & 380 \\
(B-V) color excess, $E(B-V)$ (mag) & 0.024 $\pm$ 0.015 \\
J-band extinction, $A_J$ (mag) & 0.019 $\pm$ 0.010 \\ 
\enddata
\tablecomments{Astrometric parameters originate from \textit{Gaia} DR2.}
\end{deluxetable*}

\subsection{Stellar rotation and activity} \label{subsec:activity}
\begin{figure}
    \centering
    \includegraphics[width=\linewidth]{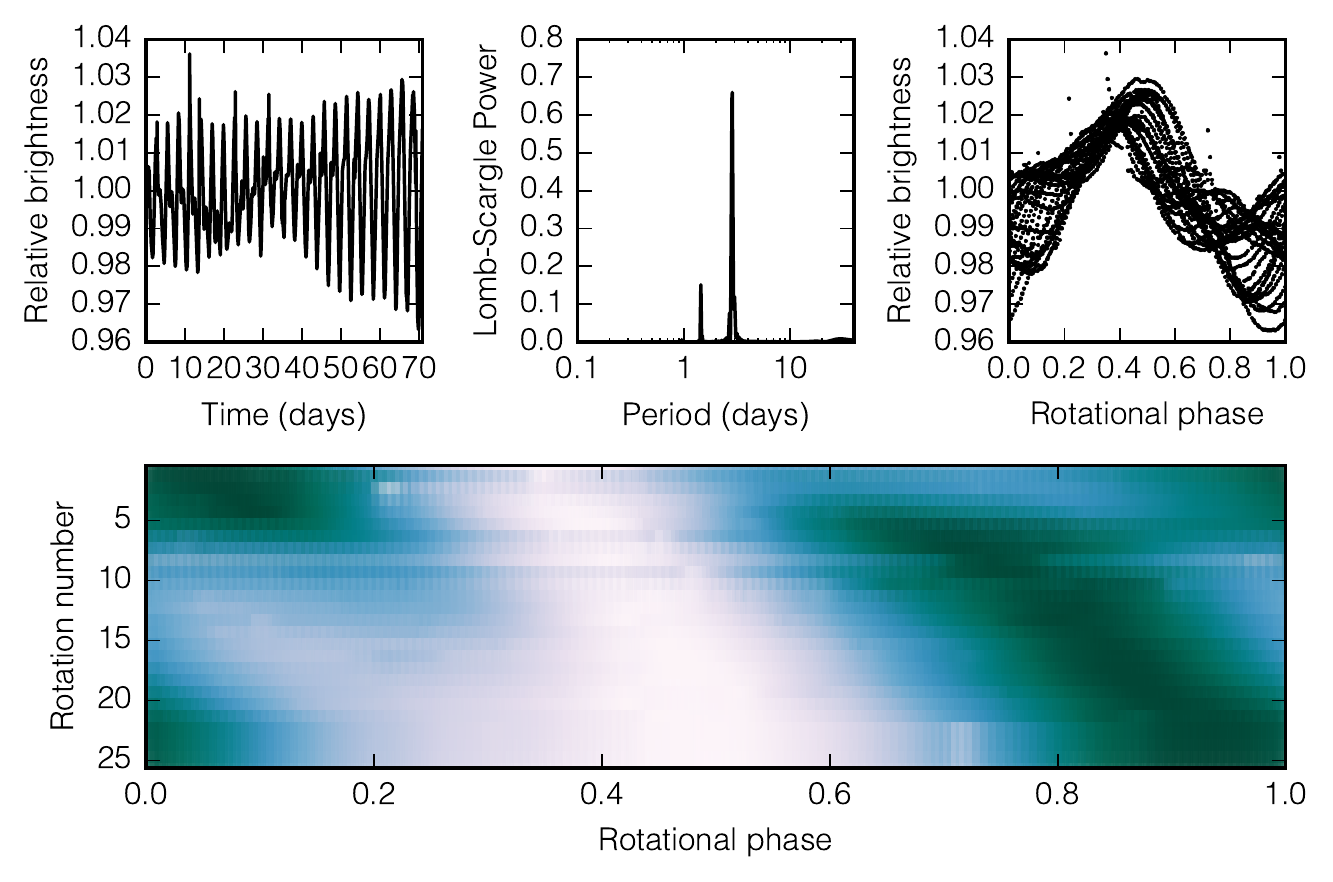}
    \caption{Stellar rotation period. \textit{Top:} from left to right, the full \textit{K2} light curve of V1298 Tau, a Lomb-Scargle periodogram of the \textit{K2} time series photometry with peak power at 2.85 days and a secondary peak at the second harmonic, and the \textit{K2} photometry phased on the rotation period. \textit{Bottom:} Waterfall diagram visualization of the brightness evolution of V1298 Tau throughout the \textit{K2} campaign.}
    \label{fig:rotation}
\end{figure}

\begin{figure}
    \centering
    \includegraphics[width=\linewidth]{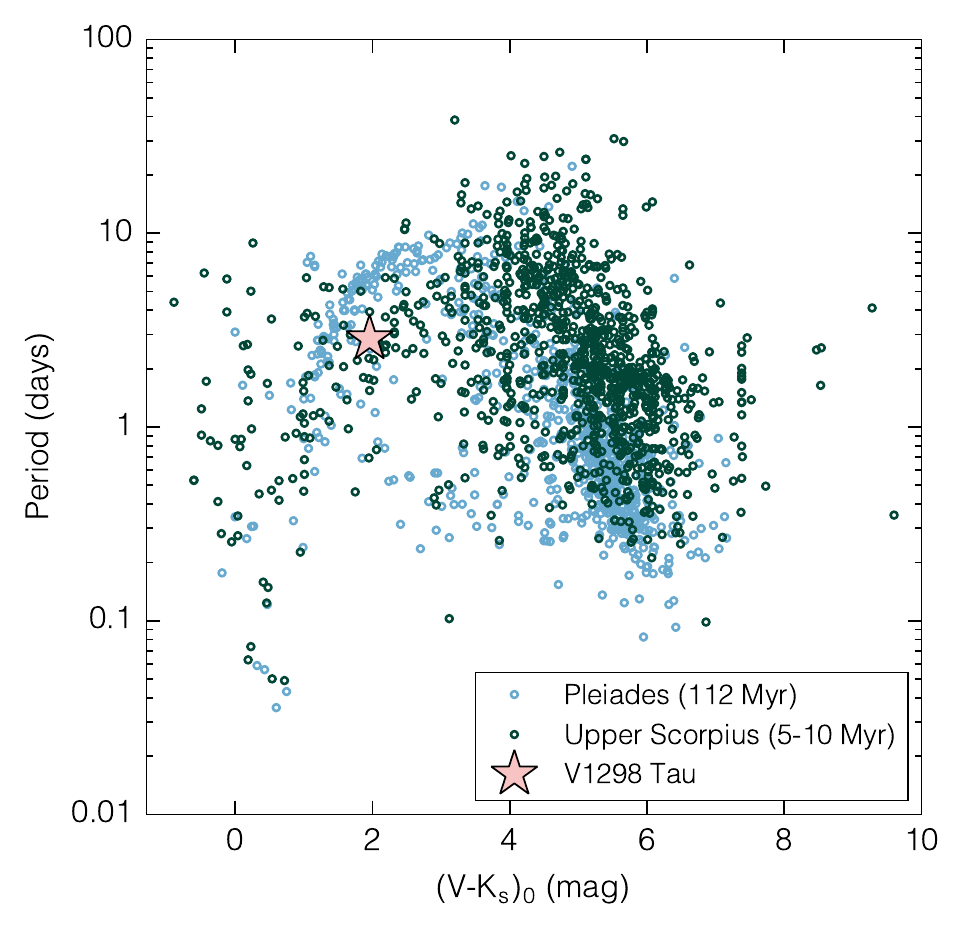}
    \caption{Relationship between colors and variability periods for members of the Upper Scorpius OB association (dark green) and the Pleiades open cluster (light blue), with V1298 Tau (pink star) shown for comparison. Note, the shortest periods at the blue end are due to pulsations rather than surface rotation. Data originate from \citet{Rebull2016,Rebull2018}.}
    \label{fig:pcolor}
\end{figure}

We infer a stellar rotation period of $P_\mathrm{rot} = 2.851\pm0.050$ days from a Lomb-Scargle periodogram \citep{Lomb1976, Scargle1982} of the \textit{K2} time series photometry (Figure~\ref{fig:rotation}). The period and uncertainty were determined from the mean and the half-width at half-maximum of a Gaussian fit to the periodogram peak, respectively. An autocorrelation function (ACF) of the \textit{K2} light curve suggests a rotation period of $P_\mathrm{rot} = 2.8605\pm0.0082$ days, consistent within 1$\sigma$. In this case the period is determined from the slope of a linear fit to the first four peaks of the ACF, with the uncertainty determined from the root mean square of the fit residuals. The period we report is in agreement with a previously published value \citep{Grankin2007}, and consistent with the period distribution among similarly young stars (Figure~\ref{fig:pcolor}). The uncertainties we quote for the rotation period reflect a measurement error and do not account for differential rotation, which can be as large as 0.2 radian day$^{-1}$ for pre-main sequence stars \citep{Waite2011}. The amplitude of brightness modulations is seen to evolve throughout the \textit{K2} observation period. Such an effect may be observed when two signals with different periods give rise to a beat pattern, which may be due to surface differential rotation, star spot emergence and decay, or two stars contained within the photometric aperture. The rotation period was also measured through Gaussian process regression, as described in \S~\ref{subsec:simulfit}, and found to be $P_\mathrm{rot} = 2.865 \pm 0.012$~days, which we ultimately adopt.

\begin{figure}
    \centering
    \includegraphics[width=\linewidth]{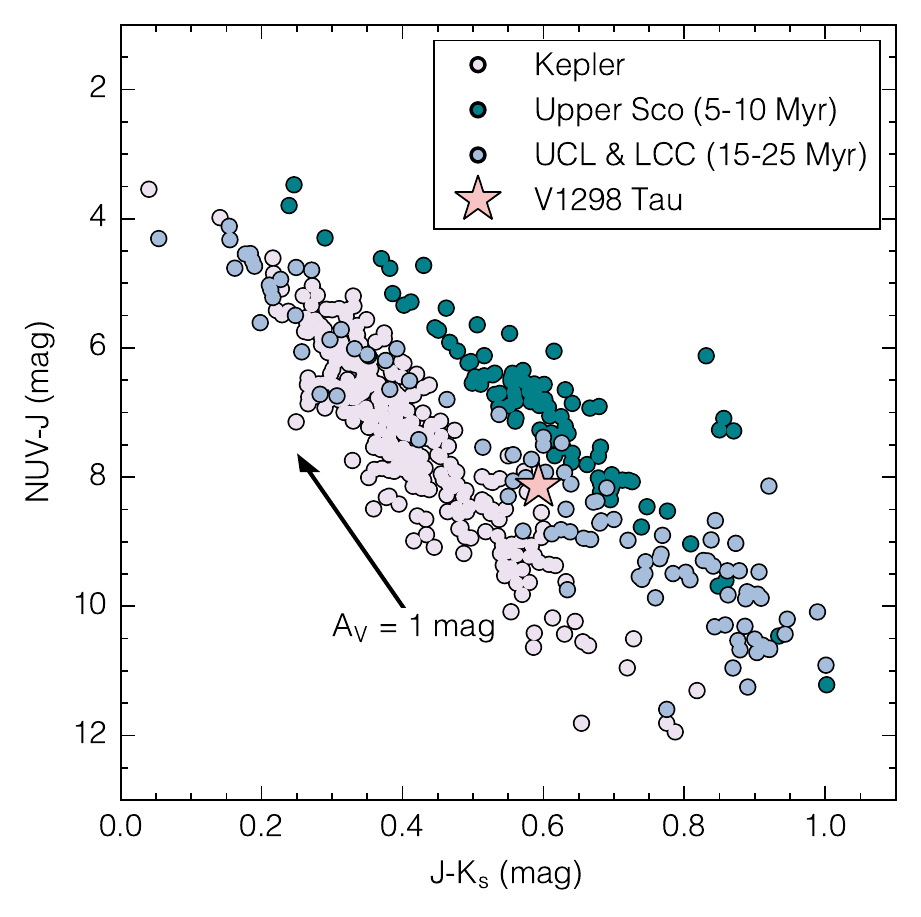}
    \caption{Near-UV and infrared color-color diagram. V1298 Tau exhibits excess UV emission, consistent with similarly young (15-25 Myr) stars in Upper Centaurus-Lupus (UCL) and Lower Centaurus-Crux (LCC), but at a lower level than stars in Upper Scorpius (5-10 Myr). The \textit{Kepler} sample is shown for comparison. The photometry have not been corrected for extinction. A reddening vector is shown for reference. Photometry originate from 2MASS \citep{Skrutskie2006} and \textit{GALEX} \citep{Martin2005}.}
    \label{fig:nuv}
\end{figure}

V1298 Tau exhibits excess ultraviolet (UV) emission, a common characteristic of similarly young stars. Using data from the \textit{GALEX} mission, \citet{FindeisenHillenbrand2010} found a far-UV excess of 3.4 $\pm$ 0.7~mag (at 4.7$\sigma$ significance) and a near-UV excess of 0.8 $\pm$ 0.3~mag (2.2$\sigma$). The level of UV excess exhibited by V1298 Tau is consistent with members of a similar color in the Lower Centaurus-Crux and Upper Centaurus-Lupus associations (15-25 Myr), while lower than members of the Upper Scorpius OB association (5-10 Myr) and higher than the \textit{Kepler} sample \citep[][see Figure~\ref{fig:nuv}]{Olmedo2015}.

From the \textit{ROSAT} X-ray flux ($f_X = 1.64 \times 10^{-15}$ W m$^{-2}$), a published MEKAL plasma fit, and hydrogren column density \citep{Boller2016}, we estimate an X-ray luminosity of $\log_{10}(L_X/$erg~s$^{-1}$) = 30.37, corresponding to a fractional X-ray luminosity of $\log_{10}(L_X/L_\mathrm{bol}) = -3.22$. The X-ray luminosity of V1298 Tau is essentially consistent with that of a saturated X-ray emitter \citep{Wright2011} and similar to that for other pre-main sequence fast-rotating stars. The predicted X-ray flux may also be calculated from the Rossby number, $Ro = P_\mathrm{rot}/\tau_c$, where $\tau_c$ is the convective turnover time. For V1298 Tau, $Ro = 0.173$, where $\tau_c = 16.518$ days was calculated from the $(V-K_s)_0$ color \citep{Wright2011}. From empirical relations \citep{Wright2011}, which suggest V1298 Tau is just below the saturated regime, the predicted fractional X-ray luminosity is $\log_{10}(L_X/L_\mathrm{bol}) = -3.47$. Using both the measured and predicted $\log_{10}(L_X/L_\mathrm{bol})$ values, the age of V1298 Tau implied by an empirical X-ray-age relation \citep{MamajekHillenbrand2008} is 12--24 Myr, consistent with the age found in a Hertzpsrung-Russell diagram.

\subsection{Transit model fitting} \label{subsec:transitfit}
We fit \citet{mandel2002} analytic transit models to the \textit{K2} photometry using a combination of the \textsc{pytransit} package \citep{Parviainen2015} and \textsc{emcee}, a Python implementation of the affine invariant Markov chain Monte Carlo (MCMC) ensemble sampler \citep{foremanmackey2013, GoodmanWeare2010}. The transit model parameters sampled were the orbital period ($P$), the time of mid-transit ($T_0$), the planet-to-star radius ratio ($R_P/R_*$), the scaled semi-major axis ($a/R_*$), and the cosine of the inclination ($\cos{i}$), and two parameter combinations of the eccentricity and longitude of periastron, ($\sqrt{e}\cos\omega$, $\sqrt{e}\sin\omega$). 

We assumed a quadratic limb darkening law, imposing Gaussian priors on the linear and quadratic coefficients. The centers and widths of the limb darkening coefficient priors ($u_1 = 0.621 \pm 0.023$, $u_2 = 0.103 \pm 0.015$) were determined from tabulated values appropriate for the temperature and surface gravity of V1298 Tau \citep{Claret2012,Claret2013}. Using an approximate formula for the mean stellar density in the case of eccentric orbits \citep{Kipping2010}, we additionally applied a Gaussian prior on the light curve derived stellar density (which is a function of period, $a/R_\star$, $e$, and $\omega$) with a center and width of 0.697~g~cm$^{-3}$ and 0.225~g~cm$^{-3}$, respectively. Model transit profiles were numerically integrated to match the 1766 second cadence of the \textit{K2} observations. The target probability density sampled was therefore,

\begin{multline}
    \ln\mathcal{L} = -\frac{1}{2}\chi^2 - \frac{1}{2}\frac{(\rho_\star- \mu_{\rho_\star})^2}{\sigma_{\rho_\star}^2} \\ -\frac{1}{2}\frac{(u_1- \mu_{u_1})^2}{\sigma_{u_1}^2} - \frac{1}{2}\frac{(u_2- \mu_{u_2})^2}{\sigma_{u_2}^2},
\end{multline}

where the first term describes the likelihood and the last three terms describe the priors on the mean stellar density and limb darkening parameters.

Convergence was assessed iteratively until the following criteria were met for each directly sampled parameter: (1) the chain length exceeded 50 times the autocorrelation length, and (2) the autocorrelation length estimate changed by $<$2\% from the prior iteration. We discarded the first $10 \times \left \langle \tau_\mathrm{acor} \right \rangle$ steps as burn-in, where $\left \langle \tau_\mathrm{acor} \right \rangle$ is the average autocorrelation length across all parameters. In addition to the fit described above, we performed a circular orbit fit with no prior directly applied to the mean stellar density. This secondary fit was used in an analysis of the host star's evolutionary stage. The radius ratio inferred from the circular orbit fit is nearly indistinguishable from that found for the eccentric fit, $(R_P/R_\star)_\mathrm{circ}$=0.0713$^{+0.0012}_{-0.0007}$. The results of the transit modeling and derived planet parameters are summarized in Table~\ref{table:lcresults} under the Fit 1 columns.

As mentioned earlier, several in-transit observations are outliers with no easily discernible nature (i.e., these observations are not obviously affected by spacecraft systematics). Most of these outliers serve to diminish the transit depth. To determine the effect of the outlying observations on the inferred planetary radius we performed transit fits that both included and excluded these outliers. We adopt the fit which excludes the putative spot-crossing events because the scatter about the lower envelope of the transit profile is smaller. When including all observations, the inferred planet size is 1$\sigma$ smaller than our adopted value ($\mathrm{R_P}~=~0.86 \pm 0.05~\mathrm{R_{Jup}}$). Future observations in the infrared, where the amplitude of stellar variability is smaller, may measure the planetary radius more securely. 

\begin{figure}
    \centering
    \includegraphics[width=\linewidth]{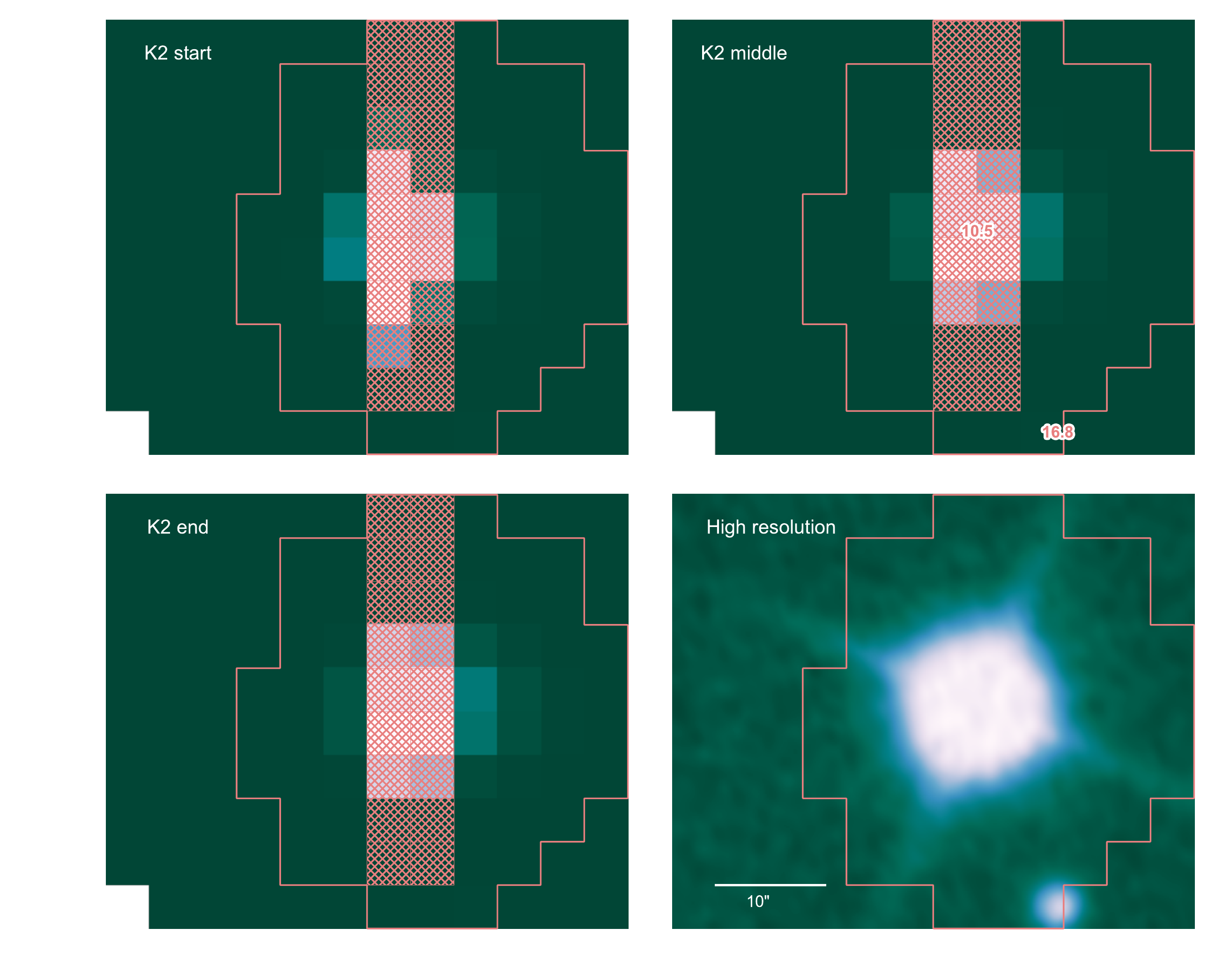}
    \caption{\textit{K2} photometric aperture for V1298 Tau. \textit{K2} target pixel files for V1298 Tau at the start (\textit{top left}), middle (\textit{top right}), and end (\textit{bottom left}) of Campaign 4. \textit{Bottom right:} High resolution image from the Palomar Observatory Sky Survey showing a faint background source to the southwest. The photometric aperture boundary is indicated in pink, with saturated columns shown as hatched regions.}
    \label{fig:aperture}
\end{figure}

An unassociated star to the southwest of V1298 Tau is partially contained within the \textit{K2} photometric aperture (Figure~\ref{fig:aperture}). We show in \S~\ref{subsec:falsepositives} that this star is not responsible for the transits, but we consider here the effect of flux dilution on the inferred planet radius. In the presence of light from another star, the ratio of the true planet radius to the observed planet radius is given by $R_\mathrm{P,true}/R_\mathrm{P,obs} = \sqrt{1+F_{\mathrm{2}}/F_{\mathrm{1}}}$, where $F_{\mathrm{2}}/F_{\mathrm{1}}$ is the optical flux ratio between the two stars \citep{Ciardi2015}. The nearby star is 6.3 magnitudes fainter than V1298 Tau in the \textit{Kepler} bandpass, and thus impacts the planet radius by $<$0.2\%, which is much smaller than the stellar radius uncertainty.

\begin{figure}
    \centering
    \includegraphics[width=\linewidth]{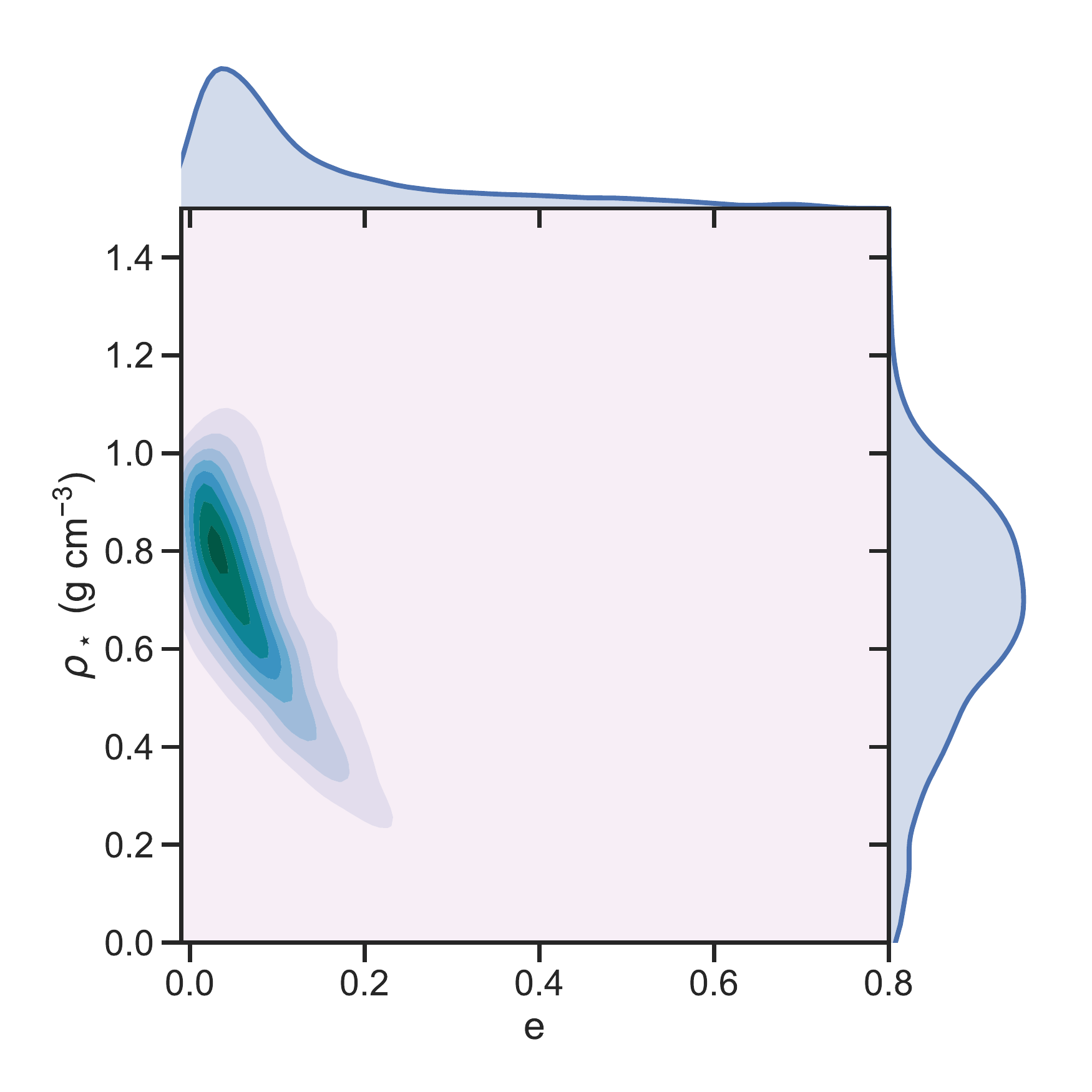}
    \caption{Constraints on eccentricity of V1298 Tau  b from the transit fit (Fit 1, as described in \S~\ref{subsec:transitfit}). Shaded contours show the joint posterior probability densities for the eccentricity of V1298 Tau b and mean stellar density from the transit fit. Marginalized probability densities are shown at top and at right.}
    \label{fig:erhopost}
\end{figure}

For planets with ingress and egress durations that are sufficiently resolved in time, the eccentricity can be derived from MCMC sampling and a loose stellar density prior \citep{DawsonJohnson2012}. From the transit fit, we find the marginalized posterior density in eccentricity places limits of $e <$ 0.16, 0.51, 0.71 at 68.3\%, 95.5\%, and 99.7\% confidence, respectively (Figure~\ref{fig:erhopost}). While transit-derived eccentricities can be robust, we stress that radial velocity monitoring or secondary eclipse timing is required to constrain the eccentricity with higher precision and confidence.

\subsection{Simultaneous variability and transit fits}
\label{subsec:simulfit}
To investigate the impact of our variability detrending on the inferred planet parameters, we also performed simultaneous fits to the stellar variability and planet transits using the \texttt{exoplanet} \citep{exoplanet:exoplanet}, \texttt{starry} \citep{exoplanet:luger18}, and \texttt{PyMC3} \citep{exoplanet:pymc3} packages. We again used the \textsc{everest 2.0} light curve for this fit. Observations with quality bit flags in the range of [1,17] were discarded.

We constructed a \texttt{PyMC3} model to describe the light curve using the following parameters: the mean out-of-transit flux ($\langle f \rangle$), quadratic limb darkening coefficients ($u_1$, $u_2$), stellar mass ($M_\star$) and radius  ($R_\star$), log of the orbital period ($\ln P$), time of mid-transit ($T_0$), radius ratio ($R_P/R_\star$), impact parameter ($b$), eccentricity ($e$), longitude of periastron ($\omega$), and several hyperparameters to describe the stellar variability using a Gaussian process (GP).

We used the ``Rotation Term'' GP kernel in \texttt{exoplanet}, which models rotationally modulated variability as a mixture of two stochastically driven, damped simple harmonic oscillators with undamped periods of $P_\mathrm{rot}$ and $P_\mathrm{rot}/2$.\footnote{\url{https://exoplanet.dfm.io/en/stable/user/api/\#exoplanet.gp.terms.RotationTerm}} The hyperparameters sampled for the GP were the log of the variability amplitude ($\ln A$), the log of the primary variability period ($\ln P_\mathrm{rot}$), the log of the quality factor minus 1/2 for the secondary oscillation ($\ln Q_0$), the log of the difference between the quality factors of the first and second modes ($\ln \Delta Q$), and the fractional amplitude of the secondary mode relative to the primary mode (a mixture term referred to here as ``mix''). In addition to the hyperparameters described above, a ``jitter'' parameter ($\ln S_2$) was introduced to account for excess white noise.

For uninformative sampling of the quadratic limb darkening coefficients, \texttt{exoplanet} uses the parameterization recommended by \citet{Kipping:2013}. Additionally, for efficient sampling of the radius ratio and impact parameter, \texttt{exoplanet} uses the joint parameterization of those parameters suggested by \citet{Espinoza2018}. For this fit, a $\beta$ distribution prior was assumed for the eccentricity, with the assumed values $a = 0.867, b = 3.03$ as recommended by \citet{Kipping:2013ecc}.

We masked observations within 12 hours of each transit in order to ensure that the transits did not influence the variability model and performed an initial optimization of the \texttt{PyMC3} model. After the initial optimization, 7-$\sigma$ outliers were masked (where $\sigma$ was defined as the RMS of the residuals of this initial fit). This step effectively masked both in-transit and out-of-transit outliers from the sampling procedure which followed.  A new optimization was performed, followed by MCMC sampling with the No U-Turns step-method \citep{hoffman2014}. We ran 4 chains with 500 tuning steps to learn the step size, 4500 tuning iterations (tuning samples were discarded), a target acceptance of 99\%, and 3000 draws for a final chain length of 12,000 in each parameter. Convergence was assessed using the Gelman-Rubin diagnostic \citep{Gelman:Rubin:1992}, which was below 1.001 for each parameter.

\begin{figure}
    \centering
    \includegraphics[width=\linewidth]{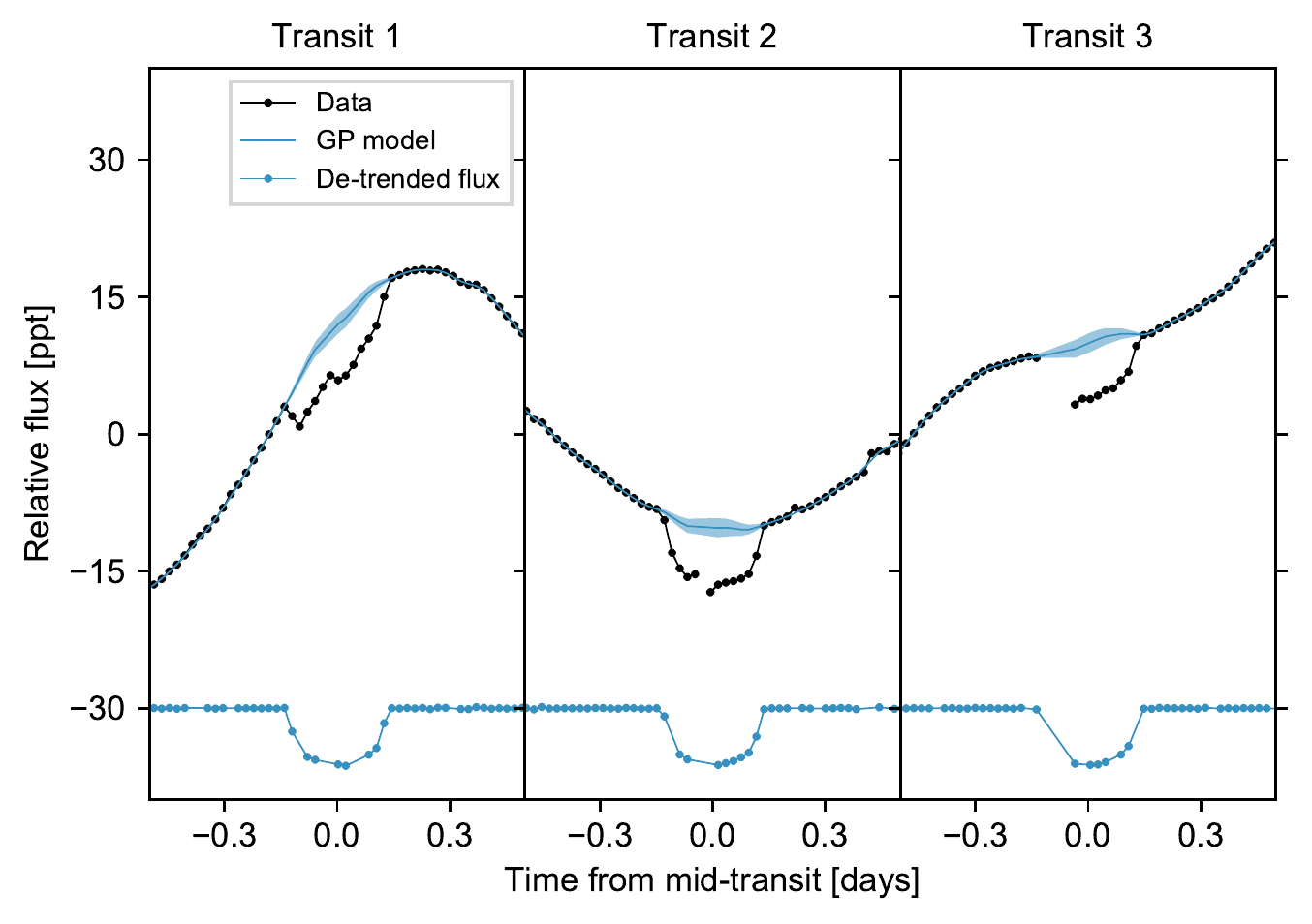}
    \caption{Transits of V1298 Tau b in the \textsc{everest 2.0} light curve (black dotted line) with the GP model from the simultaneous variability and transit fits described in \S~\ref{subsec:simulfit} shown by the blue line. The shaded bands show the 3$\sigma$ error contours of the GP model. The de-trended flux (data - median GP model - offset) is shown by the blue dotted line below each transit.}
    \label{fig:gp-intransit}
\end{figure}

\begin{figure}
    \centering
    \includegraphics[width=\linewidth]{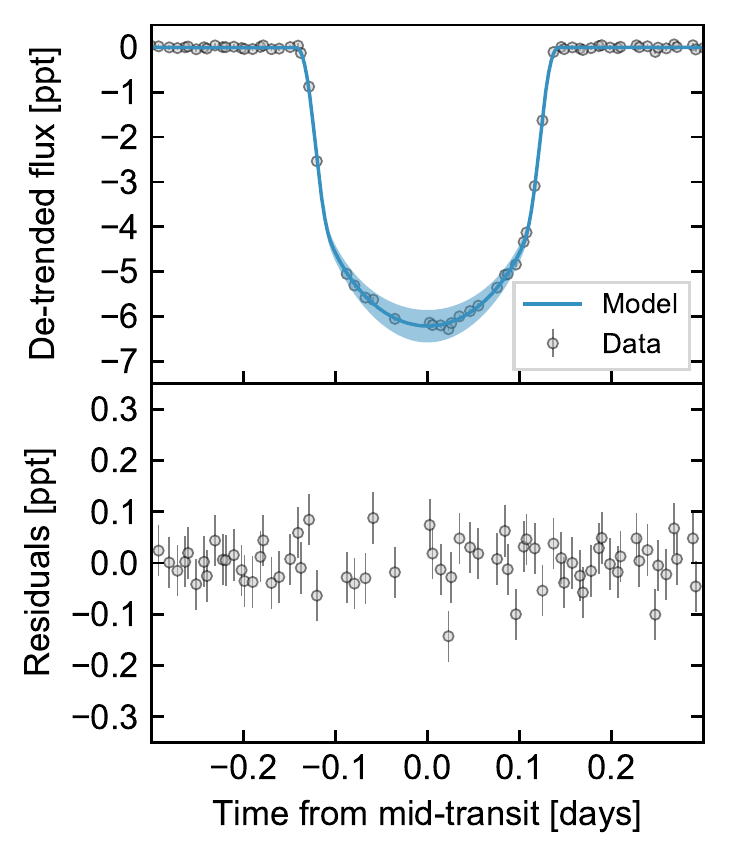}
    \caption{\textit{Top:} Phase-folded transit observations (points) and median model (solid line) from the simultaneous variability and transit fits described in \S~\ref{subsec:simulfit} (Fit 2). The 1$\sigma$ error contours are shown by the shaded bands. \textit{Bottom:} Residuals from the median model.}
    \label{fig:simulfit}
\end{figure}

The procedures described above were adapted from \texttt{exoplanet} tutorials which are published online.\footnote{\url{https://exoplanet.dfm.io/en/stable/}} Results from the simultaneous variability and transit fits, along with the priors imposed on all of the sampled parameters, are summarized in Table~\ref{table:lcresults} under the Fit 2 columns. Figure~\ref{fig:gp-intransit} shows the GP model around the transits, and Figure~\ref{fig:simulfit} shows the phase-folded transit and residuals from the simultaneous variability and transit fit.

\begin{deluxetable*}{lllll}[b!]
\tablecaption{V1298 Tau light curve modeling results. \label{table:lcresults}}
\tablecolumns{5}
\tablewidth{0pt}
\tablehead{
\colhead{Parameter} &
\colhead{Fit 1 Value} & 
\colhead{Fit 1 Prior} & 
\colhead{Fit 2 Value} & 
\colhead{Fit 2 Prior} \\
}
\startdata
$P$ (days) & 24.13889$^{+0.00043}_{-0.00044}$ & $\mathcal{G}$(24.13889, 0.00082) & 24.13861$^{+0.00102}_{-0.00090}$ & $\mathcal{G}$(24.14, 0.24)\tablenotemark{a} \\
$T_0$ (BJD) & 2457091.18842$^{+0.00039}_{-0.00038}$ & $\mathcal{G}$(2457091.18848,0.00070) & 2457067.04914$^{+0.00058}_{-0.00061}$ & $\mathcal{G}$(2457067.05, 0.01)\\
$R_P/R_\star$ & 0.07111$^{+0.00117}_{-0.00061}$ & $\mathcal{U}$[-1,1] & 0.0713$^{+0.0021}_{-0.0020}$ & $\mathcal{G}$(0.073, 0.155)\tablenotemark{b} \\
$b$ & 0.23$^{+0.16}_{-0.15}$ & $\mathcal{U}$[-1,1] & 0.29$^{+0.16}_{-0.19}$ & $\mathcal{U}$[0,1] in $r_1, r_2$\tablenotemark{c}\\
$\cos{i}$ & 0.0086$^{+0.0067}_{-0.0058}$ & $\mathcal{U}$[0,$\cos^{-1}(50^\circ)$] & 0.0114$^{+0.0057}_{-0.0072}$ & \nodata\\
$i$ (deg) & 89.51$^{+0.33}_{-0.38}$ & \nodata & 89.35$^{+0.41}_{-0.33}$ & \nodata \\
$a/R_\star$ & 28.7$^{+1.5}_{-2.3}$ & $\mathcal{U}$[-1,$\infty$) & 27.6$^{+1.5}_{-1.0}$ & \nodata \\
$\sqrt{e}\cos\omega$ & -0.01$^{+0.26}_{-0.31}$ & $\mathcal{U}$[-1,1] & 0.00$^{+0.34}_{-0.33}$ & \nodata\\
$\sqrt{e}\sin\omega$ & 0.10$^{+0.18}_{-0.18}$ & $\mathcal{U}$[-1,1] & 0.12$^{+0.16}_{-0.19}$ & \nodata\\
$e$ & 0.087$^{+0.216}_{-0.062}$ & \nodata & 0.112$^{+0.133}_{-0.083}$ & $\beta(a=0.867, b=3.03)$\\
$\omega$ (deg) & 92$^{+71}_{-72}$ & \nodata & 91$^{+70}_{-71}$ & $\mathcal{U}[-180,180]$\tablenotemark{d}\\
$u_1$ & 0.591$^{+0.020}_{-0.021}$ & $\mathcal{G}$(0.621, 0.023) & 0.71$^{+0.14}_{-0.16}$ & $\mathcal{U}$[0,1] in $q_1$\tablenotemark{e}\\
$u_2$ & 0.098$^{+0.015}_{-0.015}$ & $\mathcal{G}$(0.103, 0.015) & -0.13$^{+0.23}_{-0.19}$ & $\mathcal{U}$[0,1] in $q_2$\\ 
$\rho_{\star,\mathrm{ecc}}$ (g cm$^{-3}$)\tablenotemark{f} & 0.69$^{+0.21}_{-0.26}$ & $\mathcal{G}$(0.697,0.225) & 0.59$^{+0.19}_{-0.15}$ & \nodata\\
$\rho_{\star,\mathrm{circ}}$ (g cm$^{-3}$) & 0.77$^{+0.13}_{-0.17}$ & \nodata & 0.679$^{+0.117}_{-0.072}$ & \nodata \\
\hline
$M_\star$ ($M_\odot$) & \nodata & \nodata & 1.099$^{+0.049}_{-0.049}$ & $\mathcal{G}$(1.10, 0.05)\\
$R_\star$ ($R_\odot$) & \nodata & \nodata & 1.314$^{+0.052}_{-0.064}$ & $\mathcal{G}$(1.305, 0.07)\\
$\langle f \rangle$ (ppt) & \nodata & \nodata & 0.02$^{+0.34}_{-0.35}$ & $\mathcal{G}$(0, 10) \\
$\ln$($A$/ppt) & \nodata & \nodata & 4.98$^{+0.79}_{-0.58}$ & $\mathcal{G}$($\ln(\sigma^2)$, 5.0) \\
$\ln$($P_\mathrm{rot}$/day) & \nodata & \nodata & 1.0526$^{+0.0041}_{-0.0040}$ & $\mathcal{G}$($\ln$(2.85), 5.0) \\
$\ln$($Q_0$) & \nodata & \nodata & 0.51$^{+0.17}_{-0.15}$ & $\mathcal{G}$($\ln$(1), 10.0) \\
$\Delta Q_0$ & \nodata & \nodata & 5.50$^{+0.88}_{-0.72}$ & $\mathcal{G}$($\ln$(2), 10.0) \\
mix & \nodata & \nodata & 0.26$^{+0.21}_{-0.14}$ & $\mathcal{U}$[0,1] \\
$\ln$($S_2$) & \nodata & \nodata & -5.050$^{+0.060}_{-0.060}$ & $\mathcal{G}$($\ln(\sigma^2)$, 10.0) \\ %Note y is masked
\hline
$R_P$ ($R_\mathrm{Jup}$) & 0.904$^{+0.053}_{-0.048}$ & \nodata & 0.911$^{+0.049}_{-0.053}$ & \nodata\\
$R_P$ ($R_\oplus$) & 10.14$^{+0.58}_{-0.54}$ & \nodata & 10.22$^{+0.55}_{-0.59}$ & \nodata\\
$a$ (AU) & 0.1688$^{+0.0025}_{-0.0026}$ & \nodata & 0.1687$^{+0.0025}_{-0.0026}$ & \nodata\\
$T_{14}$ (hours) & 6.386$^{+0.048}_{-0.034}$ & \nodata & 6.420$^{+0.071}_{-0.055}$ & \nodata \\
$T_{23}$ (hours) & 5.486$^{+0.039}_{-0.072}$ & \nodata & 5.478$^{+0.068}_{-0.103}$ & \nodata \\
$T_\mathrm{eq}$ (K)\tablenotemark{g} & 657$^{+32}_{-24}$ & \nodata & 668$^{+22}_{-22}$ & \nodata \\
$S$ ($S_\oplus$) & 32.8$^{+1.9}_{-1.8}$ & \nodata & 33.2$^{+4.5}_{-4.2}$ & \nodata \\
\enddata
\tablecomments{Priors are noted for those parameters which were directly sampled. $\mathcal{G}$: Gaussian. $\beta$: Beta distribution. $\mathcal{U}$: Uniform. $\sigma$: Standard deviation in flux (191 ppt). Quoted transit parameters and uncertainties are medians and 15.87\%, 84.13\% percentiles of the posterior distributions. We recommend adopting the values from Fit 2.}
\tablenotetext{a}{Sampling performed in $\ln(P)$.}
\tablenotetext{b}{Sampling performed in $\ln(R_P/R_\odot)$.}
\tablenotetext{c}{Joint sampling of impact parameter and radius ratio performed using \citet{Espinoza2018} parameterization.}
\tablenotetext{d}{Sampling performed in ($\cos\omega,\sin\omega$).}
\tablenotetext{e}{Uninformative sampling of quadratic limb darkening coefficients performed using parametrization of \citet{Kipping:2013}.}
\tablenotetext{f}{Calculated from Equation 39 of \citet{Kipping2010}.}
\tablenotetext{g}{Calculated assuming an albedo of 0.}
\end{deluxetable*}

\subsection{Centroid motion analysis} \label{subsec:centroid}
\begin{figure}
    \centering
    \includegraphics[width=\linewidth]{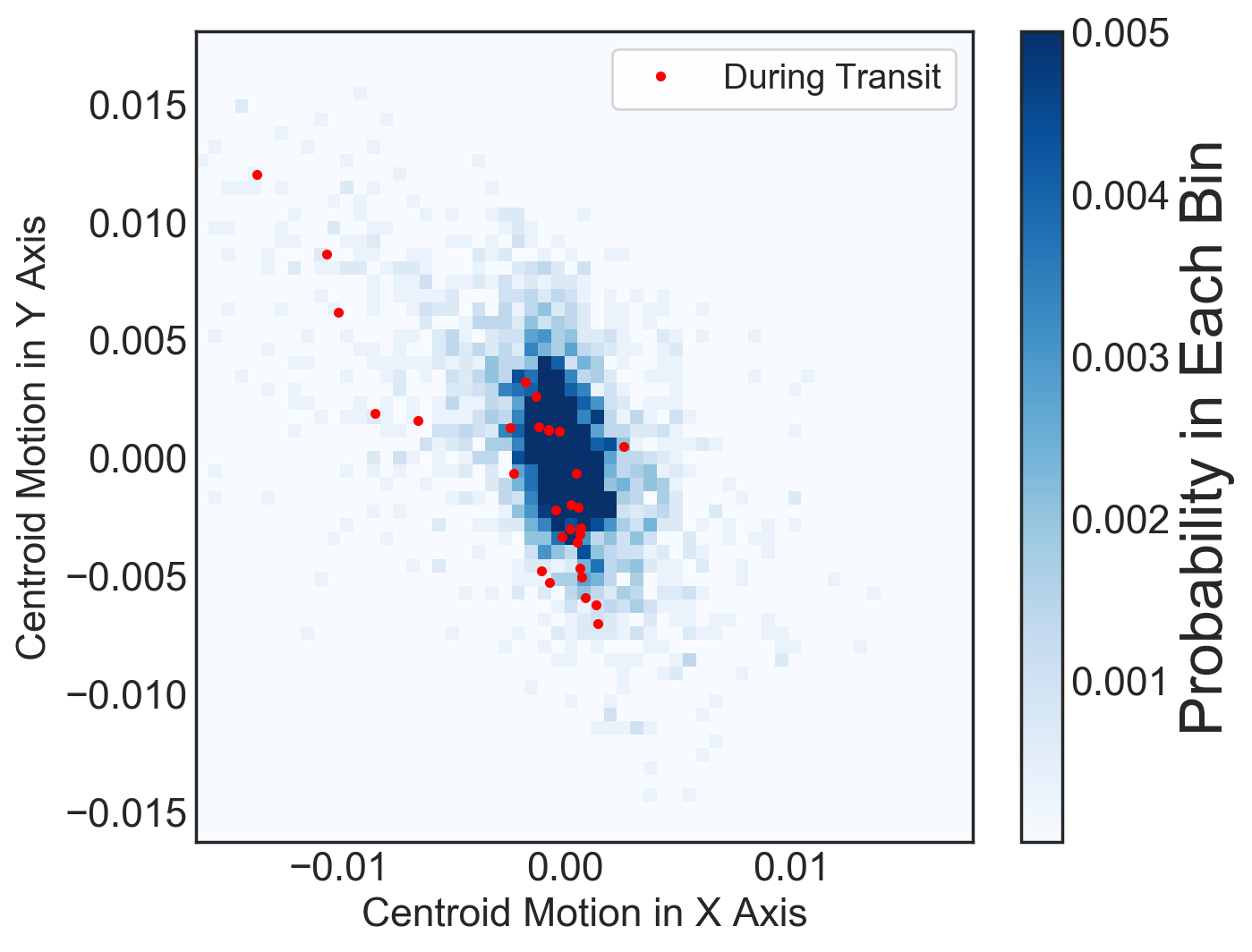}
    \caption{Centroid motion of V1298 Tau. Motion of the point spread function centroid of V1298 Tau throughout the \textit{K2} campaign (two-dimensional histogram). The centroid motions during the transits of V1298 Tau b are shown as red points, and are consistent with with the distribution of out-of-transit centroid shifts at the $\lesssim$1$\sigma$ level. The five clear outliers correspond to observations which occurred immediately after a loss of fine-pointing by the telescope.}
    \label{fig:centroid}
\end{figure}

Evidence of offsets in the point spread function (PSF) centroid during transit are indicative of a transit occurring due to a contaminant or background star with a transient nature \citep[e.g. a background eclipsing binary,][]{Thompson2018}. Based on a simple centroiding test, accounting for the \textit{K2} roll motion, the PSF centroids during the expected transit of V1298 Tau b are consistent at the $\lesssim$ 1 $\sigma$ level with the out of transit centroids (Figure~\ref{fig:centroid}). This suggests that the transit signal is not due to a background eclipsing binary and is consistent with originating from the target star.  

\subsection{Limits on companions from the transit shape} \label{subsec:transitshape}
The transit shape constrains the probability of hierarchical triple scenarios in which the observed transit signal is due to an eclipsing binary or transiting planet host which is distinct from, but gravitationally bound to, V1298 Tau. In such scenarios, the flux dilution from V1298 Tau itself requires larger radius ratios (and thus more V-shaped eclipses) between the eclipsing companions in order to reproduce the observed transit depth. To estimate the relative likelihood of these scenarios, we performed Levenberg-Marquardt least-squares fits (with free parameters $R_P/R_*$, $a/R_*$, $\cos{i}$) to the transit profile over a grid of assumed optical contrasts between V1298 Tau and the putative eclipsing companions. 

In this case the model time series is given by the equation,

\begin{equation}
f_{\text{dil}}(t) = \frac{f(t) + F_1/F_2}{1+F_1/F_2},
\end{equation}

where $f(t)$ is the normalized eclipse time series in the absence of dilution and $F_1/F_2$ is the flux ratio between the primary star and the eclipsing companions.

\begin{figure}
    \centering
    \includegraphics[width=\linewidth]{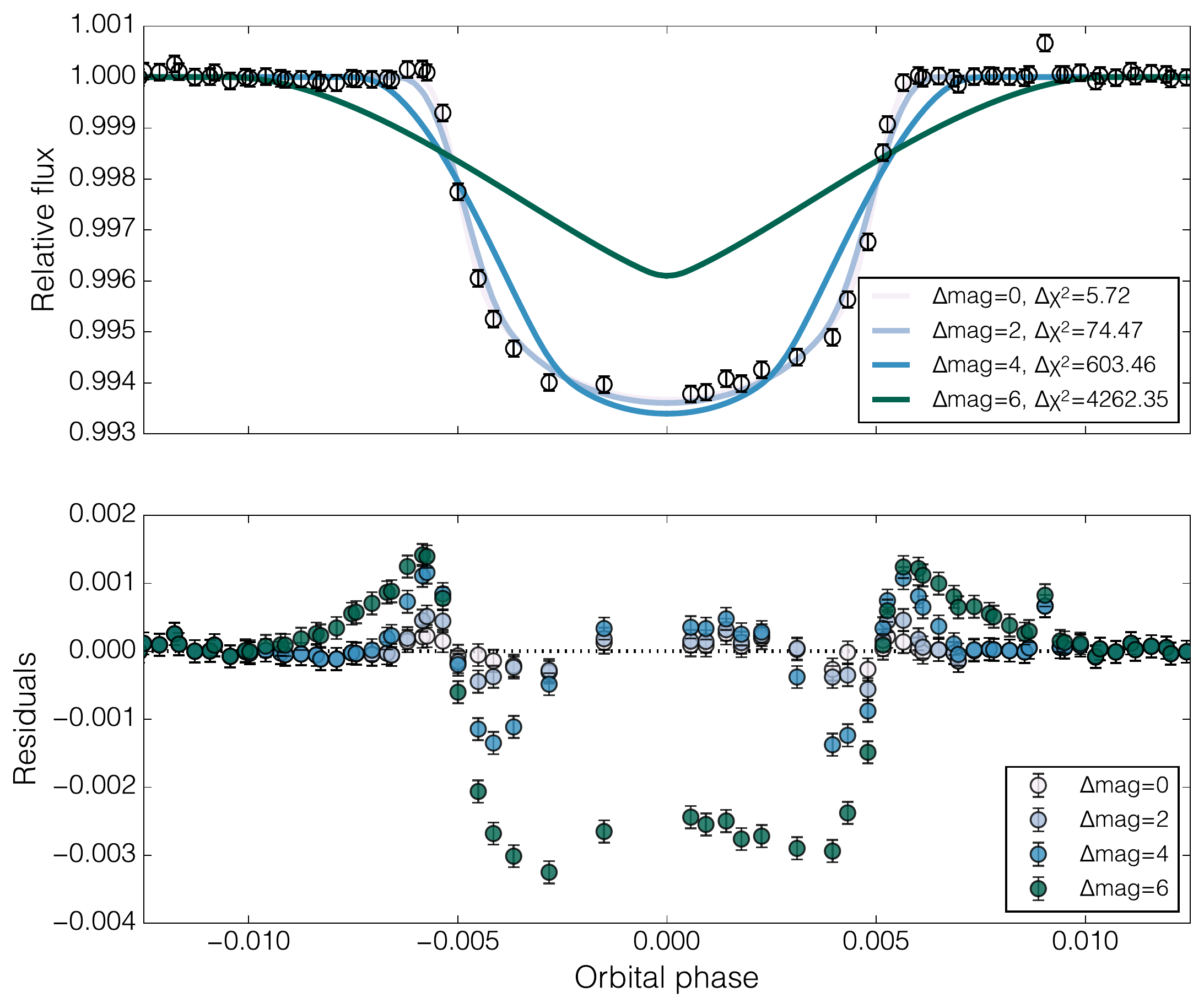}
    \caption{Constraints on companions to V1298 Tau from the transit shape. Best fit transit models after accounting for dilution from V1298 Tau. For large optical contrasts between V1298 Tau and a putative companion, models can not reproduce the observed transit shape and depth. For each assumed contrast, the $\Delta \chi^2$ value between the hierarchical triple model and the single star model is indicated. Scenarios involving a companion with $\Delta$mag $>$ 4 are ruled out.}
    \label{fig:transitshape}
\end{figure}

For each contrast value and best-fit eclipse profile, we recorded the $\chi^2$ value. Using the Akaike Information Criterion (AIC) we then evaluated the relative likelihoods of models which assume dilution from the primary star. We found that a model with a companion contrast of $\Delta$mag = 1 is 5.6$\times$10$^{-4}$ times as likely as one with $\Delta$mag = 0. By comparison, a model assuming dilution from an equal-brightness companion is found to be 0.057 times as likely as the best fit model assuming no additional companion. We also performed an MCMC fit for a diluted transit model, sampling the parameters $R_P/R_*$, $a/R_*$, $\cos{i}$, and $\Delta$mag and using the same convergence criteria utilized in the transit fit described earlier. This analysis leads to a constraint of $\Delta$mag $<$ 0.76 for a putative companion at 99.7\% confidence. Although the statistical tests described above suggest a stringent limit on the contrast of a putative companion, we adopt a more conservative limit of $\Delta$mag $<$ 4, based on the shapes of the best fit transit profiles after accounting for dilution (Figure \ref{fig:transitshape}), for plausible hiearchical triple scenarios. Using equation 7 from \citet{Ciardi2015}, which assumes the planet transits the primary star, we calculated the true planet radius would be larger by 41\%, 22\%, and 1\% for putative companions with $\Delta$mag = 0, 0.76, and 4, respectively.

\subsection{Limits on companions from imaging} \label{subsec:imaginglimits}

\begin{figure}
    \centering
    \includegraphics[width=\linewidth]{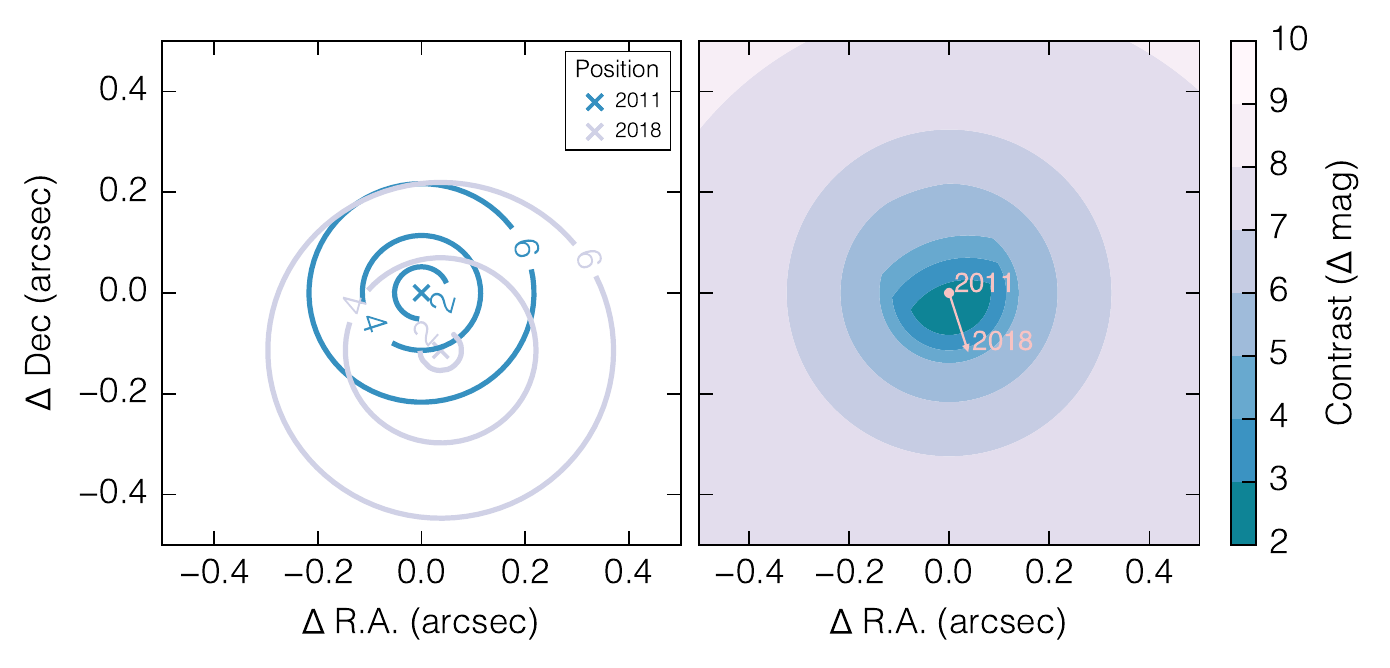}
    \caption{Multi-epoch adaptive optics imaging of V1298 Tau. \textit{Left:} Radial contrast contours in the 1'' region surrounding V1298 Tau from multi-epoch adaptive optics imaging. \textit{Right:} Combination of the individual constraints results in a minimum contrast of $\Delta K$=2.1~mag everywhere in the surrounding region, including behind the position of V1298 Tau during the \textit{K2} observations.}
    \label{fig:bglimits}
\end{figure}

\citet{Daemgen2015} observed V1298 Tau with adaptive optics and the NIRI instrument at Gemini North Observatory on 2011 October 22 11:00 UTC. Those data rule out nearly all scenarios in which V1298 Tau hosts a stellar mass companion with a projected separation of 10.85--1085~AU, where we have interpolated between a 20~Myr isochrone to convert near-infrared contrasts to mass limits \citep{Baraffe2015}. In the time between those observations and our NIRC2 imaging, V1298 Tau moved by 0.120~$\pm$~0.062 arcseconds on the sky due to its proper motion. Combining the two epochs of imaging constraints we rule out a vast swath of parameter space involving a background or foreground eclipsing binary that would have been aligned by chance with V1298 Tau during the \textit{K2} observations (Figure~\ref{fig:bglimits}). 
 
We used a galactic structure model \citep{Girardi2012} to simulate a 1-deg$^2$ field centered on V1298 Tau in order to estimate the number of foreground or background stars in the region of parameter space not excluded by our observations. We found that 0.004 sources with 3.4~$< \Delta V <$~5.8 mag are expected in the surrounding 0.1\arcsec$\times$0.1\arcsec\ region. An even smaller number of sources are expected to be eclipsing binaries. 
 
\subsection{Limits on companions from spectroscopy} \label{subsec:speclimits}
\begin{figure}
    \centering
    \includegraphics[width=\linewidth]{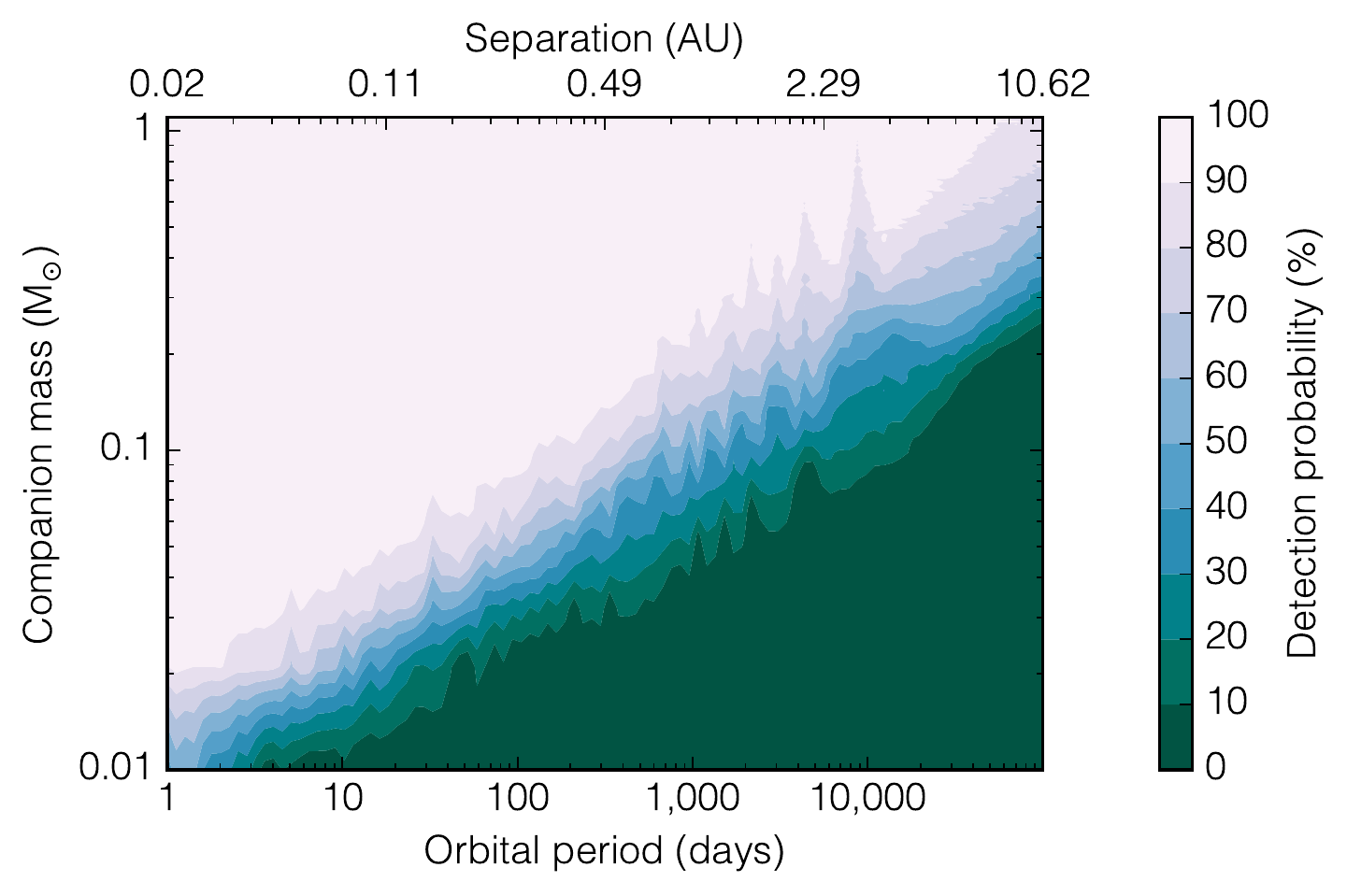}
    \caption{Radial velocity limits on companions to V1298 Tau. Detection probability in the period-companion mass plane for simulated spectroscopic binaries given the 24 year radial velocity time series. The axis at top indicates the physical separation for a 0.5~$\mathrm{M_\odot}$ companion. The 95\% detection probability contour is depicted in Figure~\ref{fig:complimits}.}
    \label{fig:rvlimits}
\end{figure}

We used a 24-year radial velocity time series \citep[][this work]{Wichmann2000,Nguyen2012}, including our newly acquired data, to search for bound companions to V1298 Tau. We assumed an uncertainty of 1~\kms for each observed radial velocity to allow for the possibility of zero-point offsets between different instruments or other systematic biases. At each point in a grid of orbital period and companion mass we simulated 1000 circular spectroscopic binary orbits, sampled at the times of the observations. For each simulated orbit, the inclination was drawn randomly from a uniform distribution in $\cos{i}$ and the phase was drawn randomly from a uniform distribution between [0,2$\pi$]. To simulate radial velocity jitter, we added Gaussian noise with an amplitude of 200~\ms\ to each model. At each point in the period-companion mass plane we then determined the detection probability as the fraction of successfully detected simulated orbits. An individual simulated orbit was considered successfully detected if the $\Delta\chi^2$ value between the simulated radial velocities and the null hypothesis of constant velocity exceeded 20. From this analysis we are able to rule out a wide range of brown dwarf and stellar companions with orbital periods between 1--10,000 days (Figure~\ref{fig:rvlimits}). 

In a search for secondary spectral lines in the HIRES spectrum \citep{kolbl2015} we found no stars brighter than 5\% the brightness of the primary and with a projected separation of $\leq$0''.4. We note that this search is only sensitive to stars with radial velocity separations $>$50 km~s$^{-1}$ (corresponding approximately to two linewidths of V1298 Tau), as the two sets of spectral lines would otherwise be indistinguishable.

\subsection{Limits on companions from astrometry} \label{subsec:astrometrylimits}
\textit{Gaia} resolves double stars, either associated or unassociated, outside of 1\arcsec\ down to optical contrasts of 6 magnitudes and irrespective of a companion's position angle \citep{Ziegler2018}. Only one other star was detected by \textit{Gaia} within the boundaries of the \textit{K2} photometric aperture. We show later that star is too faint to be a false-positive. At smaller separations, the goodness-of-fit of the \textit{Gaia} astrometric solution provides another means of assessing multiplicity. A previous study of exoplanet host stars with closely-projected companions showed that the \textit{Gaia} astrometric fit metrics can be used to reliably detect companions with separations of 0.08--1\arcsec\ and optical contrasts $<$2~mag \citep{Rizzuto2018}. Furthermore, exoplanet host stars with detected companions at projected separations of 0.05--1\arcsec\ typically have astrometric goodness-of-fit values $>$20 with excess astrometric noise often detected at 5$\sigma$ significance \citep{Evans2018}. In the case of V1298 Tau, the goodness-of-fit in the along-scan direction is 9.0, with zero detectable excess astrometric noise. While not conclusive, the \textit{Gaia} data suggest V1298 Tau is unlikely to host a stellar companion with a mass above 0.5~$M_\odot$ (corresponding to a model-derived optical contrast of 2~mag at 20~Myr \citep{Baraffe2015}) and separation greater than 0.08\arcsec (8.7 AU). 

\subsection{False-positive scenario assessment} \label{subsec:falsepositives}

\begin{figure*}
    \centering
    \includegraphics[width=0.6\textwidth]{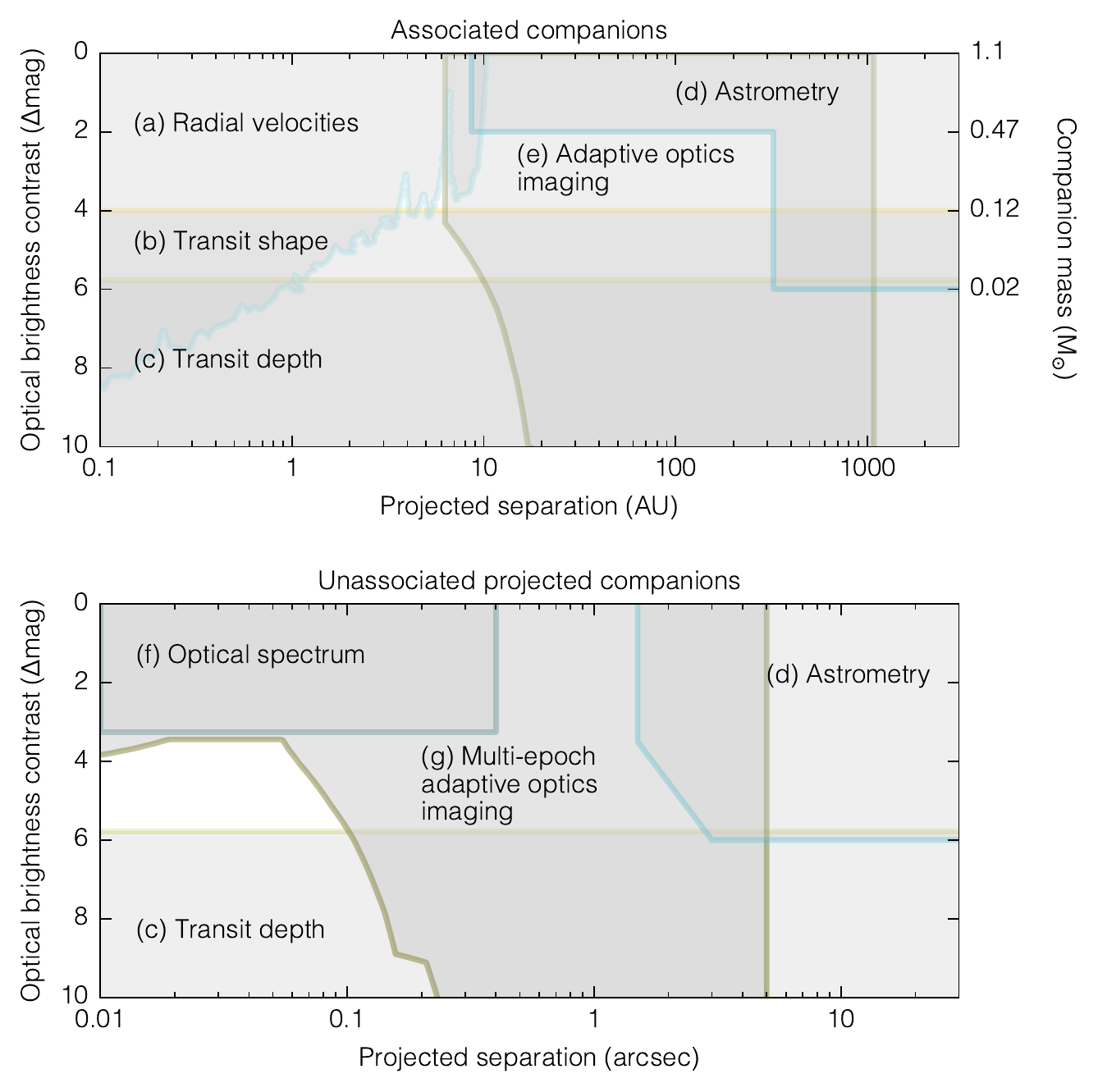}
    \caption{Constraints on astrophysical false-positive scenarios. Regions of the separation-contrast plane excluded by observations are shown for associated companions at top, and unassociated companions below. \textit{(a)} Sustained radial velocity monitoring rules out associated eclipsing binaries at close separations. \textit{(b)} The transit shape rules out certain hierarchical triple scenarios at any separation. \textit{(c)} The observed transit depth sets an upper limit of 5.7~mag to the optical contrast between V1298 Tau and a putative false-positive, whether it be associated or not. \textit{(d)} \textit{Gaia} astrometry resolves point sources beyond 1\arcsec, while the astrometric goodness-of-fit is a predictor of associated companions outside of 80 mas. \textit{(e)} Adaptive optics imaging rules out wide associated companions. \textit{(f)} A search for secondary spectral lines would have detected bright eclipsing binaries at close separations as long as the velocity separation with V1298 Tau was $>$50~\kms. \textit{(g)} Multi-epoch adaptive optics combined with the proper motion of V1298 Tau eliminates the bulk of unassociated false-positive scenarios. The probability of an unassociated eclipsing binary residing within 0.1\arcsec, where our observations are not sensitive, is $<$4/1000.}
    \label{fig:complimits}
\end{figure*}

False-positive signals in transit surveys can be due to an eclipsing binary or a planet transiting a star other than the assumed host. We consider each false-positive scenario in detail below (summarized in Figure~\ref{fig:complimits}), and conclude that the interpretation of a Jupiter-sized planet transiting V1298 Tau is the one most consistent with observations. 

An eclipsing binary can dim by a maximum of 100\%, which sets a firm upper limit of 5.7 magnitudes to the optical contrast between V1298 Tau and a putative false-positive capable of reproducing the observed transit depth. Although an unassociated star is included in the target pixel files for V1298 Tau, it is too faint to reproduce the observed transit depth and we recovered the transits from a small photometric aperture excluding that star. Inside of 0.1\arcsec, assuming an unassociated eclipsing binary would not be co-moving with V1298 Tau, we effectively ruled out background or foreground stars by leveraging the proper motion of V1298 Tau between two epochs of adaptive optics imaging separated by 6 years. At separations of 0.1--20\arcsec, adaptive optics imaging, all-sky photometric surveys, and \textit{Gaia} rule out additional stars bright enough to be a false-positive. Furthermore, a time series analysis of the \textit{K2} point spread function indicates that the centroid shifts during transits are consistent with those signals originating from V1298 Tau.

False-positive scenarios involving hierarchical triples, i.e. an eclipsing binary or a transiting planet host which is gravitationally bound to V1298 Tau, are similarly disfavored. At close separations ($<$10~AU), we used a 24-year radial velocity time series \citep{Wichmann2000, Nguyen2012} to rule out stellar and brown dwarf companions in a mass-dependent manner, allowing for isotropically distributed inclinations of a putative companion. At intermediate separations ($\sim$10--1000 AU), previously published \citep{Daemgen2015} and newly acquired adaptive optics imaging rules out nearly all stellar companions. Furthermore, the lack of astrometric noise detected with \textit{Gaia} suggests there are no companions down to $\sim$0.5~$M_\odot$ in the $\sim$10--300~AU range. At projected separations beyond 1000~AU, stellar mass and some brown dwarf companions would have been detected by \textit{Gaia} or all-sky surveys. Further discussion of false-positive scenarios is presented in Appendix \ref{appendix:falsepositives}.

\subsection{Limit on the planet mass} \label{subsec:masslimit}
\begin{figure}
    \centering
    \includegraphics[width=\linewidth]{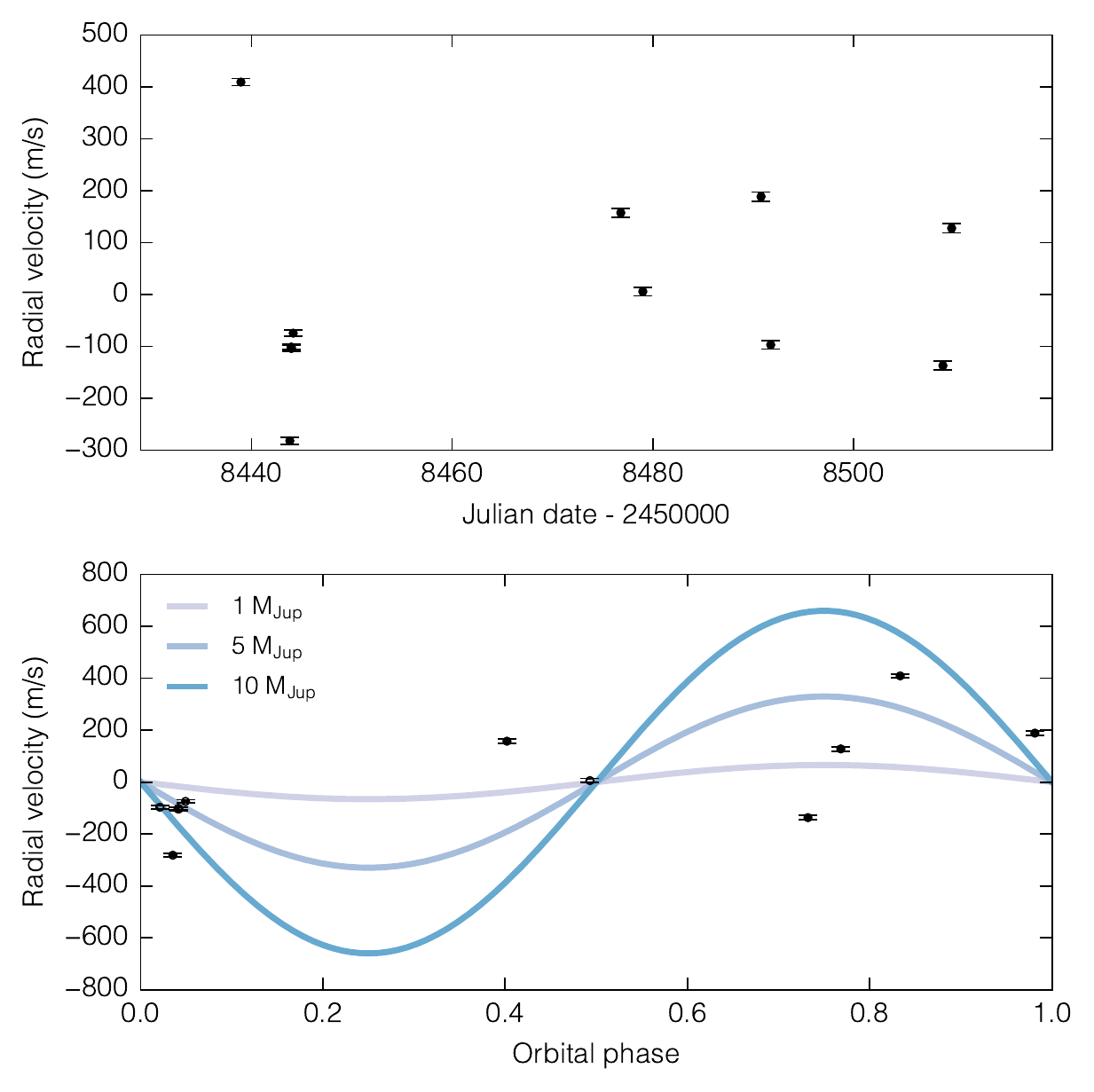}
    \caption{Precision radial velocities of V1298 Tau. \textit{Top:} Keck/HIRES precision radial velocity time series of V1298 Tau. \textit{Bottom:} PRV measurements as a function of orbital phase, with representative circular Keplerian orbits shown for reference.}
    \label{fig:prvlimits}
\end{figure}

Using the \textsc{radvel} radial velocity orbit fitting code \citep{Fulton2018}, we fit the PRV measurements of V1298 Tau in order to find an upper limit to the planet's mass (Figure~\ref{fig:prvlimits}). Due to the sparse orbital phase coverage of our data and to the high radial velocity jitter of the star, this upper limit is very likely to be conservative. We assumed a circular orbit for our fits, imposing Gaussian priors on the period and time of inferior conjunction corresponding to the values found from the transit fit (Table~2). Furthermore, the jitter amplitude was fixed to 200 \ms, corresponding approximately to the root mean square dispersion between all measurements. The parameter space defined by the period, time of inferior conjunction, radial velocity semi-amplitude, and velocity zero-point was then explored through MCMC sampling. The resulting posterior of the radial velocity semi-amplitude provides a 3$\sigma$ upper limit to the planet mass of $M_P <$ 8.3~\mjup.

\section{Discussion} \label{sec:discussion}

Without a measurement of the planet's mass, it is difficult to place V1298 Tau b in proper context with respect to the observed exoplanet population. \textit{A priori}, the most likely hypothesis from an occurrence rate standpoint is that V1298 Tau b has a mass more similar to typical \textit{Kepler} planets, with a radius that has not yet contracted due to its young age. Of the confirmed transiting exoplanets with measured masses and orbital periods in the 10--100 day range, the median mass is 11~\mearth with a 16th to 84th percentile range of 4--58~\mearth. Depending on the exact core mass and the mass and initial entropy of the H/He envelope, evolution models of low-mass planets ($<$100~\mearth) do indeed predict large radii ($\sim$10~\rearth) at ages of 20--30 Myr \citep{OwenWu2013, LopezFortney2013, Jin:etal:2014, Chen:Rogers:2016}. Interestingly, at a separation of $\approx$0.17~AU, V1298 Tau b should be largely unaffected by photo-evaporation \citep{OwenWu2013}. Thus, if V1298 Tau b does indeed have a low mass more representative of the \textit{Kepler} population, it may be a particularly valuable benchmark for planet evolution models; V1298 Tau b would provide an opportunity to learn more about the initial conditions of this common class of planet in the absence of substantial photo-evaporation.

On the other hand, given that the current limit on the planet's mass is not very restrictive, we must consider the possibility that V1298 Tau b belongs to the class of planets known as ``warm Jupiters.'' The size of V1298 Tau b is significantly smaller than the predictions of many evolutionary models for Jovian-mass planets \citep{Burrows1997, Baraffe2003, Fortney2007, Mordasini2012, OwenWu2013}, which predict radii of $\gtrsim$1.3~\rjup\ at an age of $\sim$20 Myr. Such a large radius would be possible if V1298 Tau hosts a nearly equal brightness companion, though such a companion should have been detected from the suite of follow-up observations. In fact, the models of \citet{Fortney2010} suggest that the combination of age, radius, and separation for V1298 Tau b is inconsistent with a gas-dominated composition for a planet with mass $>$0.1~\mjup\ (32 \mearth). However, the observed properties of V1298 Tau b might be reproduced by a planet with a core-dominated composition and mass in the range 0.1--0.3~\mjup (with a core mass of 25~\mearth). If the true mass of V1298 Tau b is $\gtrsim$0.3~\mjup, it may still be considered a warm Jupiter by conventional definitions.

Relative to hot Jupiters, less is known about warm Jupiters, though similar mechanisms are invoked to explain their origins. There is evidence for two distinct populations of warm Jupiters. The majority of these planets are characterized by low eccentricities ($e \lesssim 0.2$), nearby and coplanar super-Earth companions, and a dearth of external Jovian-mass companions \citep{Dong2014,Huang2016}. The remaining warm Jupiters are characterized by moderately eccentric ($e \gtrsim 0.4$) orbits, often accompanied by external Jovian-mass companions which are mutually inclined and apsidally misaligned \citep{DawsonChiang2014}. 

In general, warm Jupiters do not have eccentricities large enough to become hot Jupiters through tidal dissipation \citep{Dawson2015}. Furthermore, population synthesis studies examining the outcomes of various high-eccentricity migration scenarios have struggled to reproduce the observed ratio of hot and warm Jupiters \citep{Antonini2016, PetrovichTremaine2016}. Both of these points of evidence are seen as a major weakness in the high-eccentricity migration hypothesis. However, secular interactions with a neighboring planet could excite the eccentricity of V1298 Tau b at a later stage \citep{AndersonLai2017}. 

While the transit fit for V1298 Tau b suggests a low to moderate eccentricity, solutions with $e >$0.5 are not ruled out. However, eccentricities measured from transit photometry (particularly long-cadence \textit{Kepler} data as in this case) should be regarded with caution. Until the mass and eccentricity of V1298 Tau b are measured through radial velocities or secondary eclipse timing, it remains unclear whether the planet belongs to either of the known warm Jupiter populations. Even if V1298 Tau b is found to have a moderately low eccentricity, long-term radial velocity monitoring is required to test the hypothesis that the planet might have an external companion that is nearby and massive enough to force episodic tidal migration through secular planet-planet interactions. In the absence of such secular interactions, the final semi-major axis can be calculated from the present day semi-major axis and eccentricity. From the posterior probability densities resulting from the light curve modeling, we calculated the final semi-major axis, $a_\mathrm{final} = a(1-e^2)$, is $>$0.13~AU at 95\% confidence, and $>$0.09~AU at 99\% confidence. For V1298 Tau b to migrate inwards to $\leq$0.1~AU requires an eccentricity of $\geq$0.64.

%\hfill \break 
\section{Conclusions} \label{sec:conclusions}
We report the detection and validation of a warm Jupiter-sized planet transiting a young solar analog with an estimated age of 23 million years. The star and its planet belong to a newly characterized association, named Group 29, in the foreground of the Taurus-Auriga star-forming region \citep{Oh2017, Luhman2018}. Through a detailed analysis of the \textit{K2} data and follow-on observations, we found no evidence for a stellar companion and we securely ruled out most plausible false-positive scenarios. The young age of the planetary system challenges, but does not outright exclude, high-eccentricity migration as a viable formation channel. Formation \textit{in situ} \citep[e.g.][]{Hansen:Murray:2012, Batygin2016, Boley2016} or at a wider separation followed by planet-disk interactions \citep[e.g.][]{Terquem:Papaloizou:2007, Kley2012} would be compatible with current observations and not at odds with the star's youth.

\begin{figure}
    \centering
    \includegraphics[width=\linewidth]{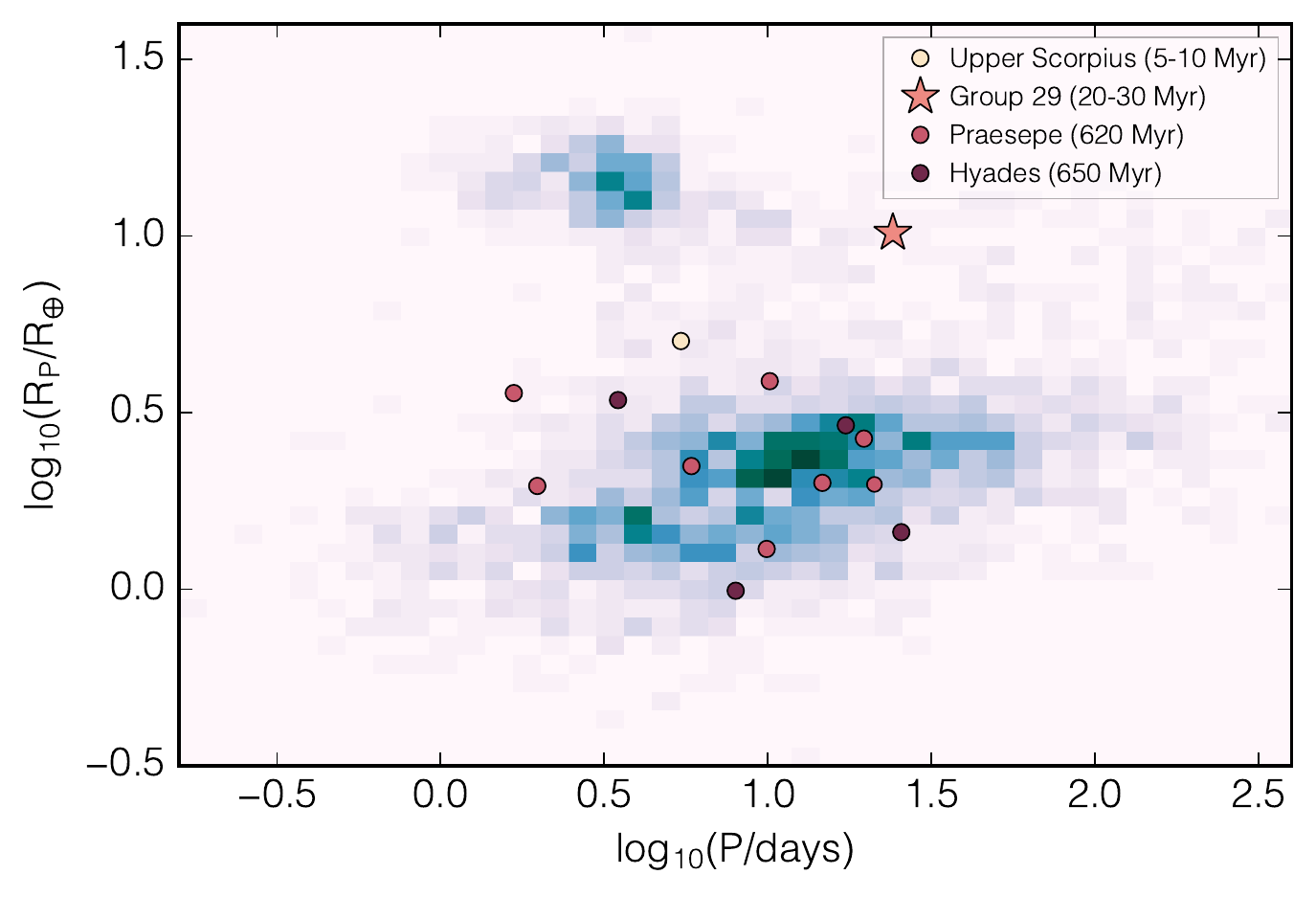}
    \caption{V1298 Tau b (pink star) compared to the sample of confirmed transiting exoplanets (two-dimensional histogram) in the period-radius domain. Transiting planets in young clusters and associations are shown for comparison.}
    \label{fig:prad}
\end{figure}

The transit profiles exhibit some anomalous features, and it is presently unclear whether the outlying observations are systematic or astrophysical in nature. Transit observations in the infrared, where the amplitude of stellar variability is lower, should yield a more secure measurement of the planet's radius. To measure the planet's mass, it is also preferable to observe in the infrared, as we have measured the optical radial velocity jitter to be $\sim$200~\ms (compared with the range of predicted Doppler semi-amplitudes of 13--230~\ms). We discuss the favorability of V1298 Tau for various follow-up observations further in Appendix~\ref{sec:favorability}.

Although several groups have published candidate exoplanet catalogs encompassing nearly every \textit{K2} campaign, the V1298 Tau system seems to have evaded detection, despite the relatively large transit depth. The most likely explanation seems to be that automated transit search pipelines are not tailored to the high-amplitude, short-timescale variability exhibited by young stars. Customized or more generalized routines which are not susceptible to under- or over-fitting stellar variability are therefore needed to adequately search the light curves of young stars for transiting planets \citep[e.g.][]{Rizzuto2017}.

Presently, there is only one other secure case of an exoplanet transiting a pre-main sequence star: the Neptune-sized planet K2-33 b in the 5 to 10 million-year-old Upper Scorpius OB association \citep{David2016,Mann2016}. V1298 Tau b is now the second youngest transiting exoplanet, and the first secure case of a transiting Jovian-sized planet orbiting a pre-main sequence star. Many of the other transiting planets found in young ($<$1 Gyr) clusters reside in low-occurrence regions of the period-radius diagram (Figure~\ref{fig:prad}), such as the radius valley or the sub-Jovian desert. V1298 Tau b seems to follow this trend, though it is as yet unclear whether this planet is a Jovian-mass resident of the ``period-valley'' \citep{Udry2003, Santerne2016}, a progenitor of the rare sub-Saturn class \citep{Petigura2017}, or perhaps contracting down to the sub-Neptune ``main-sequence.''

From an occurrence rate standpoint, the hypothesis that V1298 Tau b has a mass $<30$~\mearth is at least an order of magnitude more likely than the hypothesis that the mass is above this threshold \citep{Howard:etal:2010, Mayor:etal:2011}. Furthermore, at an age of $\sim$23 Myr, the planet's size is significantly smaller than expected for Jovian-mass planets \citep{Burrows1997, Baraffe2003, Fortney2007, Mordasini2012} but consistent with the radii of low-mass \textit{Kepler}-type planets for a range of plausible core and envelope masses \citep{OwenWu2013, LopezFortney2013, Jin:etal:2014, Chen:Rogers:2016}.

In general, \textit{Kepler} planets (i.e. super-Earths and mini-Neptunes) are believed to have formed while the gas disk was still present, given that a substantial fraction of them possess an envelope constituting $\sim$1--10\% of the total mass. However, the details of where and when the embryos formed, whether migration plays a significant role, and where the envelopes are accreted are debated \citep[e.g.][]{Terquem:Papaloizou:2007, Ida:Lin:2010, Hansen:Murray:2012, Lee:etal:2014, Lee:etal:2016, Schlichting:2018}. Detailed studies of young exoplanets such as V1298 Tau b may help to answer some of these questions.

\acknowledgments \copyright~2019. All rights reserved. We are grateful to Rodrigo Luger, Dan Foreman-Mackey, Jeffrey Smith, Marcie Smith, Konstantin Batygin, Yanqin Wu, and Eve Lee for helpful discussions, and to Scott Davidoff for guidance on figures. This work made use of the gaia-kepler.fun crossmatch database created by Megan Bedell. T.J.D. and E.E.M. gratefully acknowledge support from the Jet Propulsion Laboratory Exoplanetary Science Initiative. E.E.M. acknowledges support from the NASA NExSS program. E.A.P. is supported through a Hubble Fellowship. Part of this research was carried out at the Jet Propulsion Laboratory, California Institute of Technology, under a contract with NASA. This paper includes data collected by the {\em Kepler} mission, funded by the NASA Science Mission directorate. This work has made use of data from the European Space Agency (ESA) mission {\it Gaia} (\url{https://www.cosmos.esa.int/gaia}), processed by the {\it Gaia} Data Processing and Analysis Consortium (DPAC, \url{https://www.cosmos.esa.int/web/gaia/dpac/consortium}). Funding for the DPAC has been provided by national institutions, in particular the institutions participating in the {\it Gaia} Multilateral Agreement. Some data presented herein were obtained at W.M. Keck Observatory, which is operated as a scientific partnership among the CIT, the Univ. of California and NASA. The authors recognize and acknowledge the significant cultural role and reverence that the summit of Mauna Kea has always had within the indigenous Hawaiian community.  We are most fortunate to conduct observations from this mountain. 

\vspace{5mm}
\facilities{Gaia, Keck:I (HIRES), Keck:II (NIRC2), Kepler}

\software{\texttt{astropy} \citep{exoplanet:astropy13, exoplanet:astropy18},
          \texttt{BANYAN $\Sigma$} \citep{Gagne2018},
          \texttt{emcee} \citep{foremanmackey2013},
          \texttt{exoplanet} \citep{exoplanet:exoplanet},
          \texttt{EVEREST 2.0} \citep{Luger2018},
          \texttt{K2SC} \citep{Aigrain2016},
          \texttt{K2PHOT} \citep{Petigura2018},
          \texttt{lightkurve} \citep{lightkurve},
          \texttt{lmfit} \citep{lmfit}, 
          \texttt{matplotlib} \citep{matplotlib},
          \texttt{PyMC3} \citep{exoplanet:pymc3},
          \texttt{PyTransit} \citep{Parviainen2015},
          \texttt{RadVel} \citep{Fulton2018},
          \texttt{starry} \citep{exoplanet:luger18},
          \texttt{TERRA} \citep{petigura2013b},
          \texttt{theano} \citep{exoplanet:theano}
          }
          
\appendix
\section{False-positive scenario assessment} \label{appendix:falsepositives}
An eclipsing binary or a planet transiting a star other than V1298 Tau would be capable of producing the observed transit signal. Below, we assess the likelihood of each possible false-positive scenario in a quantitative manner when possible and qualitatively otherwise.

We first consider a scenario in which the transit signal is due to an unassociated eclipsing binary widely separated from V1298 Tau on the sky. Occasionally, a bright eclipsing binary within the telescope field of view may contaminate pixels elsewhere on the detector and lead to spurious detections of transiting planets. There are several stars, some saturated, in the vicinity of V1298 Tau on the \textit{K2} detector. We examined light curves for each of these stars (EPIC IDs 210819568, 210817793, 210818376)  and confirmed that none showed eclipses. We similarly examined light curves for neighboring stars identified using the \textsc{star.neighbors} function with \textsc{everest 2.0}, and found no eclipsing binaries among those stars (EPIC IDs 210787602, 210755820, 210852007, 210832303, 210829398, 210769653, 210828429, 210815489, 210761528).

Our group also performed a systematic search of every \textit{K2} light curve from Campaign 4 \citep{crossfield2016} for periodic transits or eclipses using the \textsc{terra} code \citep{petigura2013b}. We used the output of this search to attempt a cross-match between the ephemeris of V1298 Tau with known periodic signals detected by \textsc{terra} above a signal-to-noise of 10. We examined the \textit{K2} light curves for 50 stars having signals detected with periods within 0.5 days of the period of V1298 Tau b or half of that period. We found none of those stars to show eclipses coinciding with the transit times of V1298 Tau b. A search of the \textit{Kepler} Eclipsing Binary Catalog (http://keplerebs.villanova.edu/) for eclipsing binaries discovered from Campaign 4 of the \textit{K2} mission also yielded no candidates with period differences of $<$0.64 days relative to V1298 Tau b. 

Next we consider closely-projected eclipsing binaries that are either associated or unassociated with V1298 Tau. An eclipsing binary can dim by at most 100\%. The observed transit depth therefore sets an upper limit to the brightness difference in the \textit{Kepler} bandpass of 5.7 magnitudes between V1298 Tau and a putative eclipsing binary capable of producing the observed transit depths. The target pixel files for V1298 Tau contain an unassociated star, Gaia DR2 51886331671984640, which resides 19.8 arcseconds to the southwest. This star is more than 6 magnitudes fainter than V1298 Tau at optical wavelengths, and therefore too faint to be responsible for the periodic transits. Using the interactive \textsc{lightkurve}\footnote{\url{http://docs.lightkurve.org}} tool, we examined time series photometry for V1298 Tau from a 24''$\times$24'' square aperture as well as from a 12''$\times$8'' rectangular aperture surrounding the fainter source to the southwest. We confirmed that the fainter source is not an eclipsing binary and that the transit signal originates from the region surrounding V1298 Tau.

False-positive scenarios involving hierarchical triples, e.g. a bound eclipsing binary or a planet transiting an undetected companion to V1298 Tau, are similarly disfavored. \citet{Nguyen2012} searched for close companions to V1298 Tau through radial velocity monitoring. Those measurements rule out most scenarios in which V1298 Tau hosts a stellar mass companion with an orbital period between 1--100 days. Furthermore, dynamical stability arguments \citep{MardlingAarseth2001} and empirical evidence from hierarchical multiple star systems \citep{Tokovinin2018} suggest the ratio of the outer to inner periods in a coplanar hierarchical triple must exceed 4.7. Thus, hierarchical triple scenarios involving a coplanar stellar companion with an orbital period $<$113 days ($\sim$0.5~AU, $\sim$5~milliarcseconds) are \textit{a priori} unlikely.

Although hierarchical triple scenarios are considered unlikely, we consider the implications of such scenarios here. If V1298 Tau is in fact a binary and the transiting body orbits the primary star, the effect of flux dilution is then maximized when the undetected companion is equal in brightness. In such a scenario, the planet's true radius is larger by a factor of $\sqrt{2}$, corresponding to a radius for the transiting companion of 1.3~\rjup \citep{Ciardi2015}. Notably, such a scenario would still favor a planetary nature for the transiting companion given that low-mass stars and brown dwarfs have radii $\gtrsim$2~\rjup at ages $<$40~Myr. However, if the transiting body instead orbits an undetected companion star, then the radius is most severely underestimated when the optical contrast between a putative companion and V1298 Tau is maximized. Using evolutionary models and previously established methods \citep{David2016}, we calculated the corrected companion radius and mass in such scenarios, finding that the mass of the transiting body could exceed 0.1 $M_\odot$ if V1298 Tau hosts an undetected companion of 0.5 $M_\odot$ or lower.

\begin{figure*}
    \centering
    \includegraphics[width=\textwidth]{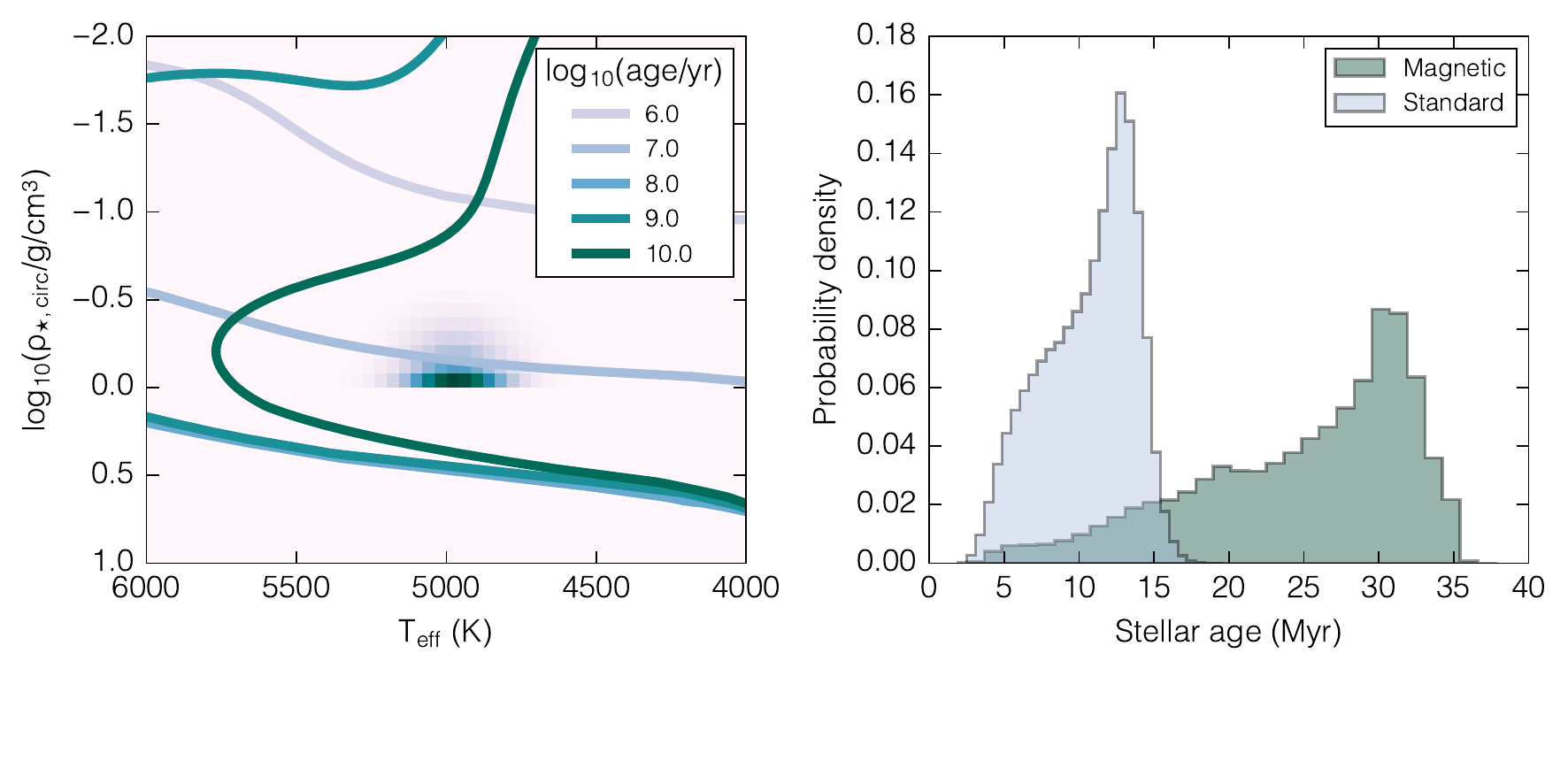}
    \caption{Age inference in a temperature-density diagram. \textit{Left:} The mean stellar density as a function of effective temperature varies with age (solid lines). The two-dimensional histogram shows a normal distribution in temperature and the mean stellar density posterior from the V1298 Tau b transit fit assuming a circular orbit. \textit{Right:} The probability distributions in age for V1298 Tau resulting from a linear interpolation of the values at left using two different evolutionary models \citep{Dotter2008, Feiden2016}.}
    \label{fig:stdensity}
\end{figure*}

The mean stellar density measured from the transits provides another means of assessing false-positive scenarios. Using the posteriors in period and $a/R_*$ from a circular orbit fit, we calculated the mean stellar density and compared it to stellar evolution models in a temperature-density diagram (Figure~\ref{fig:stdensity}). Although this analysis assumes a circular orbit and should thus be regarded with caution, we find the stellar age implied by the density posterior is consistent with our Hertzsprung-Russell diagram analysis and inconsistent with a main sequence star or a post-main sequence star of similar effective temperature and with an age less than the age of the universe.

\section{Favorability for follow-up observations} \label{sec:favorability}
We compared V1298 Tau b with the population of confirmed transiting exoplanets using data from the NASA Exoplanet Archive \citep{Akeson2013}, accessed on May 15, 2019. We first compared the favorability for atmospheric characterization between V1298 Tau b and known transiting planets. We followed the methods of \citet{Vanderburg2016b} to predict a scaled signal-to-noise for transmission spectroscopy observations using the following equations:

\begin{equation}
    S/N \propto \frac{R_P H \sqrt{Ft_{14}}}{R_\star^2}
\end{equation}

and

\begin{equation}
    H = \frac{k_\mathrm{B} T_\mathrm{eq}}{\mu g},
\end{equation}

where $R_P$ and $R_\star$ are the radii of the planet and star, respectively, $H$ is the atmospheric scale height, $F$ is the stellar flux at $H$-band, $t_{14}$ is the total transit duration, $k_\mathrm{B}$ is Boltzmann's constant, $T_\mathrm{eq}$ is the planet's equilibrium temperature, $\mu$ is the mean molecular weight of the planet's atmosphere, and $g$ is the planet's surface gravity. When available, we used the known mass and radius to calculate surface gravity. Otherwise, we calculated a predicted mass using the planet's radius and an exoplanet mass-radius relation \citep{Weiss2018}. The mean molecular weight was fixed to the Jovian value for planets with radii larger than 1.5 $R_\oplus$ and to the terrestrial value otherwise. 

\begin{figure}
    \centering
    \includegraphics[width=\linewidth]{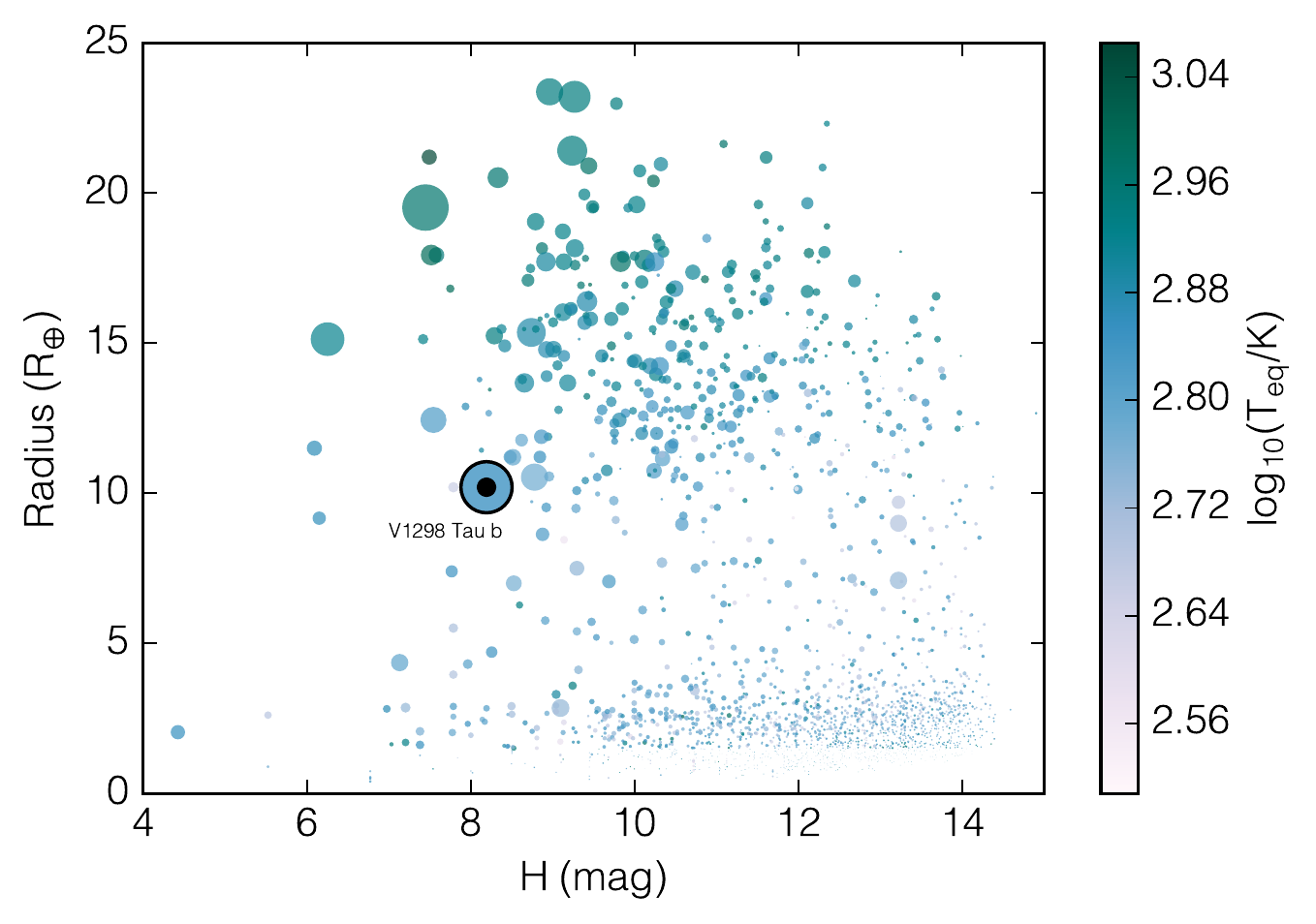}
    \caption{Favorability for transmission spectroscopy. The distribution of known transiting exoplanets as a function of $H$-band brightness and planet radius. The point size is proportional to the scaled signal-to-noise predicted for transmission spectroscopy observations. Point colors indicate the logarithm of the planet equilibrium temperature, assuming an albedo of 0.3. V1298 Tau b is shown as the point with the black border. The outer boundary represents an optimistic case of $M_P$ = 10~\mearth, while the black point at the center indicates the case of $M_P$ = 100~\mearth.}
    \label{fig:tts}
\end{figure}

The scaled transmission spectrum signal-to-noise as a function of $H$-band brightness, planet radius, and equilibrium temperature is depicted in Figure~\ref{fig:tts}. Using this simplistic methodology, we determined that V1298 Tau b ranks among the top 40 most favorable transiting exoplanets for atmospheric characterization. However, in the optimistic case that the planet in fact has a core-dominated composition with a total mass of 30~\mearth, V1298 Tau b would be in the top 5 in terms of favorability for infrared transmission spectroscopy using the metrics described. Of course, the unusually high stellar activity presents both challenges (potentially requiring multiple transit observations to disentangle stellar and planetary signals) and opportunities (for studying star-planet interactions). 

There are presently 16 known exoplanets with radii $>0.8$~\rjup, periods between 10--50 days, and well-measured masses $<10$~\mjup. Those planets have masses ranging from 0.2--3.5~\mjup. For this range of plausible planet masses, the expected radial velocity semi-amplitude ranges from 13--230~\ms. If V1298 Tau b is contracting, implying its present-day radius may be an imprecise indicator of its mass, the true radial velocity amplitude may be lower. We have measured the optical radial velocity jitter to be $\sigma_{RV} = $~216, 71, and 5~\ms over 5.2 days, 10 hours, and 30 minutes, respectively. Measuring the planet's mass may be feasible with newly commissioned infrared spectrographs, due to the lower expected intrinsic stellar variability at redder wavelengths.

V1298 Tau is also a favorable target for measuring the planet's obliquity through the Rossiter-McLaughlin (R-M) effect or Doppler tomography. Neglecting limb darkening effects, the R-M amplitude is approximately $A_{RM} \sim v\sin{i} \times (R_P/R_\star)^2 \times \sqrt{1-b^2}$, which is 117~\ms for V1298 Tau. Assuming an albedo of 0.3 and neglecting remnant heat from the planet's formation, the predicted secondary eclipse depth is $\delta_{SE}$ = ($R_P/R_\star$)$^2$ ($T_{eq}$/$T_\star$) $\approx$ 610~ppm, which is detectable with the \textit{Spitzer} space telescope. A measurement of the temperature of V1298 Tau b could help to constrain planet evolution models.

%\bibliography{V1298Tau}

\begin{thebibliography}{}
\expandafter\ifx\csname natexlab\endcsname\relax\def\natexlab#1{#1}\fi
\providecommand{\url}[1]{\href{#1}{#1}}
\providecommand{\dodoi}[1]{doi:~\href{http://doi.org/#1}{\nolinkurl{#1}}}
\providecommand{\doeprint}[1]{\href{http://ascl.net/#1}{\nolinkurl{http://ascl.net/#1}}}
\providecommand{\doarXiv}[1]{\href{https://arxiv.org/abs/#1}{\nolinkurl{https://arxiv.org/abs/#1}}}

\bibitem[{{Aigrain} {et~al.}(2016){Aigrain}, {Parviainen}, \&
  {Pope}}]{Aigrain2016}
{Aigrain}, S., {Parviainen}, H., \& {Pope}, B.~J.~S. 2016, \mnras, 459, 2408,
  \dodoi{10.1093/mnras/stw706}

\bibitem[{{Akeson} {et~al.}(2013){Akeson}, {Chen}, {Ciardi}, {Crane}, {Good},
  {Harbut}, {Jackson}, {Kane}, {Laity}, {Leifer}, {Lynn}, {McElroy}, {Papin},
  {Plavchan}, {Ram{\'{\i}}rez}, {Rey}, {von Braun}, {Wittman}, {Abajian},
  {Ali}, {Beichman}, {Beekley}, {Berriman}, {Berukoff}, {Bryden}, {Chan},
  {Groom}, {Lau}, {Payne}, {Regelson}, {Saucedo}, {Schmitz}, {Stauffer},
  {Wyatt}, \& {Zhang}}]{Akeson2013}
{Akeson}, R.~L., {Chen}, X., {Ciardi}, D., {et~al.} 2013, \pasp, 125, 989,
  \dodoi{10.1086/672273}

\bibitem[{{Anderson} \& {Lai}(2017)}]{AndersonLai2017}
{Anderson}, K.~R., \& {Lai}, D. 2017, \mnras, 472, 3692,
  \dodoi{10.1093/mnras/stx2250}

\bibitem[{{Antonini} {et~al.}(2016){Antonini}, {Hamers}, \&
  {Lithwick}}]{Antonini2016}
{Antonini}, F., {Hamers}, A.~S., \& {Lithwick}, Y. 2016, \aj, 152, 174,
  \dodoi{10.3847/0004-6256/152/6/174}

\bibitem[{{Astropy Collaboration} {et~al.}(2013){Astropy Collaboration},
  {Robitaille}, {Tollerud}, {Greenfield}, {Droettboom}, {Bray}, {Aldcroft},
  {Davis}, {Ginsburg}, {Price-Whelan}, {Kerzendorf}, {Conley}, {Crighton},
  {Barbary}, {Muna}, {Ferguson}, {Grollier}, {Parikh}, {Nair}, {Unther},
  {Deil}, {Woillez}, {Conseil}, {Kramer}, {Turner}, {Singer}, {Fox}, {Weaver},
  {Zabalza}, {Edwards}, {Azalee Bostroem}, {Burke}, {Casey}, {Crawford},
  {Dencheva}, {Ely}, {Jenness}, {Labrie}, {Lim}, {Pierfederici}, {Pontzen},
  {Ptak}, {Refsdal}, {Servillat}, \& {Streicher}}]{exoplanet:astropy13}
{Astropy Collaboration}, {Robitaille}, T.~P., {Tollerud}, E.~J., {et~al.} 2013,
  aap, 558, A33, \dodoi{10.1051/0004-6361/201322068}

\bibitem[{{Astropy Collaboration} {et~al.}(2018){Astropy Collaboration},
  {Price-Whelan}, {Sip{H o}cz}, {G{"u}nther}, {Lim}, {Crawford}, {Conseil},
  {Shupe}, {Craig}, {Dencheva}, {Ginsburg}, {VanderPlas}, {Bradley},
  {P{'e}rez-Su{'a}rez}, {de Val-Borro}, {Aldcroft}, {Cruz}, {Robitaille},
  {Tollerud}, {Ardelean}, {Babej}, {Bach}, {Bachetti}, {Bakanov}, {Bamford},
  {Barentsen}, {Barmby}, {Baumbach}, {Berry}, {Biscani}, {Boquien}, {Bostroem},
  {Bouma}, {Brammer}, {Bray}, {Breytenbach}, {Buddelmeijer}, {Burke},
  {Calderone}, {Cano Rodr{'{i}}guez}, {Cara}, {Cardoso}, {Cheedella}, {Copin},
  {Corrales}, {Crichton}, {D'Avella}, {Deil}, {Depagne}, {Dietrich}, {Donath},
  {Droettboom}, {Earl}, {Erben}, {Fabbro}, {Ferreira}, {Finethy}, {Fox},
  {Garrison}, {Gibbons}, {Goldstein}, {Gommers}, {Greco}, {Greenfield},
  {Groener}, {Grollier}, {Hagen}, {Hirst}, {Homeier}, {Horton}, {Hosseinzadeh},
  {Hu}, {Hunkeler}, {Ivezi{'c}}, {Jain}, {Jenness}, {Kanarek}, {Kendrew},
  {Kern}, {Kerzendorf}, {Khvalko}, {King}, {Kirkby}, {Kulkarni}, {Kumar},
  {Lee}, {Lenz}, {Littlefair}, {Ma}, {Macleod}, {Mastropietro}, {McCully},
  {Montagnac}, {Morris}, {Mueller}, {Mumford}, {Muna}, {Murphy}, {Nelson},
  {Nguyen}, {Ninan}, {N{"o}the}, {Ogaz}, {Oh}, {Parejko}, {Parley}, {Pascual},
  {Patil}, {Patil}, {Plunkett}, {Prochaska}, {Rastogi}, {Reddy Janga},
  {Sabater}, {Sakurikar}, {Seifert}, {Sherbert}, {Sherwood-Taylor}, {Shih},
  {Sick}, {Silbiger}, {Singanamalla}, {Singer}, {Sladen}, {Sooley},
  {Sornarajah}, {Streicher}, {Teuben}, {Thomas}, {Tremblay}, {Turner},
  {Terr{'o}n}, {van Kerkwijk}, {de la Vega}, {Watkins}, {Weaver}, {Whitmore},
  {Woillez}, {Zabalza}, \& {Astropy Contributors}}]{exoplanet:astropy18}
{Astropy Collaboration}, {Price-Whelan}, A.~M., {Sip{H o}cz}, B.~M., {et~al.}
  2018, aj, 156, 123, \dodoi{10.3847/1538-3881/aabc4f}

\bibitem[{{Baraffe} {et~al.}(2003){Baraffe}, {Chabrier}, {Barman}, {Allard}, \&
  {Hauschildt}}]{Baraffe2003}
{Baraffe}, I., {Chabrier}, G., {Barman}, T.~S., {Allard}, F., \& {Hauschildt},
  P.~H. 2003, \aap, 402, 701, \dodoi{10.1051/0004-6361:20030252}

\bibitem[{{Baraffe} {et~al.}(2015){Baraffe}, {Homeier}, {Allard}, \&
  {Chabrier}}]{Baraffe2015}
{Baraffe}, I., {Homeier}, D., {Allard}, F., \& {Chabrier}, G. 2015, \aap, 577,
  A42, \dodoi{10.1051/0004-6361/201425481}

\bibitem[{Barentsen {et~al.}(2019)Barentsen, Hedges, Vinícius, Saunders,
  gully, Hall, Sagear, KenMighell, Barclay, Bell, Zhang, Turtelboom,
  Berta-Thompson, Williams, III, Vincello, \& Sundaram}]{lightkurve}
Barentsen, G., Hedges, C., Vinícius, Z., {et~al.} 2019, KeplerGO/lightkurve:
  Lightkurve v1.0b29, \dodoi{10.5281/zenodo.2565212}.
\newblock \url{https://doi.org/10.5281/zenodo.2565212}

\bibitem[{{Barrado y Navascu{\'e}s} {et~al.}(2001){Barrado y Navascu{\'e}s},
  {Deliyannis}, \& {Stauffer}}]{Barrado2001}
{Barrado y Navascu{\'e}s}, D., {Deliyannis}, C.~P., \& {Stauffer}, J.~R. 2001,
  \apj, 549, 452, \dodoi{10.1086/319045}

\bibitem[{{Batygin} {et~al.}(2016){Batygin}, {Bodenheimer}, \&
  {Laughlin}}]{Batygin2016}
{Batygin}, K., {Bodenheimer}, P.~H., \& {Laughlin}, G.~P. 2016, \apj, 829, 114,
  \dodoi{10.3847/0004-637X/829/2/114}

\bibitem[{{Binks} \& {Jeffries}(2014)}]{BinksJeffries2014}
{Binks}, A.~S., \& {Jeffries}, R.~D. 2014, \mnras, 438, L11,
  \dodoi{10.1093/mnrasl/slt141}

\bibitem[{{Boley} {et~al.}(2016){Boley}, {Granados Contreras}, \&
  {Gladman}}]{Boley2016}
{Boley}, A.~C., {Granados Contreras}, A.~P., \& {Gladman}, B. 2016, \apjl, 817,
  L17, \dodoi{10.3847/2041-8205/817/2/L17}

\bibitem[{{Boller} {et~al.}(2016){Boller}, {Freyberg}, {Tr{\"u}mper}, {Haberl},
  {Voges}, \& {Nandra}}]{Boller2016}
{Boller}, T., {Freyberg}, M.~J., {Tr{\"u}mper}, J., {et~al.} 2016, \aap, 588,
  A103, \dodoi{10.1051/0004-6361/201525648}

\bibitem[{{Bouvier} {et~al.}(2018){Bouvier}, {Barrado}, {Moraux}, {Stauffer},
  {Rebull}, {Hillenbrand}, {Bayo}, {Boisse}, {Bouy}, {DiFolco}, {Lillo-Box}, \&
  {Calder{\'o}n}}]{Bouvier2018}
{Bouvier}, J., {Barrado}, D., {Moraux}, E., {et~al.} 2018, \aap, 613, A63,
  \dodoi{10.1051/0004-6361/201731881}

\bibitem[{{Bowler}(2016)}]{Bowler2016}
{Bowler}, B.~P. 2016, \pasp, 128, 102001,
  \dodoi{10.1088/1538-3873/128/968/102001}

\bibitem[{{Burrows} {et~al.}(1997){Burrows}, {Marley}, {Hubbard}, {Lunine},
  {Guillot}, {Saumon}, {Freedman}, {Sudarsky}, \& {Sharp}}]{Burrows1997}
{Burrows}, A., {Marley}, M., {Hubbard}, W.~B., {et~al.} 1997, \apj, 491, 856,
  \dodoi{10.1086/305002}

\bibitem[{{Capitanio} {et~al.}(2017){Capitanio}, {Lallement}, {Vergely},
  {Elyajouri}, \& {Monreal-Ibero}}]{Capitanio2017}
{Capitanio}, L., {Lallement}, R., {Vergely}, J.~L., {Elyajouri}, M., \&
  {Monreal-Ibero}, A. 2017, \aap, 606, A65, \dodoi{10.1051/0004-6361/201730831}

\bibitem[{{Chen} \& {Rogers}(2016)}]{Chen:Rogers:2016}
{Chen}, H., \& {Rogers}, L.~A. 2016, \apj, 831, 180,
  \dodoi{10.3847/0004-637X/831/2/180}

\bibitem[{{Ciardi} {et~al.}(2015){Ciardi}, {Beichman}, {Horch}, \&
  {Howell}}]{Ciardi2015}
{Ciardi}, D.~R., {Beichman}, C.~A., {Horch}, E.~P., \& {Howell}, S.~B. 2015,
  \apj, 805, 16, \dodoi{10.1088/0004-637X/805/1/16}

\bibitem[{{Ciardi} {et~al.}(2018){Ciardi}, {Crossfield}, {Feinstein},
  {Schlieder}, {Petigura}, {David}, {Bristow}, {Patel}, {Arnold}, {Benneke},
  {Christiansen}, {Dressing}, {Fulton}, {Howard}, {Isaacson}, {Sinukoff}, \&
  {Thackeray}}]{Ciardi2018}
{Ciardi}, D.~R., {Crossfield}, I.~J.~M., {Feinstein}, A.~D., {et~al.} 2018,
  \aj, 155, 10, \dodoi{10.3847/1538-3881/aa9921}

\bibitem[{{Claret} {et~al.}(2012){Claret}, {Hauschildt}, \&
  {Witte}}]{Claret2012}
{Claret}, A., {Hauschildt}, P.~H., \& {Witte}, S. 2012, \aap, 546, A14,
  \dodoi{10.1051/0004-6361/201219849}

\bibitem[{{Claret} {et~al.}(2013){Claret}, {Hauschildt}, \&
  {Witte}}]{Claret2013}
---. 2013, \aap, 552, A16, \dodoi{10.1051/0004-6361/201220942}

\bibitem[{{Crossfield} {et~al.}(2016){Crossfield}, {Ciardi}, {Petigura},
  {Sinukoff}, {Schlieder}, {Howard}, {Beichman}, {Isaacson}, {Dressing},
  {Christiansen}, {Fulton}, {L{\'e}pine}, {Weiss}, {Hirsch}, {Livingston},
  {Baranec}, {Law}, {Riddle}, {Ziegler}, {Howell}, {Horch}, {Everett}, {Teske},
  {Martinez}, {Obermeier}, {Benneke}, {Scott}, {Deacon}, {Aller}, {Hansen},
  {Mancini}, {Ciceri}, {Brahm}, {Jord{\'a}n}, {Knutson}, {Henning}, {Bonnefoy},
  {Liu}, {Crepp}, {Lothringer}, {Hinz}, {Bailey}, {Skemer}, \&
  {Defrere}}]{crossfield2016}
{Crossfield}, I.~J.~M., {Ciardi}, D.~R., {Petigura}, E.~A., {et~al.} 2016,
  \apjs, 226, 7, \dodoi{10.3847/0067-0049/226/1/7}

\bibitem[{{Curtis} {et~al.}(2018){Curtis}, {Vanderburg}, {Torres}, {Kraus},
  {Huber}, {Mann}, {Rizzuto}, {Isaacson}, {Howard}, {Henze}, {Fulton}, \&
  {Wright}}]{Curtis2018}
{Curtis}, J.~L., {Vanderburg}, A., {Torres}, G., {et~al.} 2018, \aj, 155, 173,
  \dodoi{10.3847/1538-3881/aab49c}

\bibitem[{{Daemgen} {et~al.}(2015){Daemgen}, {Bonavita}, {Jayawardhana},
  {Lafreni{\`e}re}, \& {Janson}}]{Daemgen2015}
{Daemgen}, S., {Bonavita}, M., {Jayawardhana}, R., {Lafreni{\`e}re}, D., \&
  {Janson}, M. 2015, \apj, 799, 155, \dodoi{10.1088/0004-637X/799/2/155}

\bibitem[{{David} {et~al.}(2016){David}, {Hillenbrand}, {Petigura},
  {Carpenter}, {Crossfield}, {Hinkley}, {Ciardi}, {Howard}, {Isaacson}, {Cody},
  {Schlieder}, {Beichman}, \& {Barenfeld}}]{David2016}
{David}, T.~J., {Hillenbrand}, L.~A., {Petigura}, E.~A., {et~al.} 2016, \nat,
  534, 658, \dodoi{10.1038/nature18293}

\bibitem[{{Davies} {et~al.}(2014){Davies}, {Gregory}, \&
  {Greaves}}]{Davies2014}
{Davies}, C.~L., {Gregory}, S.~G., \& {Greaves}, J.~S. 2014, \mnras, 444, 1157,
  \dodoi{10.1093/mnras/stu1488}

\bibitem[{{Dawson} \& {Chiang}(2014)}]{DawsonChiang2014}
{Dawson}, R.~I., \& {Chiang}, E. 2014, Science, 346, 212,
  \dodoi{10.1126/science.1256943}

\bibitem[{{Dawson} \& {Johnson}(2012)}]{DawsonJohnson2012}
{Dawson}, R.~I., \& {Johnson}, J.~A. 2012, \apj, 756, 122,
  \dodoi{10.1088/0004-637X/756/2/122}

\bibitem[{{Dawson} {et~al.}(2015){Dawson}, {Murray-Clay}, \&
  {Johnson}}]{Dawson2015}
{Dawson}, R.~I., {Murray-Clay}, R.~A., \& {Johnson}, J.~A. 2015, \apj, 798, 66,
  \dodoi{10.1088/0004-637X/798/2/66}

\bibitem[{{Deming} {et~al.}(2013){Deming}, {Wilkins}, {McCullough}, {Burrows},
  {Fortney}, {Agol}, {Dobbs-Dixon}, {Madhusudhan}, {Crouzet}, {Desert},
  {Gilliland}, {Haynes}, {Knutson}, {Line}, {Magic}, {Mandell}, {Ranjan},
  {Charbonneau}, {Clampin}, {Seager}, \& {Showman}}]{Deming2013}
{Deming}, D., {Wilkins}, A., {McCullough}, P., {et~al.} 2013, \apj, 774, 95,
  \dodoi{10.1088/0004-637X/774/2/95}

\bibitem[{{Dobashi} {et~al.}(2005){Dobashi}, {Uehara}, {Kandori}, {Sakurai},
  {Kaiden}, {Umemoto}, \& {Sato}}]{Dobashi2005}
{Dobashi}, K., {Uehara}, H., {Kandori}, R., {et~al.} 2005, \pasj, 57, S1,
  \dodoi{10.1093/pasj/57.sp1.S1}

\bibitem[{{Donati} {et~al.}(2016){Donati}, {Moutou}, {Malo}, {Baruteau}, {Yu},
  {H{\'e}brard}, {Hussain}, {Alencar}, {M{\'e}nard}, {Bouvier}, {Petit},
  {Takami}, {Doyon}, \& {Cameron}}]{Donati2016}
{Donati}, J.~F., {Moutou}, C., {Malo}, L., {et~al.} 2016, \nat, 534, 662,
  \dodoi{10.1038/nature18305}

\bibitem[{{Dong} {et~al.}(2014){Dong}, {Katz}, \& {Socrates}}]{Dong2014}
{Dong}, S., {Katz}, B., \& {Socrates}, A. 2014, \apjl, 781, L5,
  \dodoi{10.1088/2041-8205/781/1/L5}

\bibitem[{{Dotter} {et~al.}(2008){Dotter}, {Chaboyer}, {Jevremovi{\'c}},
  {Kostov}, {Baron}, \& {Ferguson}}]{Dotter2008}
{Dotter}, A., {Chaboyer}, B., {Jevremovi{\'c}}, D., {et~al.} 2008, \apjs, 178,
  89, \dodoi{10.1086/589654}

\bibitem[{{Espinoza}(2018)}]{Espinoza2018}
{Espinoza}, N. 2018, Research Notes of the American Astronomical Society, 2,
  209, \dodoi{10.3847/2515-5172/aaef38}

\bibitem[{{Evans}(2018)}]{Evans2018}
{Evans}, D.~F. 2018, Research Notes of the American Astronomical Society, 2,
  20, \dodoi{10.3847/2515-5172/aac173}

\bibitem[{{Feiden}(2016)}]{Feiden2016}
{Feiden}, G.~A. 2016, \aap, 593, A99, \dodoi{10.1051/0004-6361/201527613}

\bibitem[{{Findeisen} \& {Hillenbrand}(2010)}]{FindeisenHillenbrand2010}
{Findeisen}, K., \& {Hillenbrand}, L. 2010, \aj, 139, 1338,
  \dodoi{10.1088/0004-6256/139/4/1338}

\bibitem[{Foreman-Mackey {et~al.}(2019)Foreman-Mackey, Barentsen, \&
  Barclay}]{exoplanet:exoplanet}
Foreman-Mackey, D., Barentsen, G., \& Barclay, T. 2019, dfm/exoplanet:
  exoplanet v0.1.5, \dodoi{10.5281/zenodo.2587222}.
\newblock \url{https://doi.org/10.5281/zenodo.2587222}

\bibitem[{{Foreman-Mackey} {et~al.}(2013){Foreman-Mackey}, {Hogg}, {Lang}, \&
  {Goodman}}]{foremanmackey2013}
{Foreman-Mackey}, D., {Hogg}, D.~W., {Lang}, D., \& {Goodman}, J. 2013, \pasp,
  125, 306, \dodoi{10.1086/670067}

\bibitem[{{Fortney} {et~al.}(2007){Fortney}, {Marley}, \&
  {Barnes}}]{Fortney2007}
{Fortney}, J.~J., {Marley}, M.~S., \& {Barnes}, J.~W. 2007, \apj, 659, 1661,
  \dodoi{10.1086/512120}

\bibitem[{{Fortney} \& {Nettelmann}(2010)}]{Fortney2010}
{Fortney}, J.~J., \& {Nettelmann}, N. 2010, \ssr, 152, 423,
  \dodoi{10.1007/s11214-009-9582-x}

\bibitem[{{Fulton} \& {Petigura}(2018)}]{Fulton:Petigura:2018}
{Fulton}, B.~J., \& {Petigura}, E.~A. 2018, \aj, 156, 264,
  \dodoi{10.3847/1538-3881/aae828}

\bibitem[{{Fulton} {et~al.}(2018){Fulton}, {Petigura}, {Blunt}, \&
  {Sinukoff}}]{Fulton2018}
{Fulton}, B.~J., {Petigura}, E.~A., {Blunt}, S., \& {Sinukoff}, E. 2018, \pasp,
  130, 044504, \dodoi{10.1088/1538-3873/aaaaa8}

\bibitem[{{Fulton} {et~al.}(2017){Fulton}, {Petigura}, {Howard}, {Isaacson},
  {Marcy}, {Cargile}, {Hebb}, {Weiss}, {Johnson}, {Morton}, {Sinukoff},
  {Crossfield}, \& {Hirsch}}]{Fulton2017}
{Fulton}, B.~J., {Petigura}, E.~A., {Howard}, A.~W., {et~al.} 2017, \aj, 154,
  109, \dodoi{10.3847/1538-3881/aa80eb}

\bibitem[{{Furlan} {et~al.}(2017){Furlan}, {Ciardi}, {Everett}, {Saylors},
  {Teske}, {Horch}, {Howell}, {van Belle}, {Hirsch}, {Gautier}, {Adams},
  {Barrado}, {Cartier}, {Dressing}, {Dupree}, {Gilliland}, {Lillo-Box},
  {Lucas}, \& {Wang}}]{Furlan2017}
{Furlan}, E., {Ciardi}, D.~R., {Everett}, M.~E., {et~al.} 2017, \aj, 153, 71,
  \dodoi{10.3847/1538-3881/153/2/71}

\bibitem[{{Gagn{\'e}} {et~al.}(2018){Gagn{\'e}}, {Mamajek}, {Malo}, {Riedel},
  {Rodriguez}, {Lafreni{\`e}re}, {Faherty}, {Roy-Loubier}, {Pueyo}, {Robin}, \&
  {Doyon}}]{Gagne2018}
{Gagn{\'e}}, J., {Mamajek}, E.~E., {Malo}, L., {et~al.} 2018, \apj, 856, 23,
  \dodoi{10.3847/1538-4357/aaae09}

\bibitem[{{Gaia Collaboration} {et~al.}(2018){Gaia Collaboration}, {Brown},
  {Vallenari}, {Prusti}, {de Bruijne}, {Babusiaux}, {Bailer-Jones}, {Biermann},
  {Evans}, {Eyer}, \& et~al.}]{Gaia2018}
{Gaia Collaboration}, {Brown}, A.~G.~A., {Vallenari}, A., {et~al.} 2018, \aap,
  616, A1, \dodoi{10.1051/0004-6361/201833051}

\bibitem[{{Gelman} \& {Rubin}(1992)}]{Gelman:Rubin:1992}
{Gelman}, A., \& {Rubin}, D.~B. 1992, Statistical Science, 7, 457,
  \dodoi{10.1214/ss/1177011136}

\bibitem[{{Ginzburg} {et~al.}(2018){Ginzburg}, {Schlichting}, \&
  {Sari}}]{Ginzburg2018}
{Ginzburg}, S., {Schlichting}, H.~E., \& {Sari}, R. 2018, \mnras, 476, 759,
  \dodoi{10.1093/mnras/sty290}

\bibitem[{{Girardi} {et~al.}(2012){Girardi}, {Barbieri}, {Groenewegen},
  {Marigo}, {Bressan}, {Rocha-Pinto}, {Santiago}, {Camargo}, \& {da
  Costa}}]{Girardi2012}
{Girardi}, L., {Barbieri}, M., {Groenewegen}, M.~A.~T., {et~al.} 2012,
  Astrophysics and Space Science Proceedings, 26, 165,
  \dodoi{10.1007/978-3-642-18418-5_17}

\bibitem[{{Goodman} \& {Weare}(2010)}]{GoodmanWeare2010}
{Goodman}, J., \& {Weare}, J. 2010, Communications in Applied Mathematics and
  Computational Science, Vol.~5, No.~1, p.~65-80, 2010, 5, 65,
  \dodoi{10.2140/camcos.2010.5.65}

\bibitem[{{Grankin} {et~al.}(2007){Grankin}, {Artemenko}, \&
  {Melnikov}}]{Grankin2007}
{Grankin}, K.~N., {Artemenko}, S.~A., \& {Melnikov}, S.~Y. 2007, Information
  Bulletin on Variable Stars, 5752

\bibitem[{{Gray}(2005)}]{Gray2005}
{Gray}, D.~F. 2005, {The Observation and Analysis of Stellar Photospheres}
  (Cambridge University Press)

\bibitem[{{Gupta} \& {Schlichting}(2018)}]{GuptaSchlichting2018}
{Gupta}, A., \& {Schlichting}, H.~E. 2018, arXiv e-prints.
\newblock \doarXiv{1811.03202}

\bibitem[{{Hansen} \& {Murray}(2012)}]{Hansen:Murray:2012}
{Hansen}, B. M.~S., \& {Murray}, N. 2012, \apj, 751, 158,
  \dodoi{10.1088/0004-637X/751/2/158}

\bibitem[{{Hartmann} {et~al.}(1991){Hartmann}, {Jones}, {Stauffer}, \&
  {Kenyon}}]{Hartmann1991}
{Hartmann}, L., {Jones}, B.~F., {Stauffer}, J.~R., \& {Kenyon}, S.~J. 1991,
  \aj, 101, 1050, \dodoi{10.1086/115747}

\bibitem[{Hoffman \& Gelman(2014)}]{hoffman2014}
Hoffman, M.~D., \& Gelman, A. 2014, Journal of Machine Learning Research, 15,
  1593

\bibitem[{{Howard} {et~al.}(2010{\natexlab{a}}){Howard}, {Johnson}, {Marcy},
  {Fischer}, {Wright}, {Bernat}, {Henry}, {Peek}, {Isaacson}, {Apps}, {Endl},
  {Cochran}, {Valenti}, {Anderson}, \& {Piskunov}}]{Howard2010}
{Howard}, A.~W., {Johnson}, J.~A., {Marcy}, G.~W., {et~al.} 2010{\natexlab{a}},
  \apj, 721, 1467, \dodoi{10.1088/0004-637X/721/2/1467}

\bibitem[{{Howard} {et~al.}(2010{\natexlab{b}}){Howard}, {Marcy}, {Johnson},
  {Fischer}, {Wright}, {Isaacson}, {Valenti}, {Anderson}, {Lin}, \&
  {Ida}}]{Howard:etal:2010}
{Howard}, A.~W., {Marcy}, G.~W., {Johnson}, J.~A., {et~al.} 2010{\natexlab{b}},
  Science, 330, 653, \dodoi{10.1126/science.1194854}

\bibitem[{{Howell} {et~al.}(2014){Howell}, {Sobeck}, {Haas}, {Still},
  {Barclay}, {Mullally}, {Troeltzsch}, {Aigrain}, {Bryson}, {Caldwell},
  {Chaplin}, {Cochran}, {Huber}, {Marcy}, {Miglio}, {Najita}, {Smith},
  {Twicken}, \& {Fortney}}]{Howell2014}
{Howell}, S.~B., {Sobeck}, C., {Haas}, M., {et~al.} 2014, \pasp, 126, 398,
  \dodoi{10.1086/676406}

\bibitem[{{Huang} {et~al.}(2016){Huang}, {Wu}, \& {Triaud}}]{Huang2016}
{Huang}, C., {Wu}, Y., \& {Triaud}, A.~H.~M.~J. 2016, \apj, 825, 98,
  \dodoi{10.3847/0004-637X/825/2/98}

\bibitem[{{Hunter}(2007)}]{matplotlib}
{Hunter}, J.~D. 2007, Computing in Science and Engineering, 9, 90,
  \dodoi{10.1109/MCSE.2007.55}

\bibitem[{{Ida} \& {Lin}(2010)}]{Ida:Lin:2010}
{Ida}, S., \& {Lin}, D.~N.~C. 2010, \apj, 719, 810,
  \dodoi{10.1088/0004-637X/719/1/810}

\bibitem[{{Jeffries} {et~al.}(2013){Jeffries}, {Naylor}, {Mayne}, {Bell}, \&
  {Littlefair}}]{Jeffries2013}
{Jeffries}, R.~D., {Naylor}, T., {Mayne}, N.~J., {Bell}, C.~P.~M., \&
  {Littlefair}, S.~P. 2013, \mnras, 434, 2438, \dodoi{10.1093/mnras/stt1180}

\bibitem[{{Jin} {et~al.}(2014){Jin}, {Mordasini}, {Parmentier}, {van Boekel},
  {Henning}, \& {Ji}}]{Jin:etal:2014}
{Jin}, S., {Mordasini}, C., {Parmentier}, V., {et~al.} 2014, \apj, 795, 65,
  \dodoi{10.1088/0004-637X/795/1/65}

\bibitem[{{Johns-Krull} {et~al.}(2016){Johns-Krull}, {McLane}, {Prato},
  {Crockett}, {Jaffe}, {Hartigan}, {Beichman}, {Mahmud}, {Chen}, {Skiff},
  {Cauley}, {Jones}, \& {Mace}}]{JohnsKrull2016}
{Johns-Krull}, C.~M., {McLane}, J.~N., {Prato}, L., {et~al.} 2016, \apj, 826,
  206, \dodoi{10.3847/0004-637X/826/2/206}

\bibitem[{{Jones} {et~al.}(1996){Jones}, {Shetrone}, {Fischer}, \&
  {Soderblom}}]{Jones1996}
{Jones}, B.~F., {Shetrone}, M., {Fischer}, D., \& {Soderblom}, D.~R. 1996, \aj,
  112, 186, \dodoi{10.1086/117999}

\bibitem[{{Kipping}(2010)}]{Kipping2010}
{Kipping}, D.~M. 2010, \mnras, 407, 301,
  \dodoi{10.1111/j.1365-2966.2010.16894.x}

\bibitem[{{Kipping}(2013{\natexlab{a}})}]{Kipping:2013}
---. 2013{\natexlab{a}}, \mnras, 435, 2152, \dodoi{10.1093/mnras/stt1435}

\bibitem[{{Kipping}(2013{\natexlab{b}})}]{Kipping:2013ecc}
---. 2013{\natexlab{b}}, \mnras, 434, L51, \dodoi{10.1093/mnrasl/slt075}

\bibitem[{{Kley} \& {Nelson}(2012)}]{Kley2012}
{Kley}, W., \& {Nelson}, R.~P. 2012, \araa, 50, 211,
  \dodoi{10.1146/annurev-astro-081811-125523}

\bibitem[{{Kolbl} {et~al.}(2015){Kolbl}, {Marcy}, {Isaacson}, \&
  {Howard}}]{kolbl2015}
{Kolbl}, R., {Marcy}, G.~W., {Isaacson}, H., \& {Howard}, A.~W. 2015, \aj, 149,
  18, \dodoi{10.1088/0004-6256/149/1/18}

\bibitem[{{Kraus} {et~al.}(2017){Kraus}, {Herczeg}, {Rizzuto}, {Mann},
  {Slesnick}, {Carpenter}, {Hillenbrand}, \& {Mamajek}}]{Kraus2017}
{Kraus}, A.~L., {Herczeg}, G.~J., {Rizzuto}, A.~C., {et~al.} 2017, \apj, 838,
  150, \dodoi{10.3847/1538-4357/aa62a0}

\bibitem[{{Kraus} {et~al.}(2014){Kraus}, {Shkolnik}, {Allers}, \&
  {Liu}}]{Kraus2014}
{Kraus}, A.~L., {Shkolnik}, E.~L., {Allers}, K.~N., \& {Liu}, M.~C. 2014, \aj,
  147, 146, \dodoi{10.1088/0004-6256/147/6/146}

\bibitem[{{Lallement} {et~al.}(2014){Lallement}, {Vergely}, {Valette},
  {Puspitarini}, {Eyer}, \& {Casagrande}}]{Lallement2014}
{Lallement}, R., {Vergely}, J.-L., {Valette}, B., {et~al.} 2014, \aap, 561,
  A91, \dodoi{10.1051/0004-6361/201322032}

\bibitem[{{Lee} \& {Chiang}(2016)}]{Lee:etal:2016}
{Lee}, E.~J., \& {Chiang}, E. 2016, \apj, 817, 90,
  \dodoi{10.3847/0004-637X/817/2/90}

\bibitem[{{Lee} {et~al.}(2014){Lee}, {Chiang}, \& {Ormel}}]{Lee:etal:2014}
{Lee}, E.~J., {Chiang}, E., \& {Ormel}, C.~W. 2014, \apj, 797, 95,
  \dodoi{10.1088/0004-637X/797/2/95}

\bibitem[{{Libralato} {et~al.}(2016){Libralato}, {Nardiello}, {Bedin},
  {Borsato}, {Granata}, {Malavolta}, {Piotto}, {Ochner}, {Cunial}, \&
  {Nascimbeni}}]{Libralato2016}
{Libralato}, M., {Nardiello}, D., {Bedin}, L.~R., {et~al.} 2016, \mnras, 463,
  1780, \dodoi{10.1093/mnras/stw1932}

\bibitem[{{Livingston} {et~al.}(2018){Livingston}, {Dai}, {Hirano}, {Gandolfi},
  {Nowak}, {Endl}, {Velasco}, {Fukui}, {Narita}, {Prieto-Arranz}, {Barragan},
  {Cusano}, {Albrecht}, {Cabrera}, {Cochran}, {Csizmadia}, {Deeg},
  {Eigm{\"u}ller}, {Erikson}, {Fridlund}, {Grziwa}, {Guenther}, {Hatzes},
  {Kawauchi}, {Korth}, {Nespral}, {Palle}, {P{\"a}tzold}, {Persson}, {Rauer},
  {Smith}, {Tamura}, {Tanaka}, {Van Eylen}, {Watanabe}, \&
  {Winn}}]{Livingston2018}
{Livingston}, J.~H., {Dai}, F., {Hirano}, T., {et~al.} 2018, \aj, 155, 115,
  \dodoi{10.3847/1538-3881/aaa841}

\bibitem[{{Livingston} {et~al.}(2019){Livingston}, {Dai}, {Hirano}, {Gandolfi},
  {Trani}, {Nowak}, {Cochran}, {Endl}, {Albrecht}, {Barragan}, {Cabrera},
  {Csizmadia}, {de Leon}, {Deeg}, {Eigm{\"u}ller}, {Erikson}, {Fridlund},
  {Fukui}, {Grziwa}, {Guenther}, {Hatzes}, {Korth}, {Kuzuhara}, {Monta{\~n}es},
  {Narita}, {Nespral}, {Palle}, {P{\"a}tzold}, {Persson}, {Prieto-Arranz},
  {Rauer}, {Tamura}, {Van Eylen}, \& {Winn}}]{Livingston2019}
---. 2019, \mnras, 484, 8, \dodoi{10.1093/mnras/sty3464}

\bibitem[{{Lomb}(1976)}]{Lomb1976}
{Lomb}, N.~R. 1976, \apss, 39, 447, \dodoi{10.1007/BF00648343}

\bibitem[{{Lopez} \& {Fortney}(2013)}]{LopezFortney2013}
{Lopez}, E.~D., \& {Fortney}, J.~J. 2013, \apj, 776, 2,
  \dodoi{10.1088/0004-637X/776/1/2}

\bibitem[{{Luger} {et~al.}(2019){Luger}, {Agol}, {Foreman-Mackey}, {Fleming},
  {Lustig-Yaeger}, \& {Deitrick}}]{exoplanet:luger18}
{Luger}, R., {Agol}, E., {Foreman-Mackey}, D., {et~al.} 2019, aj, 157, 64,
  \dodoi{10.3847/1538-3881/aae8e5}

\bibitem[{{Luger} {et~al.}(2018){Luger}, {Kruse}, {Foreman-Mackey}, {Agol}, \&
  {Saunders}}]{Luger2018}
{Luger}, R., {Kruse}, E., {Foreman-Mackey}, D., {Agol}, E., \& {Saunders}, N.
  2018, \aj, 156, 99, \dodoi{10.3847/1538-3881/aad230}

\bibitem[{{Luhman}(2018)}]{Luhman2018}
{Luhman}, K.~L. 2018, ArXiv e-prints.
\newblock \doarXiv{1811.01359}

\bibitem[{{Malo} {et~al.}(2013){Malo}, {Doyon}, {Lafreni{\`e}re}, {Artigau},
  {Gagn{\'e}}, {Baron}, \& {Riedel}}]{Malo2013}
{Malo}, L., {Doyon}, R., {Lafreni{\`e}re}, D., {et~al.} 2013, \apj, 762, 88,
  \dodoi{10.1088/0004-637X/762/2/88}

\bibitem[{{Mamajek} \& {Hillenbrand}(2008)}]{MamajekHillenbrand2008}
{Mamajek}, E.~E., \& {Hillenbrand}, L.~A. 2008, \apj, 687, 1264,
  \dodoi{10.1086/591785}

\bibitem[{{Mandel} \& {Agol}(2002)}]{mandel2002}
{Mandel}, K., \& {Agol}, E. 2002, \apjl, 580, L171, \dodoi{10.1086/345520}

\bibitem[{{Mann} {et~al.}(2016{\natexlab{a}}){Mann}, {Gaidos}, {Mace},
  {Johnson}, {Bowler}, {LaCourse}, {Jacobs}, {Vanderburg}, {Kraus}, {Kaplan},
  \& {Jaffe}}]{Mann2016a}
{Mann}, A.~W., {Gaidos}, E., {Mace}, G.~N., {et~al.} 2016{\natexlab{a}}, \apj,
  818, 46, \dodoi{10.3847/0004-637X/818/1/46}

\bibitem[{{Mann} {et~al.}(2016{\natexlab{b}}){Mann}, {Newton}, {Rizzuto},
  {Irwin}, {Feiden}, {Gaidos}, {Mace}, {Kraus}, {James}, {Ansdell},
  {Charbonneau}, {Covey}, {Ireland}, {Jaffe}, {Johnson}, {Kidder}, \&
  {Vanderburg}}]{Mann2016}
{Mann}, A.~W., {Newton}, E.~R., {Rizzuto}, A.~C., {et~al.} 2016{\natexlab{b}},
  \aj, 152, 61, \dodoi{10.3847/0004-6256/152/3/61}

\bibitem[{{Mann} {et~al.}(2017){Mann}, {Gaidos}, {Vanderburg}, {Rizzuto},
  {Ansdell}, {Medina}, {Mace}, {Kraus}, \& {Sokal}}]{Mann2017}
{Mann}, A.~W., {Gaidos}, E., {Vanderburg}, A., {et~al.} 2017, \aj, 153, 64,
  \dodoi{10.1088/1361-6528/aa5276}

\bibitem[{{Mann} {et~al.}(2018){Mann}, {Vanderburg}, {Rizzuto}, {Kraus},
  {Berlind}, {Bieryla}, {Calkins}, {Esquerdo}, {Latham}, {Mace}, {Morris},
  {Quinn}, {Sokal}, \& {Stefanik}}]{Mann2018}
{Mann}, A.~W., {Vanderburg}, A., {Rizzuto}, A.~C., {et~al.} 2018, \aj, 155, 4,
  \dodoi{10.3847/1538-3881/aa9791}

\bibitem[{{Mardling} \& {Aarseth}(2001)}]{MardlingAarseth2001}
{Mardling}, R.~A., \& {Aarseth}, S.~J. 2001, \mnras, 321, 398,
  \dodoi{10.1046/j.1365-8711.2001.03974.x}

\bibitem[{{Marley} {et~al.}(2007){Marley}, {Fortney}, {Hubickyj},
  {Bodenheimer}, \& {Lissauer}}]{Marley2007}
{Marley}, M.~S., {Fortney}, J.~J., {Hubickyj}, O., {Bodenheimer}, P., \&
  {Lissauer}, J.~J. 2007, \apj, 655, 541, \dodoi{10.1086/509759}

\bibitem[{{Martin} {et~al.}(2005){Martin}, {Fanson}, {Schiminovich},
  {Morrissey}, {Friedman}, {Barlow}, {Conrow}, {Grange}, {Jelinsky},
  {Milliard}, {Siegmund}, {Bianchi}, {Byun}, {Donas}, {Forster}, {Heckman},
  {Lee}, {Madore}, {Malina}, {Neff}, {Rich}, {Small}, {Surber}, {Szalay},
  {Welsh}, \& {Wyder}}]{Martin2005}
{Martin}, D.~C., {Fanson}, J., {Schiminovich}, D., {et~al.} 2005, \apjl, 619,
  L1, \dodoi{10.1086/426387}

\bibitem[{{Mayor} {et~al.}(2011){Mayor}, {Marmier}, {Lovis}, {Udry},
  {S{\'e}gransan}, {Pepe}, {Benz}, {Bertaux}, {Bouchy}, {Dumusque}, {Lo Curto},
  {Mordasini}, {Queloz}, \& {Santos}}]{Mayor:etal:2011}
{Mayor}, M., {Marmier}, M., {Lovis}, C., {et~al.} 2011, arXiv e-prints,
  arXiv:1109.2497.
\newblock \doarXiv{1109.2497}

\bibitem[{{Mentuch} {et~al.}(2008){Mentuch}, {Brandeker}, {van Kerkwijk},
  {Jayawardhana}, \& {Hauschildt}}]{Mentuch2008}
{Mentuch}, E., {Brandeker}, A., {van Kerkwijk}, M.~H., {Jayawardhana}, R., \&
  {Hauschildt}, P.~H. 2008, \apj, 689, 1127, \dodoi{10.1086/592764}

\bibitem[{{Mordasini}(2013)}]{Mordasini2013}
{Mordasini}, C. 2013, \aap, 558, A113, \dodoi{10.1051/0004-6361/201321617}

\bibitem[{{Mordasini} {et~al.}(2012){Mordasini}, {Alibert}, {Georgy},
  {Dittkrist}, {Klahr}, \& {Henning}}]{Mordasini2012}
{Mordasini}, C., {Alibert}, Y., {Georgy}, C., {et~al.} 2012, \aap, 547, A112,
  \dodoi{10.1051/0004-6361/201118464}

\bibitem[{{Mordasini} {et~al.}(2017){Mordasini}, {Marleau}, \&
  {Molli{\`e}re}}]{Mordasini2017}
{Mordasini}, C., {Marleau}, G.-D., \& {Molli{\`e}re}, P. 2017, \aap, 608, A72,
  \dodoi{10.1051/0004-6361/201630077}

\bibitem[{Newville {et~al.}(2014)Newville, Stensitzki, Allen, \&
  Ingargiola}]{lmfit}
Newville, M., Stensitzki, T., Allen, D.~B., \& Ingargiola, A. 2014, {LMFIT:
  Non-Linear Least-Square Minimization and Curve-Fitting for Python},
  \dodoi{10.5281/zenodo.11813}.
\newblock \url{https://doi.org/10.5281/zenodo.11813}

\bibitem[{{Nguyen} {et~al.}(2012){Nguyen}, {Brandeker}, {van Kerkwijk}, \&
  {Jayawardhana}}]{Nguyen2012}
{Nguyen}, D.~C., {Brandeker}, A., {van Kerkwijk}, M.~H., \& {Jayawardhana}, R.
  2012, \apj, 745, 119, \dodoi{10.1088/0004-637X/745/2/119}

\bibitem[{{Nidever} {et~al.}(2002){Nidever}, {Marcy}, {Butler}, {Fischer}, \&
  {Vogt}}]{nidever2002}
{Nidever}, D.~L., {Marcy}, G.~W., {Butler}, R.~P., {Fischer}, D.~A., \& {Vogt},
  S.~S. 2002, \apjs, 141, 503, \dodoi{10.1086/340570}

\bibitem[{{Obermeier} {et~al.}(2016){Obermeier}, {Henning}, {Schlieder},
  {Crossfield}, {Petigura}, {Howard}, {Sinukoff}, {Isaacson}, {Ciardi},
  {David}, {Hillenbrand}, {Beichman}, {Howell}, {Horch}, {Everett}, {Hirsch},
  {Teske}, {Christiansen}, {L{\'e}pine}, {Aller}, {Liu}, {Saglia},
  {Livingston}, \& {Kluge}}]{Obermeier2016}
{Obermeier}, C., {Henning}, T., {Schlieder}, J.~E., {et~al.} 2016, \aj, 152,
  223, \dodoi{10.3847/1538-3881/152/6/223}

\bibitem[{{Oh} {et~al.}(2017){Oh}, {Price-Whelan}, {Hogg}, {Morton}, \&
  {Spergel}}]{Oh2017}
{Oh}, S., {Price-Whelan}, A.~M., {Hogg}, D.~W., {Morton}, T.~D., \& {Spergel},
  D.~N. 2017, \aj, 153, 257, \dodoi{10.3847/1538-3881/aa6ffd}

\bibitem[{{Olmedo} {et~al.}(2015){Olmedo}, {Lloyd}, {Mamajek}, {Ch{\'a}vez},
  {Bertone}, {Martin}, \& {Neill}}]{Olmedo2015}
{Olmedo}, M., {Lloyd}, J., {Mamajek}, E.~E., {et~al.} 2015, \apj, 813, 100,
  \dodoi{10.1088/0004-637X/813/2/100}

\bibitem[{{Owen} \& {Wu}(2013)}]{OwenWu2013}
{Owen}, J.~E., \& {Wu}, Y. 2013, \apj, 775, 105,
  \dodoi{10.1088/0004-637X/775/2/105}

\bibitem[{{Palla} \& {Stahler}(2002)}]{PallaStahler2002}
{Palla}, F., \& {Stahler}, S.~W. 2002, \apj, 581, 1194, \dodoi{10.1086/344293}

\bibitem[{{Parviainen}(2015)}]{Parviainen2015}
{Parviainen}, H. 2015, \mnras, 450, 3233, \dodoi{10.1093/mnras/stv894}

\bibitem[{{Pecaut} \& {Mamajek}(2013)}]{Pecaut2013}
{Pecaut}, M.~J., \& {Mamajek}, E.~E. 2013, \apjs, 208, 9,
  \dodoi{10.1088/0067-0049/208/1/9}

\bibitem[{{Pepper} {et~al.}(2017){Pepper}, {Gillen}, {Parviainen},
  {Hillenbrand}, {Cody}, {Aigrain}, {Stauffer}, {Vrba}, {David}, {Lillo-Box},
  {Stassun}, {Conroy}, {Pope}, \& {Barrado}}]{Pepper2017}
{Pepper}, J., {Gillen}, E., {Parviainen}, H., {et~al.} 2017, \aj, 153, 177,
  \dodoi{10.3847/1538-3881/aa62ab}

\bibitem[{{Petigura} {et~al.}(2013){Petigura}, {Howard}, \&
  {Marcy}}]{petigura2013b}
{Petigura}, E.~A., {Howard}, A.~W., \& {Marcy}, G.~W. 2013, Proceedings of the
  National Academy of Science, 110, 19273, \dodoi{10.1073/pnas.1319909110}

\bibitem[{{Petigura} {et~al.}(2017){Petigura}, {Sinukoff}, {Lopez},
  {Crossfield}, {Howard}, {Brewer}, {Fulton}, {Isaacson}, {Ciardi}, {Howell},
  {Everett}, {Horch}, {Hirsch}, {Weiss}, \& {Schlieder}}]{Petigura2017}
{Petigura}, E.~A., {Sinukoff}, E., {Lopez}, E.~D., {et~al.} 2017, \aj, 153,
  142, \dodoi{10.3847/1538-3881/aa5ea5}

\bibitem[{{Petigura} {et~al.}(2018){Petigura}, {Crossfield}, {Isaacson},
  {Beichman}, {Christiansen}, {Dressing}, {Fulton}, {Howard}, {Kosiarek},
  {L{\'e}pine}, {Schlieder}, {Sinukoff}, \& {Yee}}]{Petigura2018}
{Petigura}, E.~A., {Crossfield}, I.~J.~M., {Isaacson}, H., {et~al.} 2018, \aj,
  155, 21, \dodoi{10.3847/1538-3881/aa9b83}

\bibitem[{{Petrovich} \& {Tremaine}(2016)}]{PetrovichTremaine2016}
{Petrovich}, C., \& {Tremaine}, S. 2016, \apj, 829, 132,
  \dodoi{10.3847/0004-637X/829/2/132}

\bibitem[{{Quinn} {et~al.}(2012){Quinn}, {White}, {Latham}, {Buchhave},
  {Cantrell}, {Dahm}, {F{\H u}r{\'e}sz}, {Szentgyorgyi}, {Geary}, {Torres},
  {Bieryla}, {Berlind}, {Calkins}, {Esquerdo}, \& {Stefanik}}]{Quinn2012}
{Quinn}, S.~N., {White}, R.~J., {Latham}, D.~W., {et~al.} 2012, \apjl, 756,
  L33, \dodoi{10.1088/2041-8205/756/2/L33}

\bibitem[{{Quinn} {et~al.}(2014){Quinn}, {White}, {Latham}, {Buchhave},
  {Torres}, {Stefanik}, {Berlind}, {Bieryla}, {Calkins}, {Esquerdo}, {F{\H
  u}r{\'e}sz}, {Geary}, \& {Szentgyorgyi}}]{Quinn2014}
---. 2014, \apj, 787, 27, \dodoi{10.1088/0004-637X/787/1/27}

\bibitem[{{Randich} {et~al.}(2001){Randich}, {Pallavicini}, {Meola},
  {Stauffer}, \& {Balachandran}}]{Randich2001}
{Randich}, S., {Pallavicini}, R., {Meola}, G., {Stauffer}, J.~R., \&
  {Balachandran}, S.~C. 2001, \aap, 372, 862,
  \dodoi{10.1051/0004-6361:20010339}

\bibitem[{{Rebull} {et~al.}(2018){Rebull}, {Stauffer}, {Cody}, {Hillenbrand},
  {David}, \& {Pinsonneault}}]{Rebull2018}
{Rebull}, L.~M., {Stauffer}, J.~R., {Cody}, A.~M., {et~al.} 2018, \aj, 155,
  196, \dodoi{10.3847/1538-3881/aab605}

\bibitem[{{Rebull} {et~al.}(2016){Rebull}, {Stauffer}, {Bouvier}, {Cody},
  {Hillenbrand}, {Soderblom}, {Valenti}, {Barrado}, {Bouy}, {Ciardi},
  {Pinsonneault}, {Stassun}, {Micela}, {Aigrain}, {Vrba}, {Somers},
  {Christiansen}, {Gillen}, \& {Collier Cameron}}]{Rebull2016}
{Rebull}, L.~M., {Stauffer}, J.~R., {Bouvier}, J., {et~al.} 2016, \aj, 152,
  113, \dodoi{10.3847/0004-6256/152/5/113}

\bibitem[{{Rizzuto} {et~al.}(2017){Rizzuto}, {Mann}, {Vanderburg}, {Kraus}, \&
  {Covey}}]{Rizzuto2017}
{Rizzuto}, A.~C., {Mann}, A.~W., {Vanderburg}, A., {Kraus}, A.~L., \& {Covey},
  K.~R. 2017, \aj, 154, 224, \dodoi{10.3847/1538-3881/aa9070}

\bibitem[{{Rizzuto} {et~al.}(2018){Rizzuto}, {Vanderburg}, {Mann}, {Kraus},
  {Dressing}, {Ag{\"u}eros}, {Douglas}, \& {Krolikowski}}]{Rizzuto2018}
{Rizzuto}, A.~C., {Vanderburg}, A., {Mann}, A.~W., {et~al.} 2018, \aj, 156,
  195, \dodoi{10.3847/1538-3881/aadf37}

\bibitem[{Salvatier {et~al.}(2016)Salvatier, Wiecki, \&
  Fonnesbeck}]{exoplanet:pymc3}
Salvatier, J., Wiecki, T.~V., \& Fonnesbeck, C. 2016, PeerJ Computer Science,
  2, e55

\bibitem[{{Santerne} {et~al.}(2016){Santerne}, {Moutou}, {Tsantaki}, {Bouchy},
  {H{\'e}brard}, {Adibekyan}, {Almenara}, {Amard}, {Barros}, {Boisse},
  {Bonomo}, {Bruno}, {Courcol}, {Deleuil}, {Demangeon}, {D{\'{\i}}az},
  {Guillot}, {Havel}, {Montagnier}, {Rajpurohit}, {Rey}, \&
  {Santos}}]{Santerne2016}
{Santerne}, A., {Moutou}, C., {Tsantaki}, M., {et~al.} 2016, \aap, 587, A64,
  \dodoi{10.1051/0004-6361/201527329}

\bibitem[{{Scargle}(1982)}]{Scargle1982}
{Scargle}, J.~D. 1982, \apj, 263, 835, \dodoi{10.1086/160554}

\bibitem[{{Schlichting}(2018)}]{Schlichting:2018}
{Schlichting}, H.~E. 2018, {Formation of Super-Earths} (Springer International
  Publishing AG), 141

\bibitem[{{Schlichting} {et~al.}(2015){Schlichting}, {Sari}, \&
  {Yalinewich}}]{Schlichting2015}
{Schlichting}, H.~E., {Sari}, R., \& {Yalinewich}, A. 2015, \icarus, 247, 81,
  \dodoi{10.1016/j.icarus.2014.09.053}

\bibitem[{{Skrutskie} {et~al.}(2006){Skrutskie}, {Cutri}, {Stiening},
  {Weinberg}, {Schneider}, {Carpenter}, {Beichman}, {Capps}, {Chester},
  {Elias}, {Huchra}, {Liebert}, {Lonsdale}, {Monet}, {Price}, {Seitzer},
  {Jarrett}, {Kirkpatrick}, {Gizis}, {Howard}, {Evans}, {Fowler}, {Fullmer},
  {Hurt}, {Light}, {Kopan}, {Marsh}, {McCallon}, {Tam}, {Van Dyk}, \&
  {Wheelock}}]{Skrutskie2006}
{Skrutskie}, M.~F., {Cutri}, R.~M., {Stiening}, R., {et~al.} 2006, \aj, 131,
  1163, \dodoi{10.1086/498708}

\bibitem[{{Slesnick} {et~al.}(2006){Slesnick}, {Carpenter}, {Hillenbrand}, \&
  {Mamajek}}]{Slesnick2006}
{Slesnick}, C.~L., {Carpenter}, J.~M., {Hillenbrand}, L.~A., \& {Mamajek},
  E.~E. 2006, \aj, 132, 2665, \dodoi{10.1086/508937}

\bibitem[{{Smith} {et~al.}(2012){Smith}, {Stumpe}, {Van Cleve}, {Jenkins},
  {Barclay}, {Fanelli}, {Girouard}, {Kolodziejczak}, {McCauliff}, \&
  {Morris}}]{Smith:etal:2012}
{Smith}, J.~C., {Stumpe}, M.~C., {Van Cleve}, J.~E., {et~al.} 2012, \pasp, 124,
  1000, \dodoi{10.1086/667697}

\bibitem[{{Soderblom} {et~al.}(1993){Soderblom}, {Jones}, {Balachandran},
  {Stauffer}, {Duncan}, {Fedele}, \& {Hudon}}]{Soderblom1993}
{Soderblom}, D.~R., {Jones}, B.~F., {Balachandran}, S., {et~al.} 1993, \aj,
  106, 1059, \dodoi{10.1086/116704}

\bibitem[{{Spiegel} \& {Burrows}(2012)}]{SpiegelBurrows2012}
{Spiegel}, D.~S., \& {Burrows}, A. 2012, \apj, 745, 174,
  \dodoi{10.1088/0004-637X/745/2/174}

\bibitem[{{Stumpe} {et~al.}(2012){Stumpe}, {Smith}, {Van Cleve}, {Twicken},
  {Barclay}, {Fanelli}, {Girouard}, {Jenkins}, {Kolodziejczak}, \&
  {McCauliff}}]{Stumpe:etal:2012}
{Stumpe}, M.~C., {Smith}, J.~C., {Van Cleve}, J.~E., {et~al.} 2012, \pasp, 124,
  985, \dodoi{10.1086/667698}

\bibitem[{{Terquem} \& {Papaloizou}(2007)}]{Terquem:Papaloizou:2007}
{Terquem}, C., \& {Papaloizou}, J. C.~B. 2007, \apj, 654, 1110,
  \dodoi{10.1086/509497}

\bibitem[{{Theano Development Team}(2016)}]{exoplanet:theano}
{Theano Development Team}. 2016, arXiv e-prints, abs/1605.02688

\bibitem[{{Thompson} {et~al.}(2018){Thompson}, {Coughlin}, {Hoffman},
  {Mullally}, {Christiansen}, {Burke}, {Bryson}, {Batalha}, {Haas},
  {Catanzarite}, {Rowe}, {Barentsen}, {Caldwell}, {Clarke}, {Jenkins}, {Li},
  {Latham}, {Lissauer}, {Mathur}, {Morris}, {Seader}, {Smith}, {Klaus},
  {Twicken}, {Van Cleve}, {Wohler}, {Akeson}, {Ciardi}, {Cochran}, {Henze},
  {Howell}, {Huber}, {Pr{\v s}a}, {Ram{\'{\i}}rez}, {Morton}, {Barclay},
  {Campbell}, {Chaplin}, {Charbonneau}, {Christensen-Dalsgaard}, {Dotson},
  {Doyle}, {Dunham}, {Dupree}, {Ford}, {Geary}, {Girouard}, {Isaacson},
  {Kjeldsen}, {Quintana}, {Ragozzine}, {Shabram}, {Shporer}, {Silva Aguirre},
  {Steffen}, {Still}, {Tenenbaum}, {Welsh}, {Wolfgang}, {Zamudio}, {Koch}, \&
  {Borucki}}]{Thompson2018}
{Thompson}, S.~E., {Coughlin}, J.~L., {Hoffman}, K., {et~al.} 2018, \apjs, 235,
  38, \dodoi{10.3847/1538-4365/aab4f9}

\bibitem[{{Tokovinin}(2018)}]{Tokovinin2018}
{Tokovinin}, A. 2018, \apjs, 235, 6, \dodoi{10.3847/1538-4365/aaa1a5}

\bibitem[{{Udry} {et~al.}(2003){Udry}, {Mayor}, \& {Santos}}]{Udry2003}
{Udry}, S., {Mayor}, M., \& {Santos}, N.~C. 2003, \aap, 407, 369,
  \dodoi{10.1051/0004-6361:20030843}

\bibitem[{{van Eyken} {et~al.}(2012){van Eyken}, {Ciardi}, {von Braun}, {Kane},
  {Plavchan}, {Bender}, {Brown}, {Crepp}, {Fulton}, {Howard}, {Howell},
  {Mahadevan}, {Marcy}, {Shporer}, {Szkody}, {Akeson}, {Beichman}, {Boden},
  {Gelino}, {Hoard}, {Ram{\'{\i}}rez}, {Rebull}, {Stauffer}, {Bloom}, {Cenko},
  {Kasliwal}, {Kulkarni}, {Law}, {Nugent}, {Ofek}, {Poznanski}, {Quimby},
  {Walters}, {Grillmair}, {Laher}, {Levitan}, {Sesar}, \&
  {Surace}}]{vanEyken2012}
{van Eyken}, J.~C., {Ciardi}, D.~R., {von Braun}, K., {et~al.} 2012, \apj, 755,
  42, \dodoi{10.1088/0004-637X/755/1/42}

\bibitem[{{Vanderburg} {et~al.}(2016){Vanderburg}, {Bieryla}, {Duev},
  {Jensen-Clem}, {Latham}, {Mayo}, {Baranec}, {Berlind}, {Kulkarni}, {Law},
  {Nieberding}, {Riddle}, \& {Salama}}]{Vanderburg2016b}
{Vanderburg}, A., {Bieryla}, A., {Duev}, D.~A., {et~al.} 2016, \apjl, 829, L9,
  \dodoi{10.3847/2041-8205/829/1/L9}

\bibitem[{{Vanderburg} {et~al.}(2018){Vanderburg}, {Mann}, {Rizzuto},
  {Bieryla}, {Kraus}, {Berlind}, {Calkins}, {Curtis}, {Douglas}, {Esquerdo},
  {Everett}, {Horch}, {Howell}, {Latham}, {Mayo}, {Quinn}, {Scott}, \&
  {Stefanik}}]{Vanderburg2018}
{Vanderburg}, A., {Mann}, A.~W., {Rizzuto}, A., {et~al.} 2018, \aj, 156, 46,
  \dodoi{10.3847/1538-3881/aac894}

\bibitem[{{Vogt} {et~al.}(1994){Vogt}, {Allen}, {Bigelow}, {Bresee}, {Brown},
  {Cantrall}, {Conrad}, {Couture}, {Delaney}, {Epps}, {Hilyard}, {Hilyard},
  {Horn}, {Jern}, {Kanto}, {Keane}, {Kibrick}, {Lewis}, {Osborne},
  {Pardeilhan}, {Pfister}, {Ricketts}, {Robinson}, {Stover}, {Tucker}, {Ward},
  \& {Wei}}]{Vogt1994}
{Vogt}, S.~S., {Allen}, S.~L., {Bigelow}, B.~C., {et~al.} 1994, in \procspie,
  Vol. 2198, Instrumentation in Astronomy VIII, ed. D.~L. {Crawford} \& E.~R.
  {Craine}, 362--375

\bibitem[{{Wahhaj} {et~al.}(2010){Wahhaj}, {Cieza}, {Koerner}, {Stapelfeldt},
  {Padgett}, {Case}, {Keller}, {Mer{\'{\i}}n}, {Evans}, {Harvey}, {Sargent},
  {van Dishoeck}, {Allen}, {Blake}, {Brooke}, {Chapman}, {Mundy}, \&
  {Myers}}]{Wahhaj2010}
{Wahhaj}, Z., {Cieza}, L., {Koerner}, D.~W., {et~al.} 2010, \apj, 724, 835,
  \dodoi{10.1088/0004-637X/724/2/835}

\bibitem[{{Waite} {et~al.}(2011){Waite}, {Marsden}, {Carter}, {Hart}, {Donati},
  {Ram{\'{\i}}rez V{\'e}lez}, {Semel}, \& {Dunstone}}]{Waite2011}
{Waite}, I.~A., {Marsden}, S.~C., {Carter}, B.~D., {et~al.} 2011, \mnras, 413,
  1949, \dodoi{10.1111/j.1365-2966.2011.18366.x}

\bibitem[{{Weiss} {et~al.}(2018){Weiss}, {Marcy}, {Petigura}, {Fulton},
  {Howard}, {Winn}, {Isaacson}, {Morton}, {Hirsch}, {Sinukoff}, {Cumming},
  {Hebb}, \& {Cargile}}]{Weiss2018}
{Weiss}, L.~M., {Marcy}, G.~W., {Petigura}, E.~A., {et~al.} 2018, \aj, 155, 48,
  \dodoi{10.3847/1538-3881/aa9ff6}

\bibitem[{{Wichmann} {et~al.}(1996){Wichmann}, {Krautter}, {Schmitt},
  {Neuhaeuser}, {Alcala}, {Zinnecker}, {Wagner}, {Mundt}, \&
  {Sterzik}}]{WKS1996}
{Wichmann}, R., {Krautter}, J., {Schmitt}, J.~H.~M.~M., {et~al.} 1996, \aap,
  312, 439

\bibitem[{{Wichmann} {et~al.}(2000){Wichmann}, {Torres}, {Melo}, {Frink},
  {Allain}, {Bouvier}, {Krautter}, {Covino}, \& {Neuh{\"a}user}}]{Wichmann2000}
{Wichmann}, R., {Torres}, G., {Melo}, C.~H.~F., {et~al.} 2000, \aap, 359, 181

\bibitem[{{Wright} {et~al.}(2011){Wright}, {Drake}, {Mamajek}, \&
  {Henry}}]{Wright2011}
{Wright}, N.~J., {Drake}, J.~J., {Mamajek}, E.~E., \& {Henry}, G.~W. 2011,
  \apj, 743, 48, \dodoi{10.1088/0004-637X/743/1/48}

\bibitem[{{Yee} \& {Jensen}(2010)}]{YeeJensen2010}
{Yee}, J.~C., \& {Jensen}, E.~L.~N. 2010, \apj, 711, 303,
  \dodoi{10.1088/0004-637X/711/1/303}

\bibitem[{{Yu} {et~al.}(2015){Yu}, {Winn}, {Gillon}, {Albrecht}, {Rappaport},
  {Bieryla}, {Dai}, {Delrez}, {Hillenbrand}, {Holman}, {Howard}, {Huang},
  {Isaacson}, {Jehin}, {Lendl}, {Montet}, {Muirhead}, {Sanchis-Ojeda}, \&
  {Triaud}}]{Yu2015}
{Yu}, L., {Winn}, J.~N., {Gillon}, M., {et~al.} 2015, \apj, 812, 48,
  \dodoi{10.1088/0004-637X/812/1/48}

\bibitem[{{Yu} {et~al.}(2017){Yu}, {Donati}, {H{\'e}brard}, {Moutou}, {Malo},
  {Grankin}, {Hussain}, {Collier Cameron}, {Vidotto}, {Baruteau}, {Alencar},
  {Bouvier}, {Petit}, {Takami}, {Herczeg}, {Gregory}, {Jardine}, {Morin},
  {M{\'e}nard}, \& {Matysse Collaboration}}]{Yu2017}
{Yu}, L., {Donati}, J.-F., {H{\'e}brard}, E.~M., {et~al.} 2017, \mnras, 467,
  1342, \dodoi{10.1093/mnras/stx009}

\bibitem[{{Yuan} {et~al.}(2013){Yuan}, {Liu}, \& {Xiang}}]{Yuan2013}
{Yuan}, H.~B., {Liu}, X.~W., \& {Xiang}, M.~S. 2013, \mnras, 430, 2188,
  \dodoi{10.1093/mnras/stt039}

\bibitem[{{Zhang} {et~al.}(2018){Zhang}, {Liu}, {Best}, {Magnier}, {Aller},
  {Chambers}, {Draper}, {Flewelling}, {Hodapp}, {Kaiser}, {Kudritzki},
  {Metcalfe}, {Wainscoat}, \& {Waters}}]{Zhang2018}
{Zhang}, Z., {Liu}, M.~C., {Best}, W.~M.~J., {et~al.} 2018, \apj, 858, 41,
  \dodoi{10.3847/1538-4357/aab269}

\bibitem[{{Ziegler} {et~al.}(2018){Ziegler}, {Law}, {Baranec}, {Morton},
  {Riddle}, {De Lee}, {Huber}, {Mahadevan}, \& {Pepper}}]{Ziegler2018}
{Ziegler}, C., {Law}, N.~M., {Baranec}, C., {et~al.} 2018, \aj, 156, 259,
  \dodoi{10.3847/1538-3881/aad80a}

\end{thebibliography}

\end{document}